%% file: arXiv.tex
\documentclass[journal,twocolumn]{IEEEtran}
\input{input.tex}
\usepackage[table]{xcolor}
\usepackage{epstopdf}
\usepackage{makecell}
\usepackage{array,colortbl,multirow,multicol,booktabs,ctable}
\usepackage{enumerate}

\usepackage{algorithmicx}
\usepackage{algorithm}
\usepackage{amsmath}
\usepackage[noend]{algpseudocode}
\usepackage{float}
\usepackage{soul}
\usepackage{hyperref}
\usepackage{color}
\usepackage{makeidx}
\usepackage{bbm}
\usepackage{graphicx}
\usepackage{caption}
\usepackage{array}
\usepackage{epsfig}
\usepackage{comment}
\usepackage{textcomp}
\usepackage{subcaption}
\usepackage{cite}
\usepackage{balance}

\newcommand{\sref}[1]{{Section}~\ref{#1}}

\newcommand{\pinv}[1]{\ensuremath{#1^{\dagger}}} 	% Moore-Penrose pseudo-inverse

\def\BibTeX{{\rm B\kern-.05em{\sc i\kern-.025em b}\kern-.08em T\kern-.1667em\lower.7ex\hbox{E}\kern-.125emX}}
\begin{document}

\title{Blockage Prediction Using Wireless Signatures: \\ Deep Learning Enables Real-World  Demonstration}

\author{Shunyao Wu, Muhammad Alrabeiah,  Chaitali Chakrabarti, and Ahmed Alkhateeb\\  \thanks{The authors are with the School of Electrical, Computer and Energy Engineering, Arizona State University, (Email: vincentw, malrabei, chaitali, alkhateeb@asu.edu). Part of this work will be submitted to the IEEE International Conference on Communications (ICC) 2022 \cite{Wu2021}.   This work is supported in part by the National Science Foundation under Grant No. 2048021.}}
\maketitle

\begin{abstract} Overcoming the link blockage challenges is essential for enhancing the reliability and latency of millimeter wave (mmWave) and sub-terahertz (sub-THz) communication networks. Previous approaches relied mainly on either (i) multiple-connectivity, which under-utilizes the network resources, or on (ii) the use of out-of-band and non-RF sensors to predict link blockages, which is associated with increased cost and system complexity. In this paper, we propose a novel solution that relies only on in-band mmWave wireless measurements to proactively predict future dynamic line-of-sight (LOS) link blockages.  The proposed solution utilizes deep neural networks and special patterns of received signal power, that we call \textit{pre-blockage wireless signatures} to infer future  blockages. Specifically, the developed machine learning models attempt to predict: (i) If a future blockage will occur? (ii) When will this blockage happen? (iii) What is the type of the blockage? And (iv) what is the direction of the moving blockage? To evaluate our proposed approach, we build a large-scale real-world dataset comprising  nearly $0.5$ million data points (mmWave measurements) for both indoor and outdoor blockage scenarios. The results, using this dataset, show that the  proposed approach can successfully predict the occurrence of future dynamic blockages with more than 85\% accuracy. Further, for the outdoor scenario with highly-mobile vehicular blockages, the proposed model can predict the exact time of the future blockage with less than $80$ms error for blockages happening within the future $500$ms.  These results, among others, highlight the promising gains of the proposed proactive blockage prediction solution which could potentially enhance the reliability and latency of future wireless networks. \end{abstract}


\section{INTRODUCTION}

\IEEEPARstart{R} {}ealizing the high data rate gains of millimeter wave (mmWave) and sub-terahertz (sub-THz) communications requires overcoming a number of key challenges \cite{Rappaport2019,Rappaport2013a} One of these important challenges is the line-of-sight (LOS) link blockages which could cause sudden link failures, impacting the reliability and latency of the mobile networks. This is particularly important at mmWave/sub-THz systems because of (i) their reliance on LOS communications for sufficient receive signal power and (ii) the high penetration loss of these high-frequency signals (high sensitivity to blockages). One promising approach for addressing this challenge is by \textit{proactively} predicting link blockages before they happen \cite{Alkhateeb2018}, thereby allowing the network to make proactive beam switching/basestation hand-off decisions. This has been made possible by the powerful capabilities of machine learning models that can potentially utilize side information and prior observations to predict future blockages \cite{Alkhateeb2018, Alrabeiah2020a}. In this paper, \textbf{we investigate the potential of leveraging \textit{wireless mmWave/sub-THz in-band signatures} to proactively predict future link blockages} and attempt to answer four main questions: (i) Can in-band wireless signals be utilized to predict future blockages? (ii) Can these signals also predict when a blockage will happen in the future? (iii) Can these signals be used to predict the type of the blockage? And (iv) can these in-band wireless signals  be leveraged for blockage direction prediction?

\subsection{Related Work}
Initial solutions for the mmWave blockage challenges focused on multiple-connectivity where a user simultaneously maintains multiple links with multiple base stations \cite{giordani2016multi,polese2017improved,petrov2017dynamic,aziz2016architecture}. For instance,  \cite{giordani2016multi} developed a multiple-connectivity scheme that enables efficient and highly adaptive cell selection in the presence of mmWave channel variability. In \cite{polese2017improved}, an uplink control signaling system combined with a local coordinator is proposed for multiple-connectivity, which enables rapid path switching. A methodology that combines concepts from queuing theory and stochastic geometry for dynamic multiple connectivity is presented in \cite{petrov2017dynamic}. In this integrated framework,  both user- and network-centric performance indicators were quantified to highlight the effectiveness of thr proposed framework in  representative mmWave scenarios. Further, mmWave systems assisted by  sub-6GHz links are developed in  \cite{Mismar2021,aziz2016architecture} to realize fast link switching and efficient multiple-connectivity. While multiple-connectivity is a promising approach for enhancing the mmWave network reliability, it has several drawbacks: (i) These approaches require maintaining multiple simultaneous links which consumes and under-utilizes the network resources, and (ii) these multiple-connectivity solutions \textit{react} to the blockages after they happen and thus they still incur high network latency and do not completely address the reliability and latency challenges associated with LOS link blockages in mmWave and THz systems. 

Motivated by the capabilities of machine learning models in learning from prior observations and predicting outcomes, \cite{Alkhateeb2018} proposed to leverage machine learning model to proactively predict future blockages before they happen. In particular, the authors in \cite{Alkhateeb2018} developed a recurrent neural networks that uses the sequence of beams of the previous few instances to proactively predict future LOS link blockages. The solution in  \cite{Alkhateeb2018}, however, is mainly useful for stationay blockages. To predict future dynamic/moving blockages, the authors in \cite{alrabeiah2020deep,Ali2020} proposed to make use of the sub-6GHz channels to infer close-to-happen link blockages (which originates from the channe mapping concept in space and frequency \cite{Alrabeiah2019a}). Further, \cite{charan2020} proposed to leverage the rich environment senosry data obtained for example from visual sensors (cameras) to proactively prediction link blockages well before they happen. However, the work in \cite{charan2020} requires extra-sensory data (images), and it does not take full advantage of the wireless data. This motivates the development of proactive blockage prediction solutions using only in-band wireless data, which is the focus of this paper. Note that it is important here to differentiate between future blockage prediction and current blockage status prediction/identification (also called LOS/NLOS link identification) which has also been studied in recent papers \cite{choi2017deep,huang2020machine,Alrabeiah2020a}. For example, the work in \cite{choi2017deep,huang2020machine} leveraged machine/deep learning to identify whether the current receive signal corresponds to a LOS or NLOS link. Further,  \cite{Alrabeiah2020a} used a camera feed along with sub-6 GHz channels to identify currently blocked mmWave links. While idenitfying the current link status has important applications, our focus in this paper is on proactive/future blockage prediction, which could bring promising reliability and latency gains for mmWave/sub-THz networks.

\subsection{Contribution} 

In this paper, we propose a novel approach that leverages what we call \textit{pre-blockage wireless signature} to proactively predict future dynamic blockages before they happen. The pre-blockage wireless signature refers to the in-band mmWave receive power patterns (for example power fluctuations over time) that result from the signal reflection/scattering on the moving blockage as it approaches the LOS link. To validate the feasibility of the proposed approach, we build large-scale indoor and outdoor real-world datasets, develop efficient deep learning models, and perform extensive experimental evaluation under realistic wireless communication settings. The main contribution of the paper can be summarize as follows. 
\begin{itemize}
	\item We propose a new approach that leverages only the in-band mmWave signals to proactively predict future dynamic LOS link blockages before they happen. This approach does not require any extra sensors or out-of-band measurements and can work transparently to the normal operation of the communication  systems.
	
	\item We develop efficient machine learning models based on recurrent and convolutional neural networks that utilize the receive in-band signals to achieve four machine learning tasks: (i) Predicting the occurrence of future blockages, (ii) predicting when, in the future, will the the blockage happens, (iii) classifying the type of the blockage, and (iv) classifying the moving blockage direction. 
	
	\item We build a large-scale dataset for indoor and outdoor dynamic blockage scenarios using real-world millimeter wave measurements. Our dataset, that we make publicly available, includes around 460 thousand receive power/beam training measurement samples, forming more than 1400 unique blockage sequences.  The dataset contains several candidate blockages such as humans, bikes, different kinds of vehicles and can be leveraged to investigate multiple machine learning tasks. 	
\end{itemize}

Using the developed dataset, we validate the feasibility of the proposed proactive blockage prediction approach and draw important insights into its performance in indoor and outdoor wireless communication scenarios. Overall, the results highlight the potential of using the developed solution in accurately predicting future blockages, which could enhance the reliability and latency performance of future mmWave and sub-THz networks.

\subsection{Organization}
The rest of the paper is organized as follows: The system and channel models are introduced in \sref{sec:sys_ch}. Section \ref{sec:prob_formul} defines and formulates the four blockage prediction problems.  The proposed machine learning based blockage prediction approach is then presented in Section \ref{sec:deep_learning}. To evaluate the proposed solutions, we built a hardware testbed and collected a large-scale dataset that is described in Sections \ref{subsec:setup} and \ref{subsec:datasets}. Some analysis for this dataset is then presented in \sref{sec:anal_dev_data}. Finally, the experimental results for evaluating the proposed blockage prediction solutions are presented in  \ref{sec:results}. 

%Notation
\textbf{Notation}: We use the following notation throughout this paper: $\bA$ is a matrix, $\ba$ is a vector, $a$ is a scalar, and $\cA$ is a set. $|\bA|$ is the determinant of $\bA$, $\|\bA \|_F$ is its Frobenius norm, whereas $\bA^T$, $\bA^H$, $\bA^*$, $\bA^{-1}$, $\pinv{\bA}$ are its transpose, Hermitian (conjugate transpose), conjugate, inverse, and pseudo-inverse respectively. $[\bA]_{r,:}$ and $[\bA]_{:,c}$ are the $r$th row and $c$th column of the matrix $\bA$, respectively. $\mathrm{diag}(\ba)$ is a diagonal matrix with the entries of $\ba$ on its diagonal. $\bI$ is the identity matrix and $\mathbf{1}_{N}$ is the $N$-dimensional all-ones vector. $\bA \otimes \bB$ is the Kronecker product of $\bA$ and $\bB$, and $\bA \circ \bB$ is their Khatri-Rao product. $\cN(\bm,\bR)$ is a complex Gaussian random vector with mean $\bm$ and covariance $\bR$. $\bbE\left[\cdot\right]$ is used to denote expectation.

%\clearpage
\section{System Model}
\label{sec:sys_ch}

\begin{figure}[t]
	\centering
	\includegraphics[width=\linewidth]{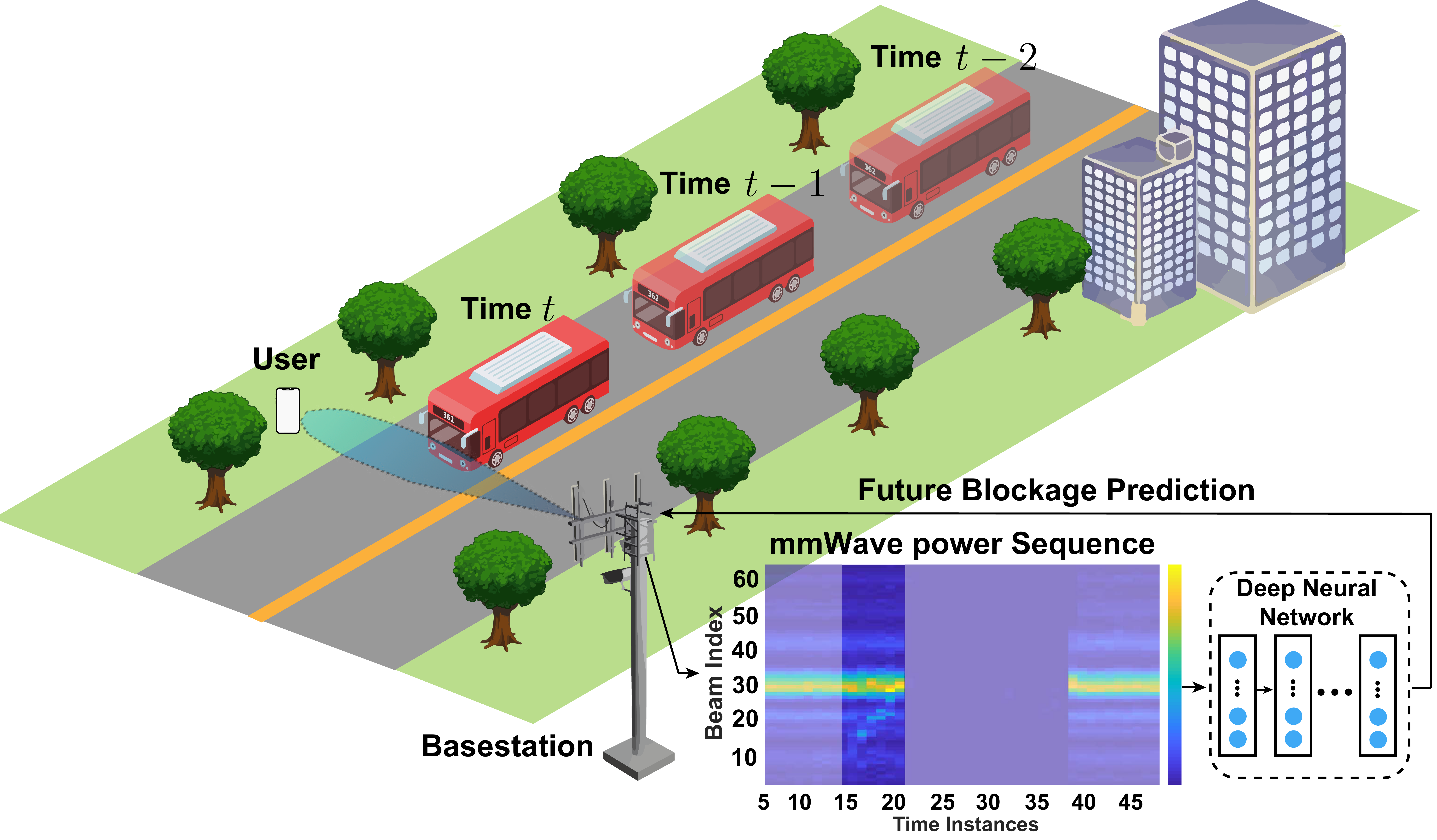}
	\caption{This figure illustrates the overall system model where a mmWave/sub-THz basestation utilizes the received mmWave/sub-THz signal power to enable the proposed proactive dynamic link blockage prediction approach.}
	\label{fig:system diagram}
\end{figure}

The communication system considered in this work is described in \figref{fig:system diagram}. It assumes a mmWave base station serving a static user that is in the vicinity of a possible moving blockage. The base station employs an $M_\mathrm{A}$-element antenna array while the user is single-antenna. To reduce the cost and power consumption, the base station uses a fully-analog beamforming architecture. Further, to simplify the system operation, the base station adopts a  predefined beam-steering codebook $\boldsymbol{\mathcal F} = \{\mathbf f_m\}_{m=1}^{M}$, with $\mathbf f_m\in \mathbb C^{M\times 1}$ defined as
\begin{equation}\label{eq:array_steering}
	\mathbf f_m = \ba \left(\overline{\phi}_m, \overline{\theta}_m\right), 
\end{equation}
where  $\ba$ is the array response vector and $\phi_m, \theta_m$ are the steering azimuth and elevation angles of the $m$th beamforming vector. As will be discussed in \sref{subsec:setup}, the  hardware prototype described in this paper adopts a phased array with a uniform linear array that can steer the beam in the azimuth direction. For this prototype, the adopted beamforming codebook has $M=64$ beamforming vectors, with azimuth angles uniformly quantized in the range $[-\pi/4, \pi/4]$. It is worth mentioning here, though, that the developed blockage prediction approaches in this paper can be applied to more general array architectures. 

To account for the variations of the channel over time, we adopt a block fading channel model where the channel is assumed to be constant over a block of duration $\tau_B$. Further, we adopt an OFDM signal transmission model of $K$  subcarriers. Given these models, we define $\bh_k[t] \in \mathbb C^{M\times 1}$ as the downlink channel at the $k$th subcarrier and as a time instance (block) $t$.  At time instance $t$, if a beamforming vector $\bff_m$ is adopted for the basestation downlink transmission, then the received signal at the user at subcarrier $k$ can be expressed as 
\begin{equation}\label{sig_model}
	r_{k,m}[t] = \bh_k[t]^T \bff_m s_k[t] + n_k[t]
\end{equation}
where $s_k[t]$ is the symbol transmitted on the $k$th subcarrier and $t$th time instance and $n_k[t] \sim \mathcal {CN}(0,\sigma^2)$ is a noise sample.

\section{Problem Formulation} %\hl{time interval 1.blockage happen or not by observising channel power 2.when the blockage happens}
\label{sec:prob_formul}
Proactively identifying the Line of Sight (LOS) link status has significant advantages at both the physical and network layer levels. In this paper, we address the following four problems. (i) How to use the received mmWave signal power information to predict whether there is a blockage in the near future or not? (ii) In case there is a blockage, how to use the received mmWave signal power information to predict when that blockage will occur? (iii) Given the knowledge of an impending blockage, is it possible to predict its size (link blockage time interval)? and, finally, (iv) Could the direction along which the blockage is moving be predicted? A more formal description of each problem is included below

\textbf{Problem 1. Blockage Prediction:} The first problem concerns the proactive prediction of whether a blockage occurs in the near future or not. This problem is formulated as follows. First, we define $t\in \mathbb Z$ as the index of the discrete time instance and $x[t]\in{0, 1}$ as the link status (blocked or unblocked) at the $t$th time instance. $x[t]=1$ indicates a blocked link, i.e., it means that the LOS path between the transmitter and receiver is blocked, while $x[t]=0$ indicates an unblocked link. Further, if at time instance $t$, the single-antenna user transmits a pilot signal to the base station and the base station received this signal using its $M$ beams in the codebook $\boldsymbol{\mathcal F}$, then we define 
\begin{equation} \label{eq:rec}
	\mathbf r[t] = \left[ |r_{1}[t]|^2, \dots, |r_{M}[t]|^2\right]^T
\end{equation} 
as the receive power vector of the $M$ beams at the $t$th instance. Note that $|r_{m}[t]|^2$ is the total power over the $K$ subcarriers, i.e., $|r_{m}[t]|^2=\sum_{k=1}^K |r_{k,m}[t]|^2$. The sequence of these receive power vectors for the $T_{ob}$ previous time instances (observation window), $t-T_{ob}+1$, ... $t$, is then combined in $\mathcal{S}_{ob} $ defined as
\begin{equation}
	\mathcal{S}_{ob} = \{ \mathbf r[t+n] \}_{n=-T_{ob}+1}^{0}
\end{equation} 
\noindent Now, given the observed sequence $\mathcal S_{ob}$, our goal is to predict whether or not a link blockage will occur within a future time interval of $T_P$ instances. We use $b_{T_P}$ to indicate whether there is a blockage occurrence within that future interval or not. More formally, $b_{T_p}$ is defined as follows
\begin{equation} \label{equ:p1_label}
	b_{T_P} = 
	\begin{cases}
		0, &  x[t+n_{p}] = 0 \quad \forall n_{p} \in \{ 1,\dots,T_P \} \\
		1, & \text{otherwise}
	\end{cases}       
\end{equation}
where $1$ indicates the occurrence of a blockage and $0$ is the absence of blockage. The goal of this problem is then to predict $b_{T_p}$ with high accuracy. If $\hat {b}_{T_p}$ denotes the predicted link status, {the objective of Problem 1} is then to maximize the prediction success probability, i.e., 
\begin{equation}
	\textbf{Problem 1:} \ \ \text{maximize} \ \mathbb P_1( \hat {b}_{T_p}= b_{T_p} | \mathcal S_{ob}),
\end{equation} 
by designing an efficient processing technique for the observed sequence of receive power vectors $S_{ob}$.

\begin{figure*}[t]
	\centering
	\includegraphics[width=2.1 \columnwidth]{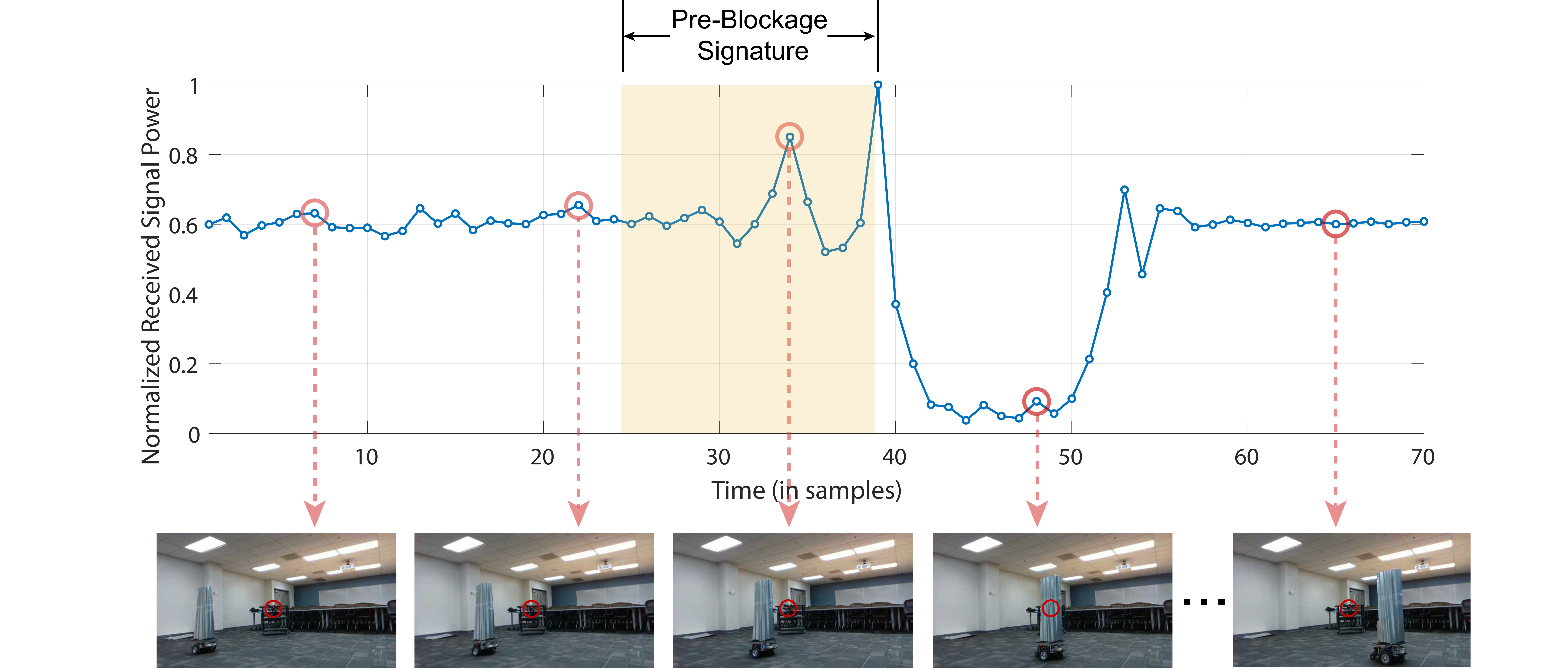}
	\caption{This figure shows an example of indoor pre-blockage signature: The upper subfigure shows the received signal power vs time. The bottom panel shows images captured by the camera.}
	\label{fig:ori_ch_pow}
\end{figure*}

\textbf{Problem 2. Blockage-instance prediction:} Given the observed sequence $\mathcal S_{ob}$ and the knowledge that there is a blockage occurring within the future $T_p$ instances, the second problem deals with identifying the time instance in the future when the blockage would occur. In particular, we want to predict the time instance $n_{p}$ at which $ x[t+n_{p}] = 1$. This represents the exact instance when the blockage occurs within a window of $T_p$ instances. Similar to {{Problem 1}}, given the observed sequence and an incoming blockage, and denoting the predicted instance as $\hat n_{p}$, the objective  of Problem 2 can be defined as follows 
\begin{equation}
	\textbf{Problem 2:} \ \ \text{maximize} \ \mathbb P_2(\hat{n}_{p} = n_{p}|\mathcal S_{ob}, b_{T_p}=1).
\end{equation}  

\textbf{Problem 3. Blockage-severity prediction:} Moving objects in the wireless environment take different shapes, sizes, and speeds, and, hence, the duration of link blockages varies. This problem focuses on proactively identifying the type of the incoming blockage from the perspective of how severe its blockage is. It proposes to define a discrete \textit{blockage-severity index} by which objects are classified. The index is directly derived from the average blockage time interval measured in number of instances (or blocks $\tau_B$) for each object\footnote{Note that this average duration is a function in various visual aspects of the object such as speed, shape, and size.}. Formally and for a specific wireless environment, let $\mathcal V$ define the set of all physical objects that could be present in the environment, e.g., car, bus, tree,... etc\footnote{These objects are labeled based only on discernible visual traits, something similar to that in \cite{ImageNet}}, and let $\bar{\rho}_v$ define the average blockage time interval of object $v\in\mathcal V$. The blockage-severity index for object $v$ could be defined as follows
\begin{equation} \label{equ:p3_label}
	b_{\text{sev},v} = \left\{ \begin{array}{ll}
		1, \quad & \hat{\rho}_v \in T_{D_1},\\
		2, \quad & \hat{\rho}_v \in T_{D_2}, \\
		\vdots & \\
		N_\mathrm{class}, \quad & \hat{\rho}_v \in T_{D_{N_\mathrm{class}}}, \\
	\end{array} \right.
\end{equation}
where $N_\mathrm{class}$ is the total number of classes based on intervals, and $T_{D_1},\dots,T_{D_{N_\mathrm{class}}}$ are consecutive time intervals measured in terms of time instance. Those intervals and their number of classes $N_\mathrm{class}$ are determined by partitioning set $\{\rho_v\}_{v=1}^{|\mathcal V|}$  into $N_\mathrm{class}$ partitions using some distortion metric. More on the partitioning process in a certain wireless environment will be presented in Sections \ref{sec:dev_p3} and \ref{sec:blk_div}. Now, if $\hat{b}_{\text{sev},v}$ denotes the predicted blockage severity index, we can define problem 3 as follows
\begin{equation}
	\textbf{Problem 3:} \ \ \text{maximize} \ \mathbb P_3(\hat{b}_{\text{sev},v} = {b}_{\text{sev},v}|\mathcal S_{ob}, b_{T_p}=1).
\end{equation}  

\textbf{Problem 4. Blockage-direction prediction:} Another interesting dimension of proactive blockage prediction is the prediction of the blockage direction, meaning predicting the direction along which the blocking object is moving. This could be mathematically described as follows. Let $b_{\text{dir}}\in\mathcal D$ be a variable indicating the motion direction of a blocking object where the motion direction is a value from a finite set of possible directions $\mathcal D = \{0,1,\dots,G\}$. Each integer in $\mathcal D$ represents a pre-defined motion direction that depends on the wireless environment and transmitter/receiver positions. For instance, for a transmitter and receiver that are positioned on two opposite sidewalks of a city street, $\mathcal D = \{0,1\}$ where 0 indicates a vehicle traveling left to right with respect to the receiver and 1 indicates the opposite direction. The goal is then to predict $b_{\text{dir}}$ given $\mathcal S_{ob}$ and $b_{T_p}=1$. Similar to the previous problems, the objective of Problem 4 can be written as follows 
\begin{equation}
	\textbf{Problem 4:} \ \ \text{maximize} \ \mathbb P_4( \hat {b}_{\text{dir}} = b_{\text{dir}}| \mathcal S_{ob}).
\end{equation}  
where $\hat b_{\text{dir}}$ is the predicted motion direction.

\begin{figure*}[t]
	\centering
	\includegraphics[width=.99\linewidth]{ 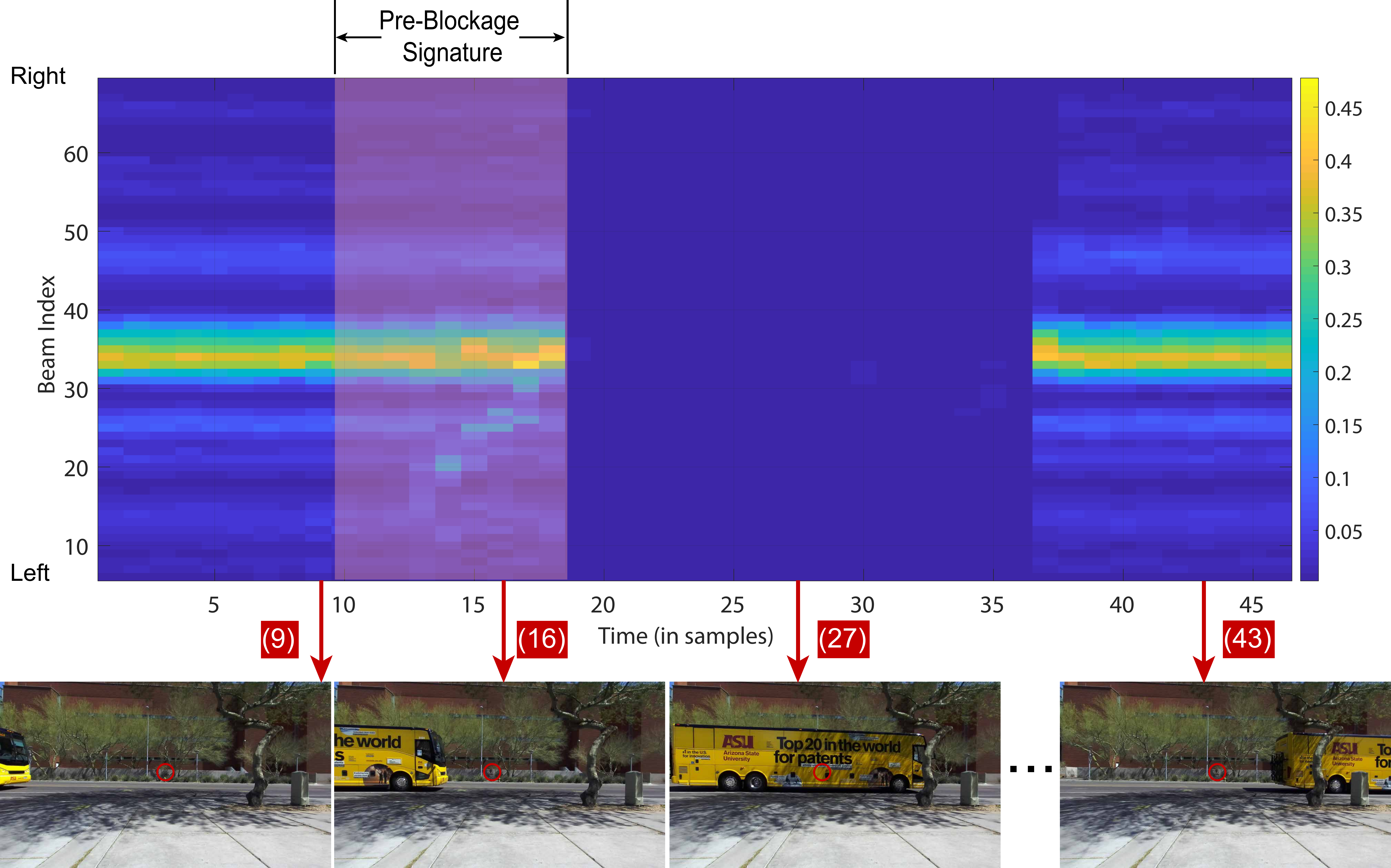}
	\caption{An example of the outdoor pre-blockage signature: The  heatmap image on the top shows the received signal power vs time. The pre-blockage signature is highlighted in red. The bottom set of images are captured by the camera at different time instances from the basestation (receiver) perspective.}
	\label{fig:mat_emp}
\end{figure*}

All four problems are inherently proactive, and this calls for the ability to sense the environment and identify patterns that characterize blocking objects and their behaviors. We propose a machine learning-based approach to tackle the problems. The algorithm is trained to identify patterns in $\mathcal S_{ob}$ and use them to perform a task of interest. Formally, the proposed approach could be described as a function learning problem with a training dataset. More specifically, a function $f_{\Theta} (\mathcal S_{ob})$ is learned by training a machine learning algorithm to estimate a set of parameters $\Theta$ from a training dataset; in this paper, we use $f_{\Theta_1}$, $f_{\Theta_2}$, $f_{\Theta_3}$, $f_{\Theta_4}$ for {Problems 1, 2, 3, 4}, respectively. The training is carried out to maximize the corresponding success probability, i.e., $\mathbb P_1(.)$, $\mathbb P_2(.)$, $\mathbb P_3(.)$, and $\mathbb P_4(.)$.

\section{Proactive Blockage Prediction  using  \\ Wireless Signatures} 
\label{sec:deep_learning}

In this section, we develop our proposed approach for proactively predicting future link blockages using what we call \textit{pre-blockage wireless signatures}. First, we present the key idea in \sref{subsec:key_idea} and then describe the details of the proposed solution in Sections \ref{subsec:model} and \ref{subsec:preprocess}.

\subsection{Key Idea} \label{subsec:key_idea}
Any wireless environment can be broadly broken down into two categories of objects, stationary and dynamic. Both shape the characterization of the wireless channel \cite{goldsmith2005}, and as a result,  the behavior of these objects (whether stationary or dynamic) affect the behavior of the wireless channel. Building on that, the key blockage prediction approaches in this paper are based on the following simple observation: Consider a fixed transmitter and receiver in a certain environment with a LOS path. If an object moves in this environment till it blocks this LOS path, then, during the movement, the object acts like a scatterer for the signal propagating from the transmitter to the receiver. The received signal during this interval will experience a constructive and destructive interference from the LOS ray and the ray scattered on the moving object. Further, the contribution of the moving blockage/scatterer will change as the scatterer approaches the LOS link and before it blocks the link. \textbf{We call this receive signal pattern that precedes the occurrence of a blockage and reflects the behavior of the blocking object the \textit{pre-blockage wireless signature}}.

The pre-blockage signature can be seen in different wireless environments and various wireless communication settings. \figref{fig:ori_ch_pow} is one such example. It depicts normalized received power versus time (in samples) for an indoor wireless environment and a receiver with a quasi-Omni beam pattern. The figure shows how the motion of the blocking object (shown in the bottom row of images) contributes to the power fluctuations at the receiver. What is interesting there is the power pattern right before the link gets blocked, i.e., the region highlighted in red. This region is an example of the pre-blockage signature. 

Such a signature could also be observed in outdoor environments with receivers implementing phased-antenna arrays with narrow beams. \figref{fig:mat_emp} displays an example for the received power across multiple beams as a function of time for a communication system deployed on two opposite sidewalks. The upper sub-figure shows the normalized received power in the form of a two-dimensional heat-map, where the y-axis is the index of the beamforming vector ($m\in\{1,\dots, M\}$) and the x-axis is time (in samples). The heat map shows the effect of the incoming blockage---the campus shuttle in the bottom row of images---on the received power across the codebook beams. In particular, the highlighted red region in the bottom left block displays fluctuations in clear color variation for the beam indices 1 to 30, indicating received power.

\begin{figure*}[t!]
	\centering
	\begin{subfigure}[]{.424\linewidth}
		\centering
		\includegraphics[width=1\linewidth]{ 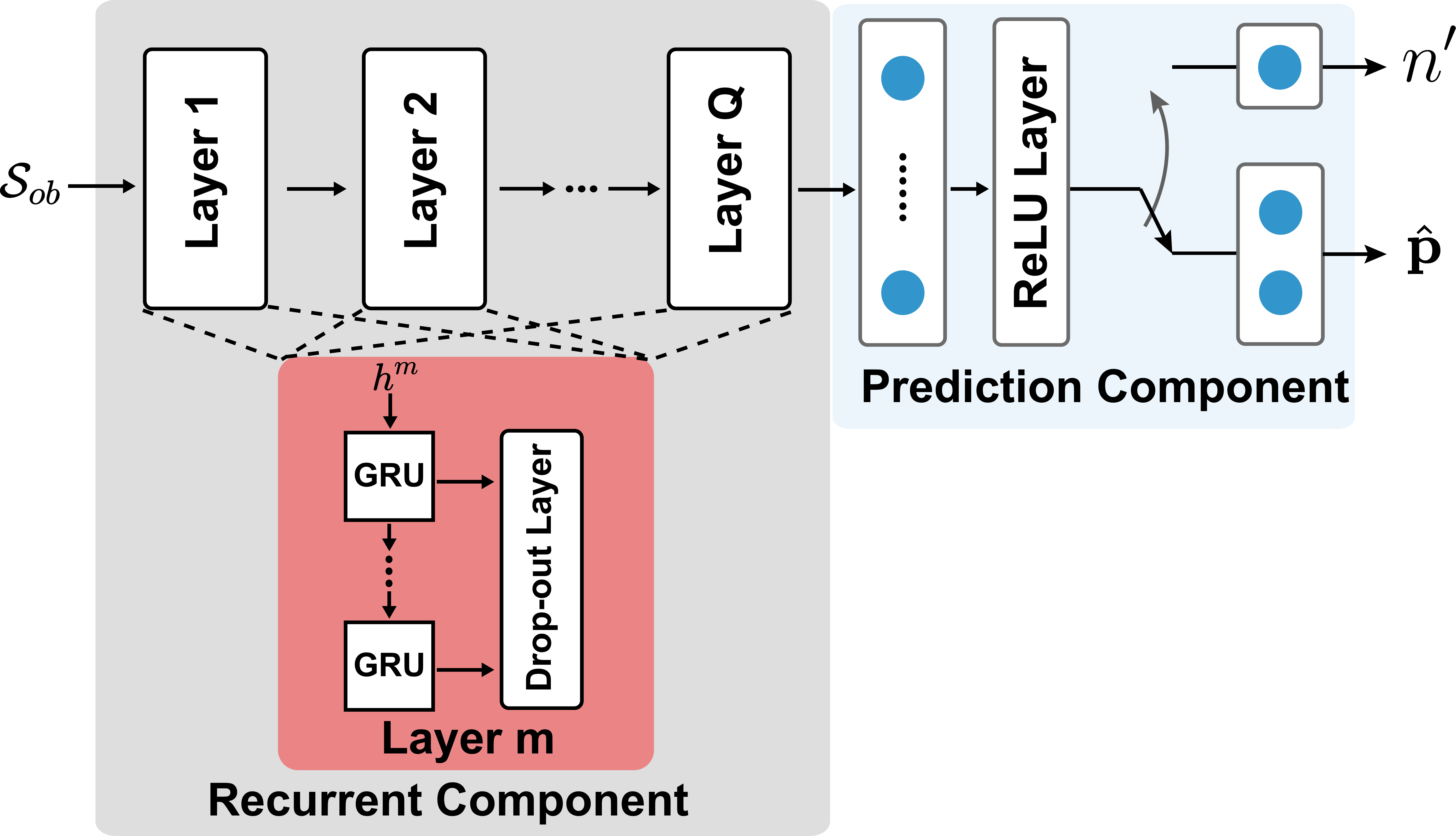}
		\caption{}
		\label{fig:RNN_arch}
	\end{subfigure}
	\begin{subfigure}[]{.53\linewidth}
		\centering
		\includegraphics[width=1\linewidth]{ 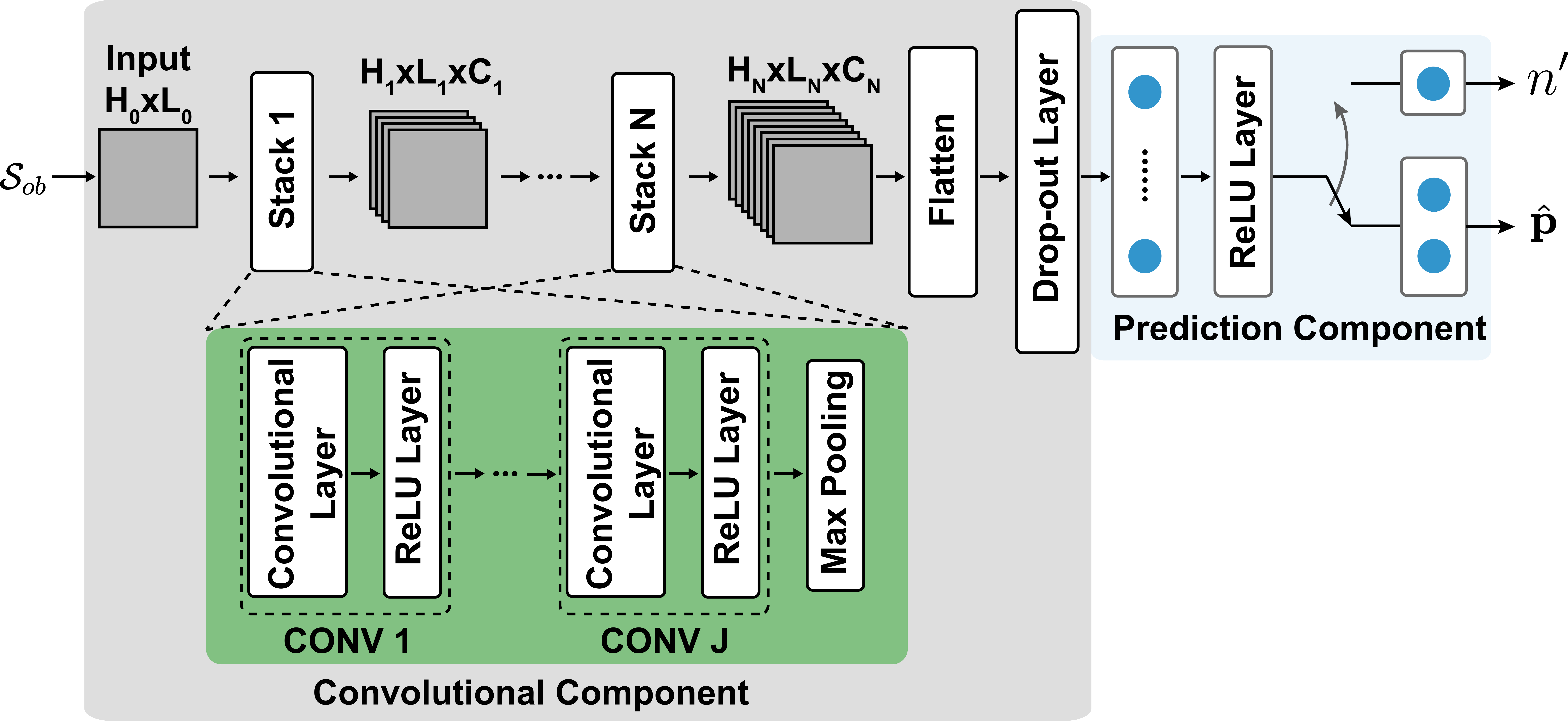}
		\caption{}
		\label{fig:CNN_arch}
	\end{subfigure}
	\caption{(a) The proposed RNN architecture to predict the link blockage status. It consists of two main components: (i) the recurrent component, and (ii) the prediction component. (b) The overall CNN architecture to predict the link status as well as type and moving direction of the blockage. The two main components are: (i) the convolutional component, and (ii) the prediction component.}
\end{figure*}

These two examples, indoor case in \figref{fig:ori_ch_pow} and outdoor case in \figref{fig:mat_emp}, show the existence of the pre-blockage signature. This signature conveys information not only on the blockage occurrence but also its type and direction. Based on that, we propose to utilize this signature in predicting future link blockages in mmWave/THz systems. Further, and given the complex nature of these signatures, we leverage machine learning (and in particular deep learning models) to learn and utilize them. Deep learning has recently established itself as a powerful learning approach for many applications, such as computer vision \cite{resnet}\cite{Liu2019}\cite{Yolo} and speech recognition \cite{SpeechRecog, SpeechRecog2}. With this motivation, we adopt deep neural networks as the core of the proposed solutions for the blockage prediction problems in Section \ref{sec:prob_formul}. The following two subsections present our deep neural network-based approach starting from the network choice, architecture, and processing pipeline.

\subsection{Deep Learning Model}  \label{subsec:model}
We take advantage of the recent advances in Convolutional Neural Networks (CNNs) and Recurrent Neural Networks (RNNs) \cite{DLBook}. The former has been setting the state-of-the-art in computer vision \cite{resnet,Liu2019,Yolo} while the latter has made some breakthroughs in natural language processing \cite{SpeechRecog,SpeechRecog2,ma2016}. As such we propose a solution based on each algorithm below.

\textbf{Recurrent Neural Network Architecture:}  Learning the pre-blockage signature from a sequence of observed received signal power vectors requires a neural network that could process input data samples over time such as recurrent neural networks. We design a Gated Recurrent Unit (GRU) network \cite{2014gru} with $Q$-layers that takes in a sequence of observed received signal powers (i.e., $\mathcal S_{ob}$) and learns to predict the link status $b_{T_p}$. \figref{fig:RNN_arch} depicts the schematic of such a network. Each layer in the network comprises of $T_{ob}$ GRUs, where $T_{ob}$ is the observation time interval. The output of the GRU of the last layer is fed to a Fully Connected (FC) layer followed by either a classifier for {{Problems 1, 3, 4}} or a regressor for {{Problem 2}}. The classifier outputs a probability vector ($\hat{\mathbf p}$) of whether the link status is blocked or not in $T_P$ future time instances. For {{Problem 2}}, the regressor outputs the predicted time instance $\hat{n}^{\prime}$ indicating the time instance when the blockage will occur. 

\textbf{Convolutional Neural Network Architecture:}  
With $M$ beams, the observed received signal power has a structure of a matrix rather than a vector. A convolutional neural network~\cite{simonyan2014CNN} is applied to recognize the pre-blockage signature in the power matrix and used to predict the link status as well as the type of the blockages (high impact or low impact). The architecture of the CNN is shown in \figref{fig:CNN_arch}. In this figure, $H$ and $L$ denote the number of rows and columns of the received power matrix while $C$ denotes the number of channels of the intermediate output of the convolutional layer. Our CNN architecture consists of a convolutional component followed by a prediction component. In the convolutional component, the first stack (Stack 1) takes the input observation power matrix (i.e., $\mathcal S_{ob}$) whose dimension is $H_0 \times L_0$ and passes the output with the dimension $H_1 \times L_1 \times C_1$ to the next stack. We design N similar stacks and the last stack (Stack N) is followed by a flatten layer (a flatten layer converts the 2-D matrix to 1-D vector). Each stack contains $J$ convolutional blocks, each of which consists of a convolutional layer and a ReLU layer, and a max-pooling layer that occurs at the end of the stack. The prediction component is identical to the one in the RNN architecture. We specify all the parameters of our neural networks in Section \ref{subsec:net_train}.

\subsection{Pre-processing Pipeline and Loss Functions} \label{subsec:preprocess}
The input data are first pre-processed before feeding them to the neural network models to improve the learning efficiency  \cite{EffBackProp}. For that, we choose to standardize the inputs by subtracting the mean $\mu$ of the dataset and dividing by its standard deviation $\sigma$. Let $\mathbf A\in\mathbb R^{U\times M}$ be the dataset matrix with $U$ rows and $M$ columns to represent power sample data of $U$ data points. Each row represents a data point and $M$ is the size of the codebook as well as the number of received power samples of each data point (for each beam, we calculate the total received power over subcarriers). Formally, the standardized dataset $\mathbf {\hat{A}}$ is calculated as 

%XXX A is just one data point
\begin{equation}
	\mathbf {\hat{A}}_{u,m} = \frac{\mathbf {A}_{u, m} - \mu}{\sigma},
\end{equation}
for all $u  \in \{1, ..., U\}$ and $m \in \{1, ..., M\}$, with the mean and standard deviation given by 
\begin{equation}
	\mu = \frac{1}{M \times U} \sum_{u=1}^{U}\sum_{m=1}^{M}\mathbf{A}_{u, m},
\end{equation} 
and
\begin{equation}
	\sigma =\left(\frac{1}{M\times U}\sum_{u=1}^{U}\sum_{m=1}^{M}(\mathbf{A}_{u, m} - \mu)^2 \right)^{\frac{1}{2}}.
\end{equation}

\begin{figure}[t]
	\centering
	\includegraphics[width=\linewidth]{ 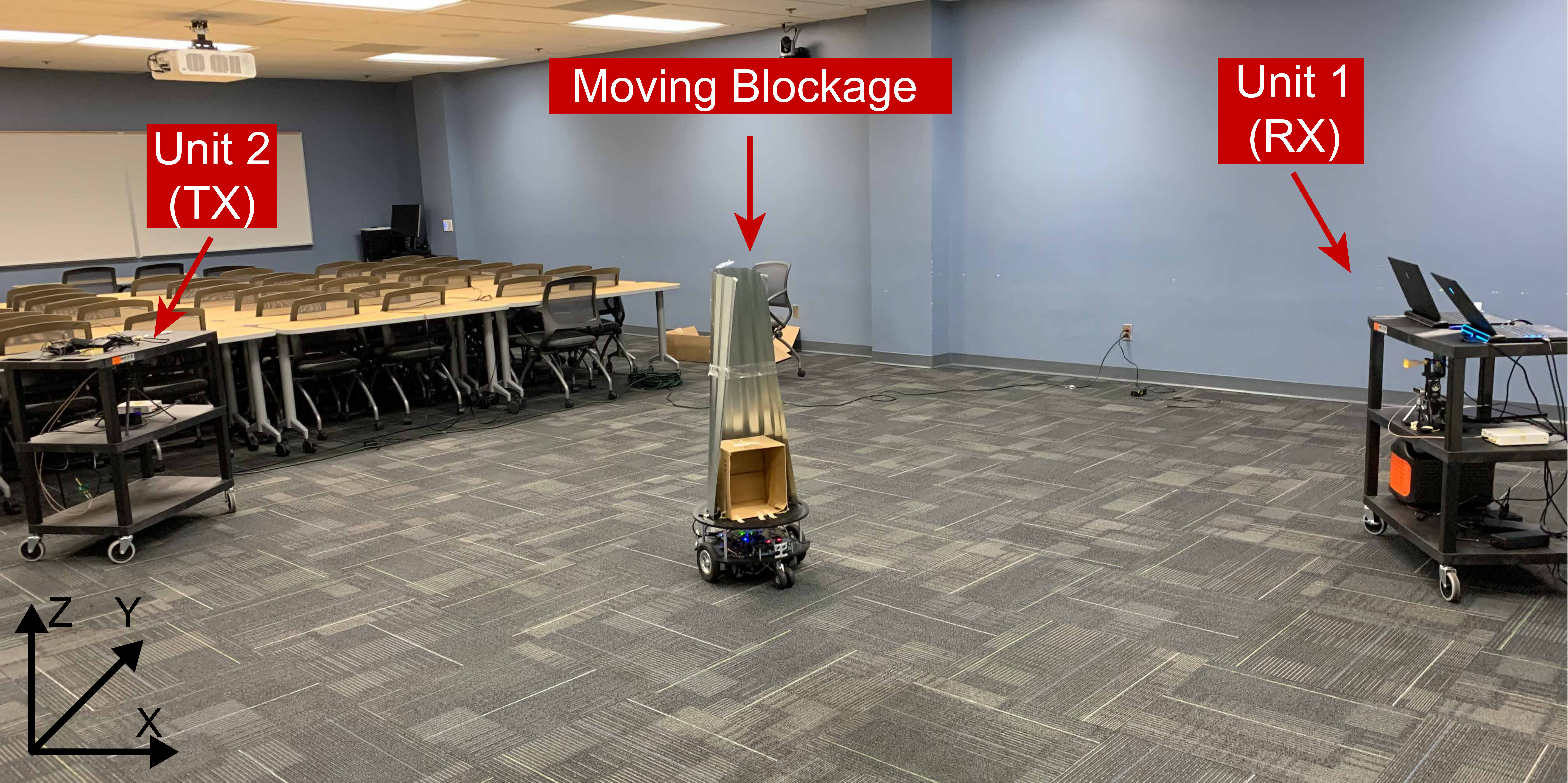}
	\caption{ The figure shows the data collection setup for Scenario 16 (indoor) using the DeepSense Testbed 2. In this scenario, a blockage moves along a trajectory between the transmitter (Unit 2) and the receiver (Unit 1).}
	\label{fig:exp_setup}
\end{figure}

\textbf{Training loss:} For the {{prediction problems (Problems 1, 3, 4)}}, the future link-status, blockage type and moving direction prediction is posed as a classification problem, in which the classifier attempts to determine whether the link status, the type of blockages or the moving direction of the blockages within the future time interval. As such, the network training is performed with a cross entropy loss function \cite{DLBook} computed over the outputs of the network
\begin{equation}
	l_{\text{CH}} =  \sum_{k = 1}^{K} p_{k}\log{\hat{p}_{k}},
\end{equation}
where $K$ is the total number of categories in our problems, $\mathbf p = [p_1, p_2, ...,p_K]^T$ is the one-hot vector, $p_{k}$ represents the $k$th categorical variable as binary, the category with highest probability is encoded as 1 others are encoded as 0's. For example, in {Problem 1}, we have two link status: ($K$ = 2) $b_{T_p} = 0$ for absence of blockage and $b_{T_p} = 1$ for occurrence of blockage, and $[1,0]^T$ is for the case when $b_{T_p} = 0$ and $[0,1]^T$ is for the case when $b_{T_p} = 1$.  $l_{\text{CH}}$ is the training loss computed for one data point.

For the {{regression problem (Problem 2)}}, we pose it as a problem of predicting the blockage instance. Our model tries to determine the exact time instance at which the blockage occurs. We use Mean Square Error (MSE) loss as training function. In formal terms, we aim to minimize the difference between the predicted instance and groundtruth instance \cite{DLBook}:
\begin{equation}
	l_{\text{MSE}} =  (n^{\prime (u)}- \hat{n}^{\prime (u)})^2
	%\frac{\sum_{u = 1}^{B} (n^{\prime (u)}- \hat{n}^{\prime (u)})^2} {B},
\end{equation}
where $n^{\prime (u)}$ and $\hat{n}^{\prime (u)}$ are ground truth time instance and predicted time instance, respectively.

\begin{figure}[t]
	\centering
	\begin{subfigure}{0.49\textwidth}
		\centering
		\includegraphics[width=\linewidth]{ 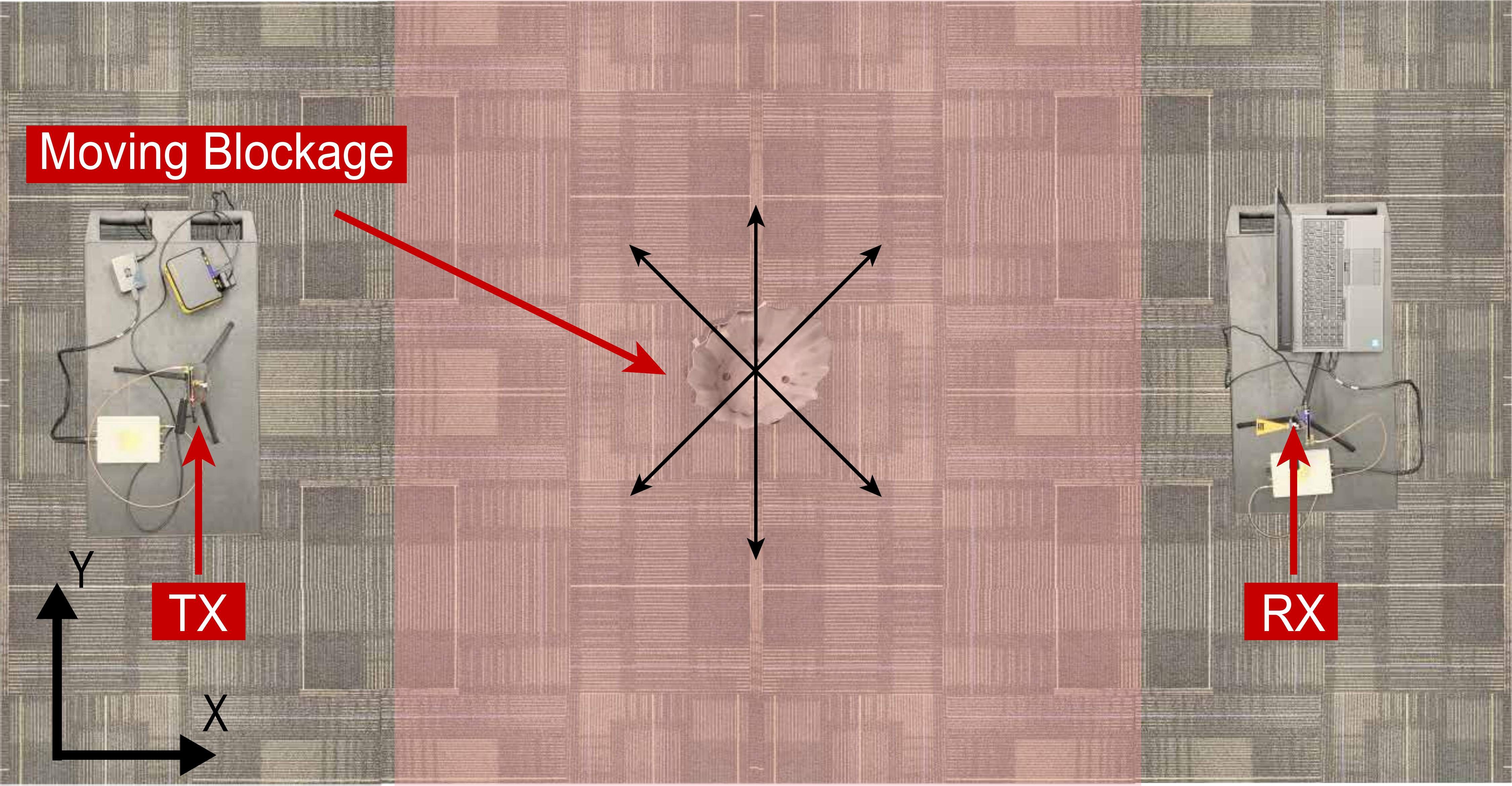}
		\caption{ }
		\label{fig:top_view}
	\end{subfigure}
	\begin{subfigure}{0.49\textwidth}
		\centering
		\includegraphics[width=\linewidth]{ 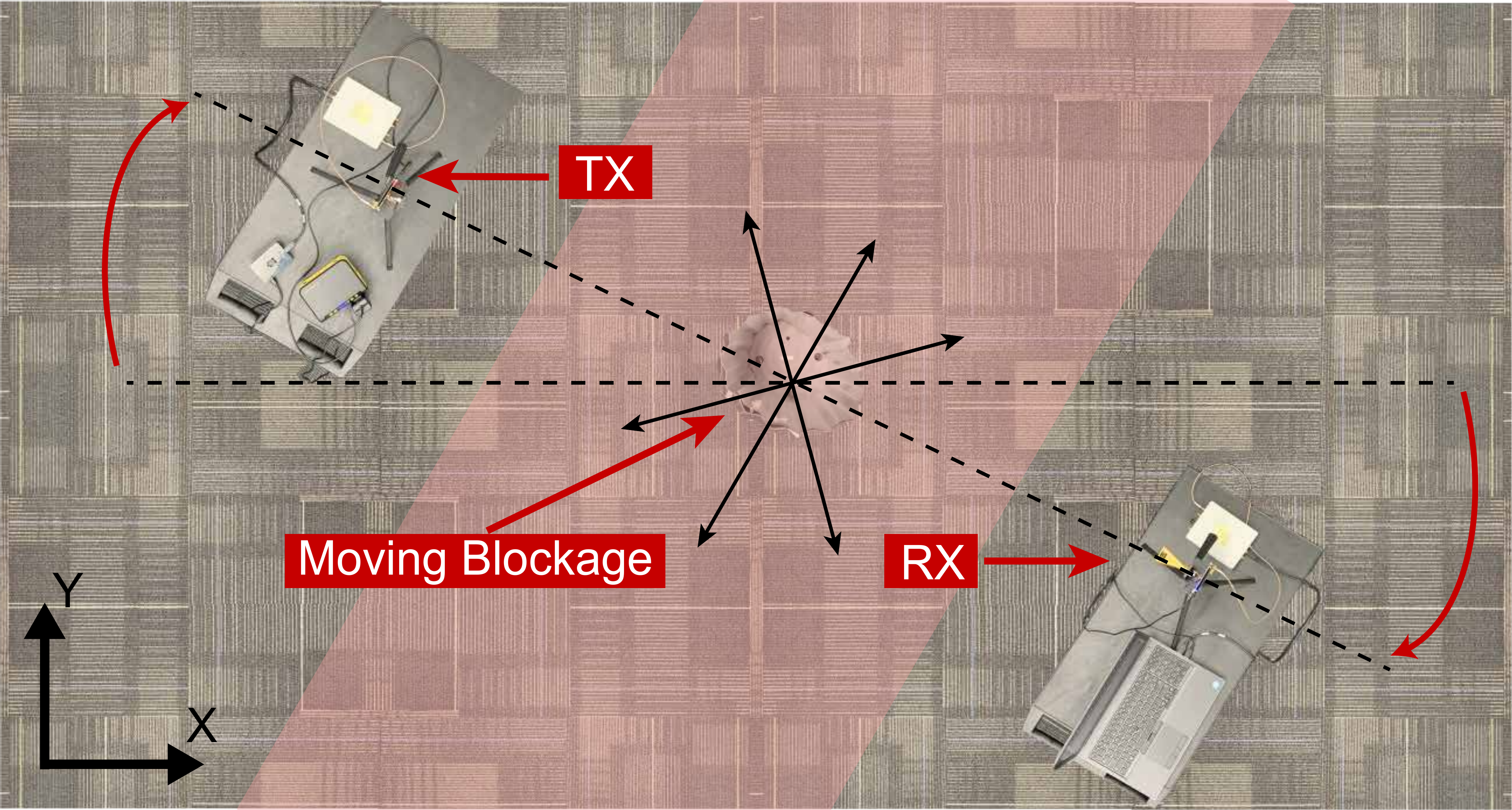}
		\caption{ }
		\label{fig:top_view_rot}
	\end{subfigure}
	%\vspace{-0.2cm}
	\caption{A top-view of the experimental set up for Scenario 16. (a) Trajectory of the moving blockage and the relative positions of transmitter (TX), receiver (RX) and moving blockage. (b) Rotated TX-RX setup.}
	\label{fig:ind_setup}
\end{figure}

%%%%%%%%%%%%%%%%%%%%%%
\section{Experimental Setup and Scenarios}\label{subsec:setup}
Studying the value of pre-blockage signatures and deep learning in tackling the proposed four problems mandates the construction of measurement-based development datasets. They should be composed of data samples that reflect the pre-blockage signature discussed in Section \ref{subsec:key_idea} in different wireless communication settings. To construct such datasets, we follow the footsteps of the DeepSense 6G dataset \cite{DeepSense} and build a couple of DeepSense-compatible testbeds. The new testbeds are deployed in two different wireless environments to collect real-world multimodal measurements and construct what will be henceforth called \textit{seed datasets}. These measurements introduce two new scenarios, namely indoor scenario (Scenario 16) and outdoor scenario (Scenarios 17-22), to the DeepSense 6G dataset. The following three sections will discuss the new testbeds, the communication scenarios where they are deployed, and provide a short analysis of the collected seed datasets.
\begin{figure*}[t!]
	\centering
	\includegraphics[width=0.8\linewidth]{ 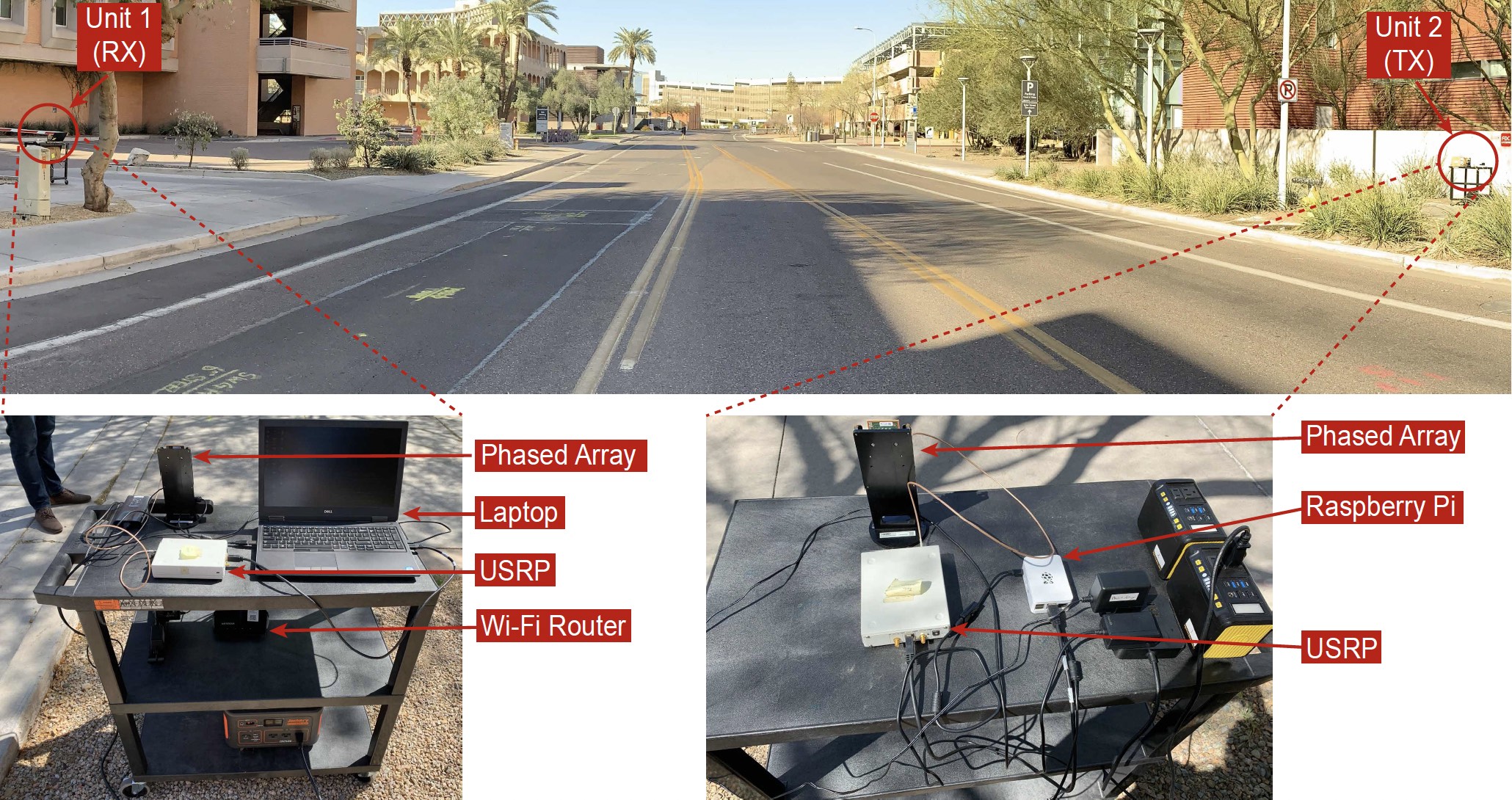}
	\caption{This figure depicts the data collection setup for Scenarios 17-22 (outdoor). The upper figure illustrates the street view and the location of Units 1 and 2. The lower left subfigure  shows the hardware setup at Unit 1 and the right subfigure  shows the hardware setup at  Unit 2.}
	\label{fig:exp_setup_outdoor}
\end{figure*}

\begin{figure*}[t]
	\centering
	\begin{subfigure}{0.32\textwidth}
		\centering
		\includegraphics[width=\linewidth]{ 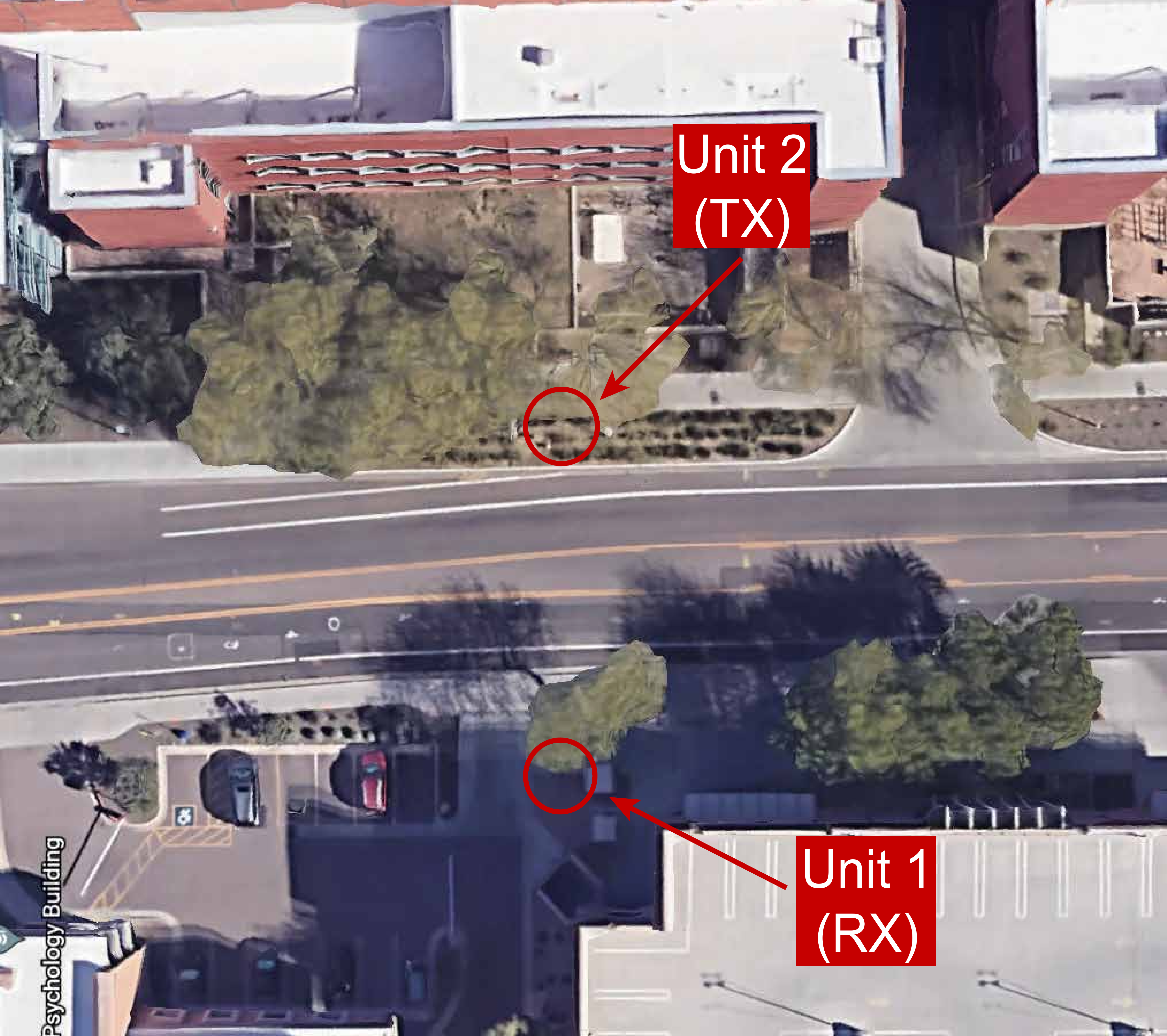}
		\caption{ }
		\label{fig:top_view1}
	\end{subfigure}
	\begin{subfigure}{0.32\textwidth}
		\centering
		\includegraphics[width=\linewidth]{ 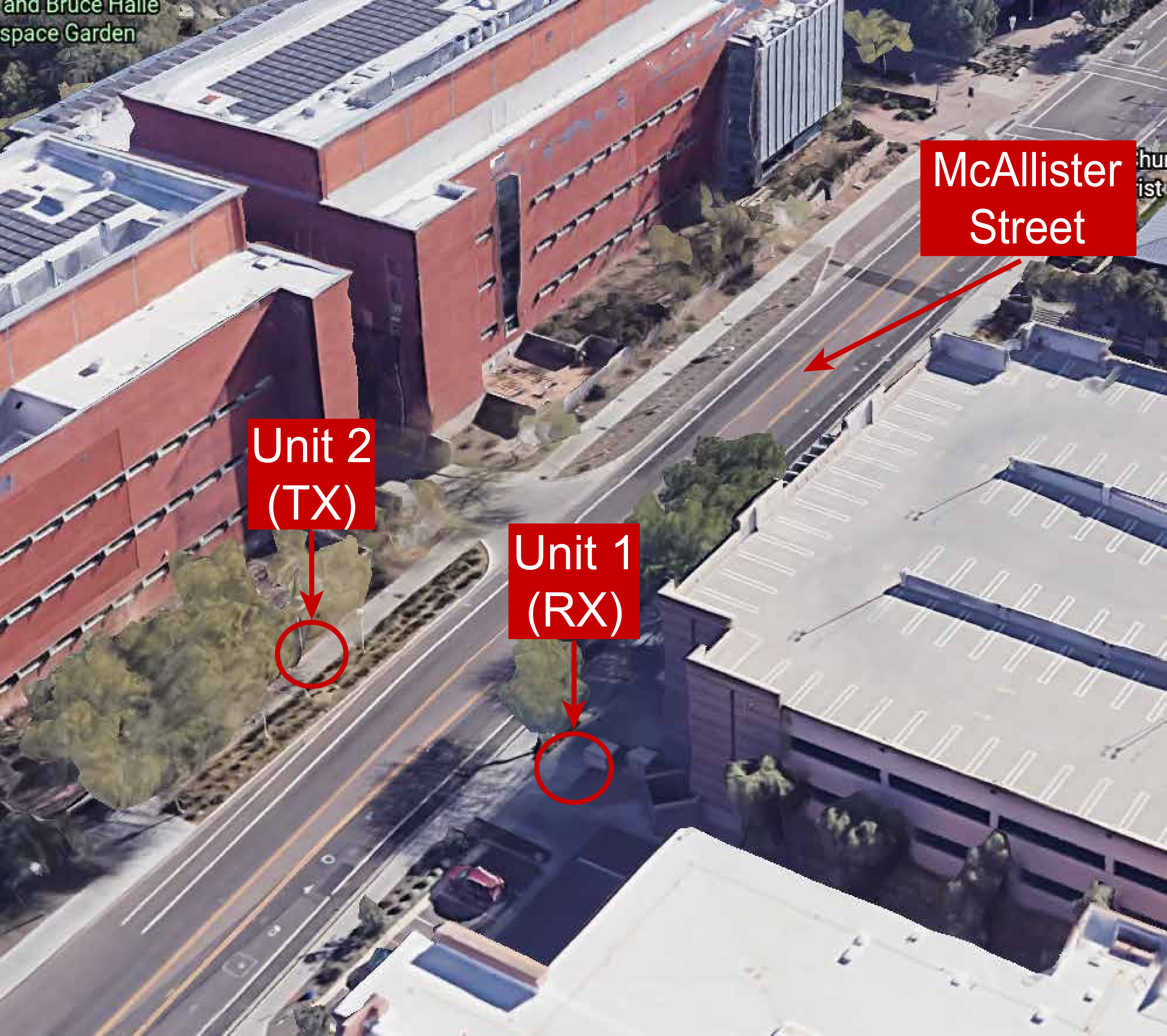}
		\caption{ }
		\label{fig:rx_view2}
	\end{subfigure}
	\begin{subfigure}{0.32\textwidth}
		\centering
		\includegraphics[width=\linewidth]{ 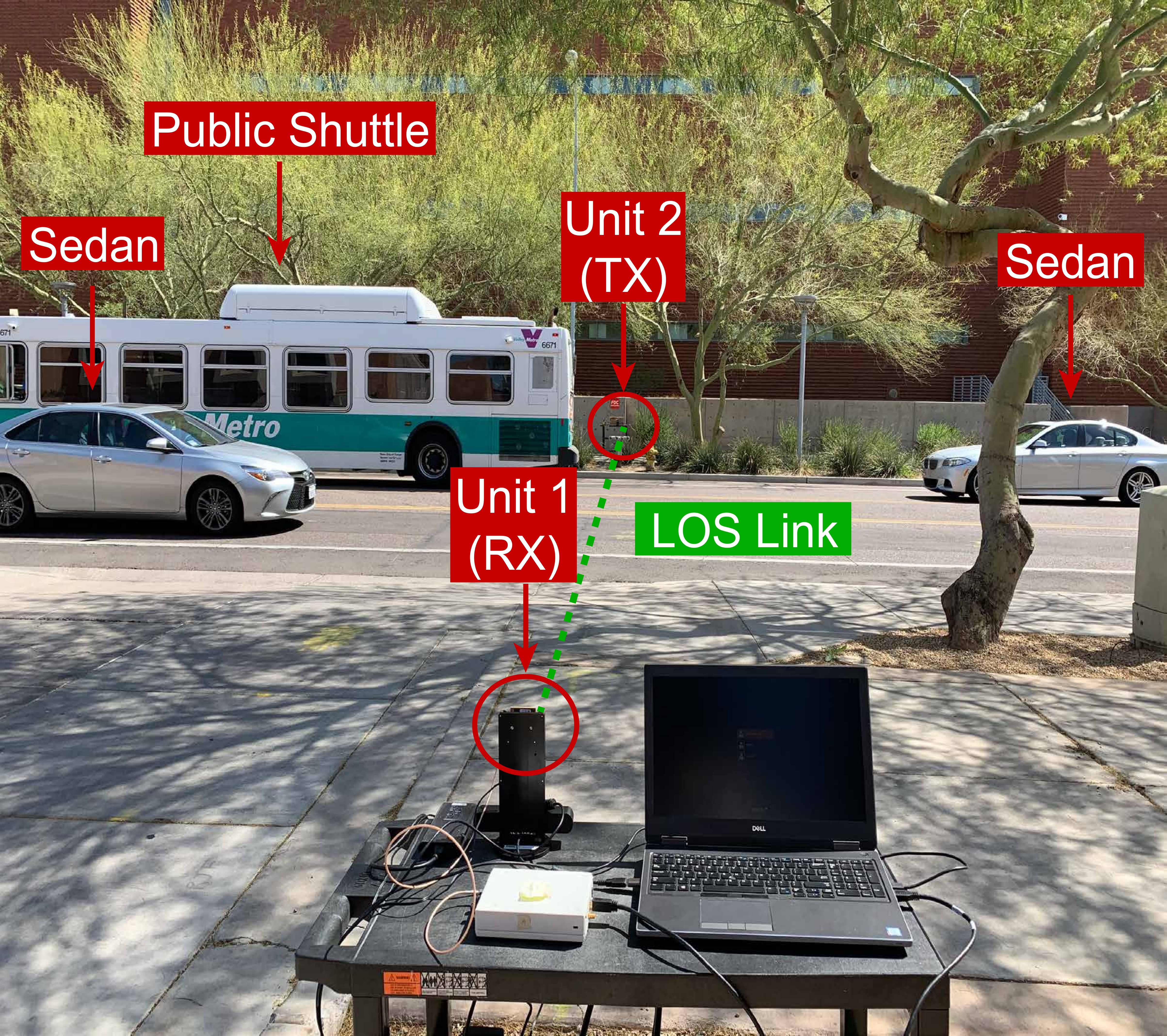}
		\caption{ }
		\label{fig:rx_view}
	\end{subfigure}
	%\vspace{-0.2cm}
	\caption{The data collection setup of Scenarios 17-22 (outdoor). The left subfigure shows the exact locations of Unit 1 and Unit 2; the middle subfigure shows the full view of the street and surroundings; the right subfigure shows the traffic and street view from Unit 1 perspective}
	\label{fig:scenario16}
\end{figure*}

\subsection{Testbed Description}
The DeepSense 6G dataset framework \cite{DeepSense} defines a generic structure  for sensing/communication datasets where a number of units, each equipped with a set of sensors, collect co-existing sensory/communication data. Following this structure, we build two new DeepSense testbeds, namely Testbed 2 and 3, each of which is composed of two stationary units. The following is a description of the two testbeds
\begin{itemize}
	\item \textbf{DeepSense Testbed 2:} It has two stationary units, namely Unit 1  and Unit 2, as shown in \figref{fig:exp_setup}. Unit 1 hosts a 60GHz mmWave receiver equipped with a 10-degree beamwidth (20dBi) horn antenna and an RGB camera, while Unit 2 hosts a 60GHz mmWave transmitter with an omni-directional antenna. The 60GHz transmitter and receiver are controlled by two laptops and the transmit/receive signals from the laptops and the transmitter/receiver pass through two 2901 NI USRPs. The two units coordinate the data collection using an IEEE 802.11-based control/communication channel.  The transmit/receive waveform adopts an OFDM structure with 20MHz bandwidth and 64 subcarriers. The total transmit power from Unit 2 is limited to {30dBm}.  
	
	\item \textbf{DeepSense Testbed 3:} It has a similar architecture to that of DeepSense Testbed 2 and only differs in the type of the mmWave TX and RX equipment. Unit 1 and 2 at Testbed 3 host 60GHz mmWave phased arrays, each of which has a {16-element} uniform linear array (ULA). Each unit has one of those phased arrays and a laptop to control the array. The  transmit/receive signals from the laptops and the transmitter/receiver phased arrays  pass through two 2901 NI USRPs. The transmit/receive waveform adopts an OFDM structure with 20MHz bandwidth/64 subcarriers and the total transmit power from Unit 2 (before the array gain) is around 20dBm.  
	
\end{itemize}

\subsection{Scenarios} 
Using the described testbeds, we introduce the following two new scenarios to the DeepSense 6G dataset. 

\textbf{1) Scenario 16--Conference Room Moving Blockage:}
\label{subsubsec:indoor}
This is an indoor scenario where the DeepSense Testbed 2 and a moving metal blockage in the shape of a cylinder are deployed in a large conference room, as depicted in \figref{fig:exp_setup}. The transmitter and receiver are placed facing each other (have LOS connection) with a separating distance of $\approx 8$ m between them. The blockage moves back and forth between the transmitter and receiver in two straight trajectories spaced 1 m apart and at a speed of 0.0625 m/s. In order to get different signal propagation conditions and increase the variance in the collected measurements, the testbed is rotated around the center of the blockage path on the x-y plane as shown in \figref{fig:ind_setup}, and the blockage is set to move again between the TX and RX in two trajectories with the same 1 m spacing. The testbed collects data samples at a rate of 1.13 samples/s. Each data sample contains (i) an RGB image and (ii) the receive power value, both collected from Unit 1.  The important aspects of the scenario and its data are summarized in  \tabref{tbl:scen_15}, which follows the standard format of the DeepSense 6G dataset \cite{DeepSense}.

\begin{table}[t]
	\caption{Indoor Scenario (Scenario 16--Conference Room)}
	\label{table}
	\centering
	\setlength{\tabcolsep}{5pt}
	\renewcommand{\arraystretch}{1.4}
	\begin{tabular}{|c|c|}
		\hline\hline
		\textbf{Testbed}             & 2                            \\ \hline
		\textbf{Number of Instances} & 11060 (from 158 trajectories)                        \\ \hline
		\textbf{Number of Units}     & 2 \\ \hline
		\textbf{Total Data Modalities}     & \thead{RGB images, received power} \\ \hline \hline
		\multicolumn{2}{|c|}{\textbf{Unit 1}} \\ \hline
		\textbf{Type} & Stationary \\ \hline
		\textbf{Hardware elements} & \thead{RGB camera, mmWave horn-antenna receiver} \\ \hline
		\textbf{Data Modalities} & \thead{RGB images, received power}  \\
		\hline \hline
		\multicolumn{2}{|c|}{\textbf{Unit 2}} \\ \hline
		\textbf{Type} & Stationary \\ \hline
		\textbf{Hardware elements} & mmWave omni-directional transmitter \\ \hline
		%	\textbf{Container Number} & 2 \\ \hline
		\textbf{Data Modalities} & None \\
		\hline\hline
	\end{tabular}
	\label{tbl:scen_15}
	%\vspace{-4mm}
\end{table}

{\textbf{2) Scenario 17-22--McAlister  Moving Blockage:} }
\label{subsubsec:outdoor}
Different from the previous scenario, Scenario 17-22 are outdoor wireless environment representing a two-way city street, as shown in \figref{fig:exp_setup_outdoor}. The DeepSense Testbed 3 is deployed in this environment where the two units are placed on the opposite sides of the street.  The transmitter (Unit 2)  constantly transmits = using one antenna element of the phased array to realize omni-directional transmission.  The receiver continuously scans the surrounding environment using a receive mmWave beamsteering codebook of 64 beams and by measuring the receive power with each beam. The testbed collects data samples at a rate of 12 samples/s. Each data sample includes an RGB image and  a 64-element receive power vector, both collected by Unit 1. An example of the collected data points is shown in \figref{fig:mat_emp}, where we plot the 64 receive powers of 48 data points as well a few samples of the RGB images.  As shown in \figref{fig:mat_emp}, the vehicles crossing the street represent LOS blockages and affect the receive signal power. It is worth mentioning here that since this is a city street, vehicles of various sizes and travel speeds pass in both directions. This creates a diverse pool of blockages for the LOS link between the transmitter and receiver as shown in \figref{fig:rx_view}, which results in diverse received power maps  (i.e., diverse power fluctuations across all 64 beam over time instances) and a diverse dataset. The diversity of the dataset will be discussed in more detail in \sref{sec:blk_div}. The important aspects of the scenario and its data are summarized in  \tabref{tbl:scen_16} following the standard format of the DeepSense 6G dataset \cite{DeepSense}.

\begin{table}[t]
	\caption{Outdoor Scenario (Scenarios 17-22--McAlister Stationary)}
	\label{table}
	\centering
	\setlength{\tabcolsep}{5pt}
	\renewcommand{\arraystretch}{1.4}
	\begin{tabular}{|c|c|}
		\hline\hline
		\textbf{Testbed}             & 3               \\ \hline
		\textbf{Number of Instances}  & \thead{Scenario 17: 70000 - Scenario 18: 120000  \\ Scenario 19: 160000 -Scenario 20: 70837 \\ Scenario 21: 20000 -Scenario 22: 20000 \\ Combined:  460837 (1407 trajectories)   }                    \\ \hline                     \hline
		\textbf{Number of Units}     & 2 \\ \hline
		\textbf{Total Data Modalities}     & \thead{RGB images, \\ 64-dimensional received power vector, \\ GPS locations} \\ \hline \hline
		\multicolumn{2}{|c|}{\textbf{Unit 1}} \\ \hline
		\textbf{Type} & Stationary \\ \hline
		\textbf{Hardware elements} & \thead{RGB camera, mmWave phased array receiver, \\ GPS receiver} \\ \hline
		\textbf{Data Modalities} & \thead{RGB images, \\ 64-dimensional received power vector, \\ GPS location }  \\
		\hline \hline
		\multicolumn{2}{|c|}{\textbf{Unit 2}} \\ \hline
		\textbf{Type} & Stationary \\ \hline
		\textbf{Hardware elements} & \thead{mmWave omni-directional transmitter, \\ GPS receiver} \\ \hline
		%	\textbf{Container Number} & 2 \\ \hline
		\textbf{Data Modalities} & GPS location \\
		\hline\hline
	\end{tabular}
	\label{tbl:scen_16}
	%\vspace{-4mm}
\end{table}

As we briefly highlighted in the description of the two scenarios, since the receiver in DeepSense Testbed 2 (used in Scenario 16) adopts a fixed horn-antenna, it can only receive the power from one direction. This implies that at any time instance $t$, $\mathbf r[t]$ (defined in \eqref{eq:rec}) is simply a scalar because $M=1$. Therefore, a data sample in Scenario 16 consists of an RGB image and a received power value. For Scenario 17-22 that adopts the DeepSense Testbed 3, the receive employs a phased array with $M_\mathrm{A}=16$ elements and applies a receive codebook of 64 beams at each data sample. Therefore, the receive signal at any time instance $t$, $\mathbf r[t]$ is a 64-element vector ($M=64$). As such, a data sample in that scenario comprises an RGB image and a 64-dimensional vector of received power. In the next section, we explain how we use the seed(raw) datasets collected in the two scenarios to generate the development datasets for the four blockage prediction problems defined in \sref{sec:prob_formul}. 

\section{Development Datasets} 
\label{subsec:datasets}
Recall that Scenario 16 represents the indoor scenario and Scenarios 17-22 represent the outdoor scenarios. The seed datasets provided by indoor scenario and outdoor scenario, are used to generate the development datasets. These datasets are the grounds on which the proposed DNNs are developed and evaluated for the four problems introduced in Section \ref{sec:prob_formul}. The generation process goes through two main stages. The first stage is common to both seed datasets. It recognizes the fact that the seed datasets provide per instance data pairs; each pair of received power and image samples represents co-existing data captured (or measured) at the same time instance. Therefore, it constructs time sequences from those per instance data pairs, each of which contains a single blockage event, see the examples in \figref{fig:ori_ch_pow} and \figref{fig:mat_emp}. It is important to note here that the received power sequence varies in structure between the two scenarios. Indoor scenario provides a sequence of scalar power values as it implements a horn antenna. On the other hand, outdoor scenario provides sequences of power vectors, which creates a 2-dimensional matrix.

The second processing stage is more problem-specific, and it aims to produce the final development datasets. A detailed description of that second stage and the final development datasets are provided in the following four subsections.

\begin{figure*}[t]
	\centering
	\begin{subfigure}{0.49\textwidth}
		\centering
		\includegraphics[width=\linewidth]{ 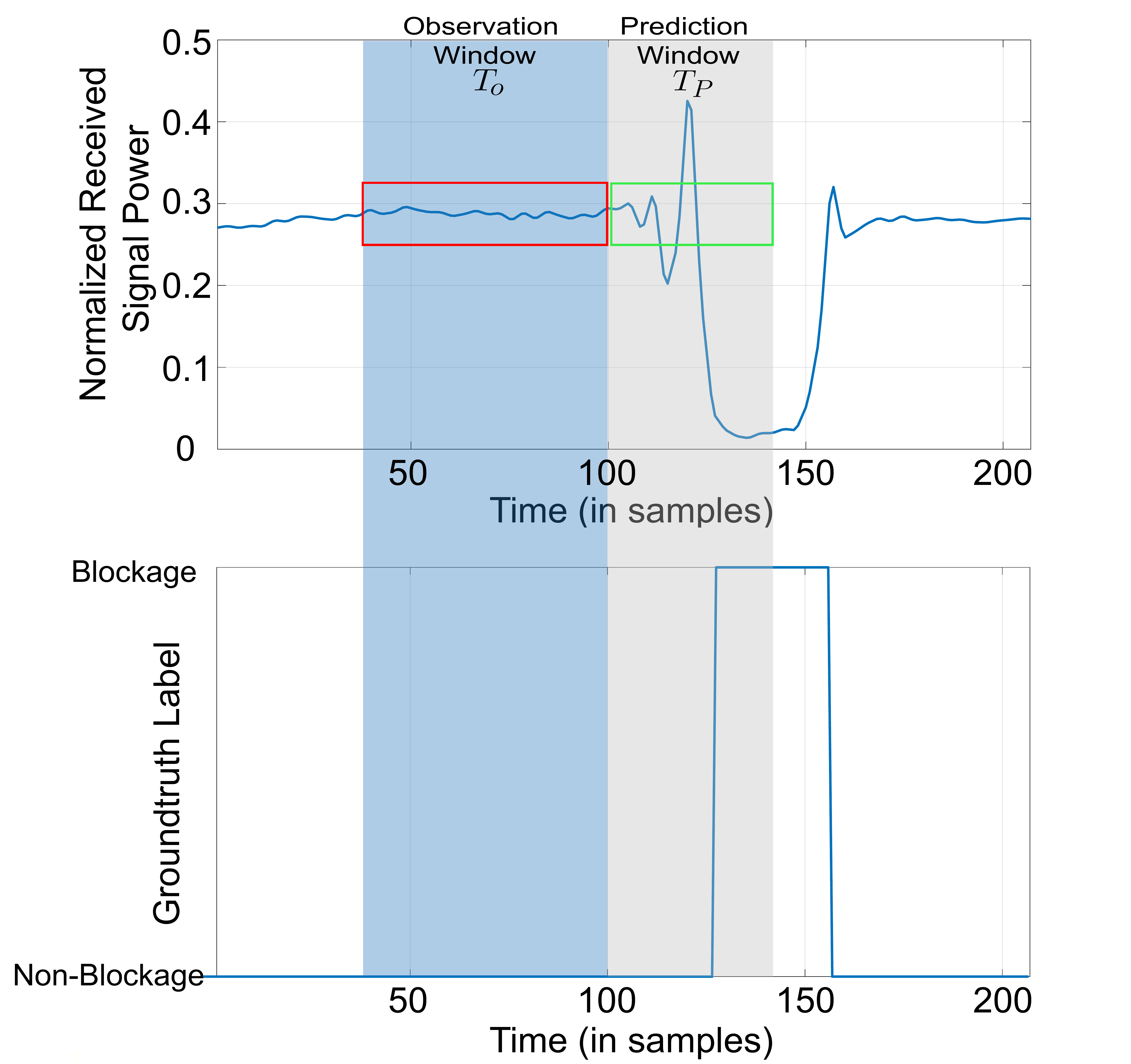}
		\caption{ }
		\label{fig:indoor}
	\end{subfigure}
	\begin{subfigure}{0.49\textwidth}
		\centering
		\includegraphics[width=\linewidth]{ 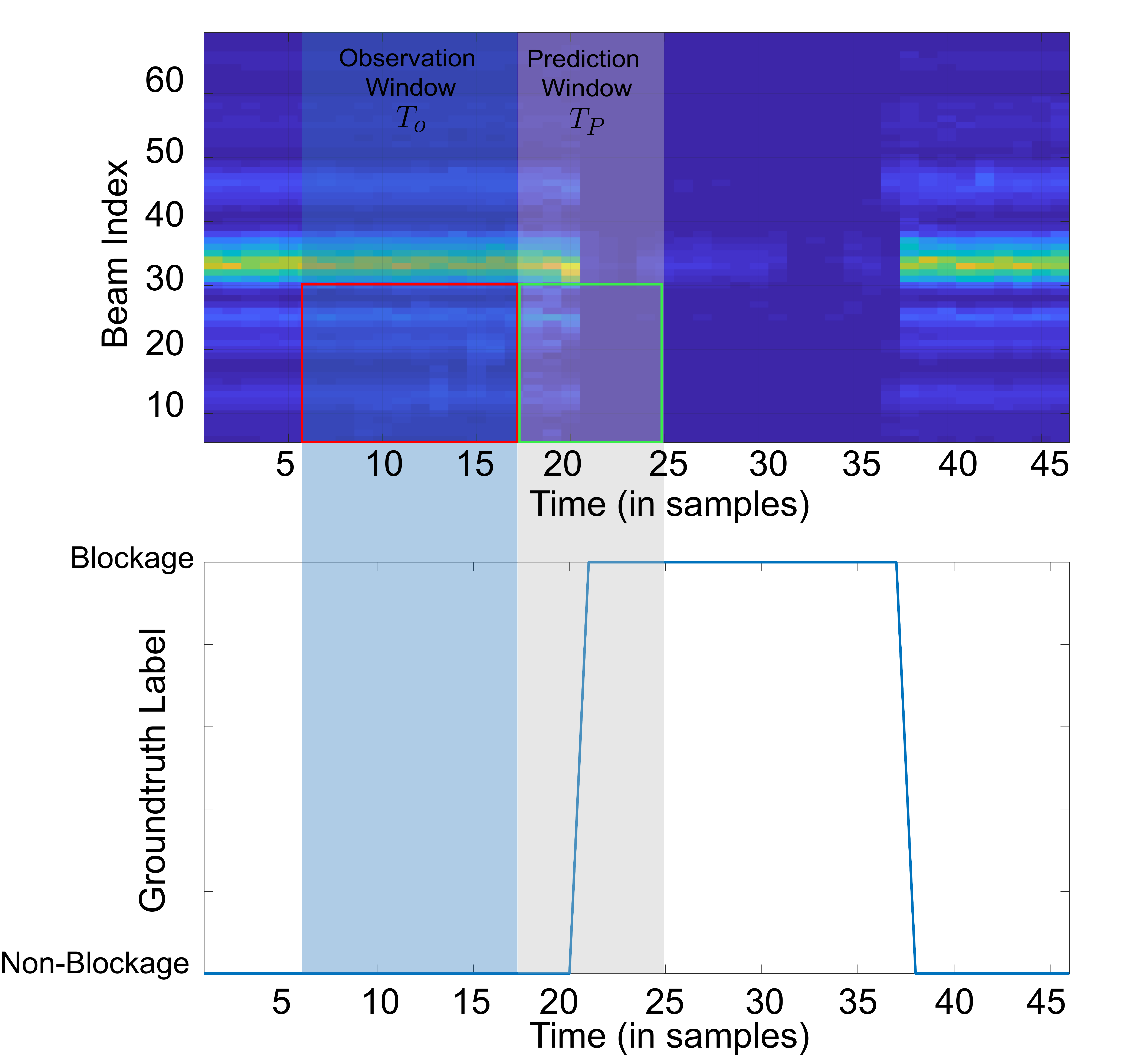}
		\caption{ }
		\label{fig:outdoor}
	\end{subfigure}
	\caption{Sequence generation using a sliding window. (a) Indoor dataset: the top image shows the received signal power raw sequence with an observation window $T_{ob}$ and a prediction window $T_p$; the bottom image shows the corresponding link status. (b) Outdoor dataset: the top image shows the heatmap of the received signal power raw matrix with an observation window $T_{ob}$ and a prediction window $T_p$; the bottom image shows the corresponding link status.}
	\label{fig:slicing}
\end{figure*} 

\subsection{Development Datasets for Problem 1}\label{sec:dev_p1}
The second stage for this problem applies equally to the time sequences of indoor and outdoor scenarios. Two development datasets for {Problem 1} are generated and termed as $\mathcal Y_{P_1}^{(16)}$ and $\mathcal Y_{P_1}^{(17)}$. The following two subsections show how the second processing stage generates the two datasets.
%The first problem aims to address the task of proactive blockage prediction given observed received power ($\mathcal S_{\text{ob}}$). As we consider two different scenarios for our evaluation, two development datasets are constructed for this problem. The following two subsections detail how the second processing stage generates the two datasets.

\subsubsection{Indoor Scenario}\label{sec:sce_15_dev_p1}
Every blockage trajectory provides us with a single time sequence of received-power and image pairs. We utilize the RGB images and manually annotate the time sequences with link-status labels. More specifically, a link-status label of `1' is assigned to a time instance when the LOS link is blocked, and a label of `0' is assigned otherwise. We discard the images and retain the received power sequence, denoted $\mathcal S_{l}$, and the link-status sequence, denoted $\mathcal X$. We refer to the pair $(\mathcal S_l, \mathcal X)$ as a raw sequence pair. The collection of raw sequence pairs are then pooled into one set denoted $\mathcal Y^{(16)}=\{(\mathcal S_l, \mathcal X)_u\}_{u=1}^{U_{16}}$ where $U_{16} = 158$ is the total number of raw sequence pairs indoor scenario provides. 

The number of raw sequence pairs in $\mathcal Y^{(16)}$ is relatively small for the generation of a deep learning development dataset, and, hence, we augment \footnote{This process is specific to the indoor scenario.} $\mathcal Y^{(16)}$ by adding samples obtained by dropping $Y^{(16)}$ three times at dropping rates of 1/2, 1/3, and 1/4. This choice of augmentation mimics an increase in the blockage speed by 2, 3, and 4 times the original speed. Combining dropped sequence pairs with the $158$ original pairs results in the augmented set $\mathcal Y^{(16)}$ having a total of $4\times 158= 632$ pairs.

A data point in the development dataset $\mathcal Y_{P_1}^{(16)}$ of Problem 1 consists of an observation received-power sequence $\mathcal S_{ob}$ and its accompanying label $b_{T_p}$, see \sref{sec:prob_formul}. We generate a pair of $(\mathcal S_{\text{ob}},b_{T_p})$ by applying a sliding window of size $T_{\text{ob}}+T_P$ on the sequences of each pair in $\mathcal Y^{(16)}$, shown in \figref{fig:slicing}. The sliding window assumes a fixed observation interval of $T_{\text{ob}} = 10$ and a variable prediction interval $T_P\in\{1,\dots,40\}$ for development datasets of all problems in indoor scenario. At any time instance $t$ and a choice of $T_p$, we extract the received power sequence from time instance $t-T_{\text{ob}} + 1$ to $t$ and assign it to $\mathcal S_{\text{ob}}$, illustrated by a red stripe in \figref{fig:slicing}-(a). The label of $\mathcal S_{\text{ob}}$ is obtained using $T_P$ and Eqn.\eqref{equ:p1_label}. In particular, we extract the link-status sequence from time instance $t+1$ to $t+T_P$ (i.e., $x[t+1], \dots,x[t+T_P]$) and use Eqn.\eqref{equ:p1_label} to get $b_{T_P}$. This is illustrated by the green stripe in \figref{fig:slicing}-(a). Applying this sliding window method on all $632$ raw sequence pairs in $\mathcal Y^{(16)}$ results in an unbalanced development dataset $\mathcal Y_{P_1}^{(16)}$; the number of data points without blockage (i.e., $b_{T_P} = 0$) is far more than those points with blockage. Therefore, we reduce the number of points without blockage to be equal to that of points with blockage using random sampling. The final result of this step is a development dataset of $U_1 = 1264$ data points, i.e., $\mathcal Y_{P_1}^{(16)} = \{(\mathcal S_{\text{ob}},b_{T_p})_u\}_{u=1}^{U_1}$.

\subsubsection{Outdoor Scenario}\label{sec:sce_16_dev_p1}
Similar to the procedure in \sref{sec:sce_15_dev_p1}, the time sequences resulting from the first processing stage is manually labeled to generate a collection of raw sequence pairs denoted as $\mathcal Y^{(17)} = \{(\mathcal S_l, \mathcal X)_u\}_{u=1}^{U_{17}}$ where $U_{17} = 1407$ is the total number of sequence pairs for the outdoor scenarios. Using the sliding window method introduced in \sref{sec:sce_15_dev_p1}, we generate a second development dataset from the set $\mathcal Y^{(17)}$. We set $T_{\text{ob}}=16$ and $T_P\in\{1,\dots,10\}$ for development datasets of outdoor scenarios. However, the center beams have high power and this affects the detection of the Pre-blockage signature, so we discard the middle 10 high power beams, resulting in received-power vector with 54 beams at any time instance $t+n$, expressed as $\mathbf r_{c}[t+n]\in\mathbb R^{54\times1}$. For any pair $(\mathcal S_l, \mathcal X)\in\mathcal Y^{(17)}$, the sliding window method is used to extract power sequence from instance $t-T_{\text{ob}} + 1$ to $t$ makes up the observed sequence, i.e., $\mathcal S_{\text{ob}} = \{\mathbf r_{c}[t+n]\}^0_{n=-T_{\text{ob}}+1}$ at any time instance $t+n$, and the label $b_{T_P}$ is determined using the Eqn.\eqref{equ:p1_label} and the link-status sequence from time instance $t+1$ to $t+T_P$. \figref{fig:slicing}-(b) depicts the sliding window method for this scenario. Similar to indoor scenario, to make the dataset balanced, we add equal number of data points with non-blockage. The development dataset for {Problem 1} in outdoor scenario is expressed as $\mathcal Y_{P_1}^{(17)}=\{(\mathcal S_{\text{ob}},b_{T_P})_u\}_{u=1}^{U_2}$, where $U_2 = 2814$ is the total number of data points. 

\subsection{Development Dataset for Problem 2} \label{sec:dev_p2}
The two development datasets for {Problem 2} are termed $\mathcal Y_{P_2}^{(16)}$ and $\mathcal Y_{P_2}^{(17)}$ and the second-stage generation process is discussed in the following two subsections.

\subsubsection{Indoor Scenario}\label{sec:sce_15_dev_p2}
The development dataset of {Problem 2} is expressed as $\mathcal Y_{P2}^{(16)} = \{  \mathcal (S_{ob},n_{p})_{u} \}_{u=1}^{U_3}$. We use the same sliding window method to generate the $\mathcal S_{\text{ob}}$ as mentioned in \sref{sec:sce_15_dev_p1}. Instead of pairing the observed received-power sequence $S_{\text{ob}}$ with a blockage indicator $b_{\text{T}_P}$, we pair $S_{\text{ob}}$ with the exact instance when the blockage occurs, i.e., $n_{p}$, see \sref{sec:prob_formul}. In this problem, we only focus on data points with LOS-to-NLOS transition to construct our dataset, which leaves us with a total of $U_3 = 632$ data points.

\subsubsection{Outdoor Scenario}\label{sec:sce_16_dev_p2}

We use $\mathcal Y_{P2}^{(17)} = \{  \mathcal (S_{ob},n_{p})_{u} \}_{u=1}^{U_4}$ to represent the input dataset for {Problem 2} in outdoor scenario, $S_{ob}^{(17)}$ represents the 2 dimension center-beam-cut observation received-power matrix. Similar to indoor scenario, $n_{p}$ is the exact occurrence instance of the blockage in outdoor scenario, $U_{4} = 1407$ are the total number of samples that are input to our model for {Problem 2}. 

\begin{figure}[t]
	\centering
	\includegraphics[width=\linewidth]{ 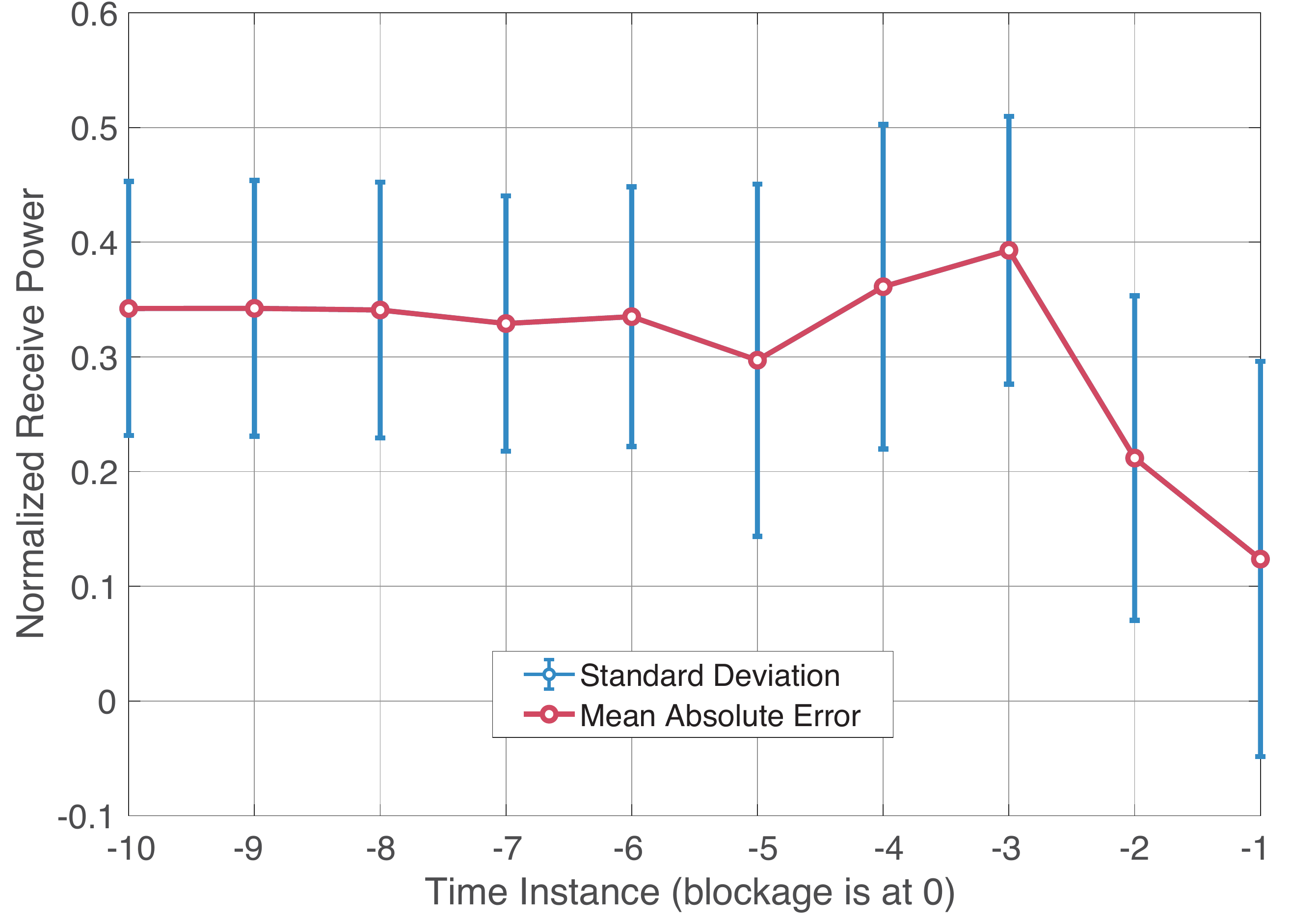}
	\caption{Received power in the indoor scenario. Mean and variance of the received power as a function of the proximity to blockage; the variance increases for the samples close to blockages.}
	\label{fig:indoor_dataset_stat}
\end{figure}

\begin{figure}[t]
	\centering
	\begin{subfigure}{0.5\textwidth}
		\centering
		\includegraphics[width=\linewidth]{ 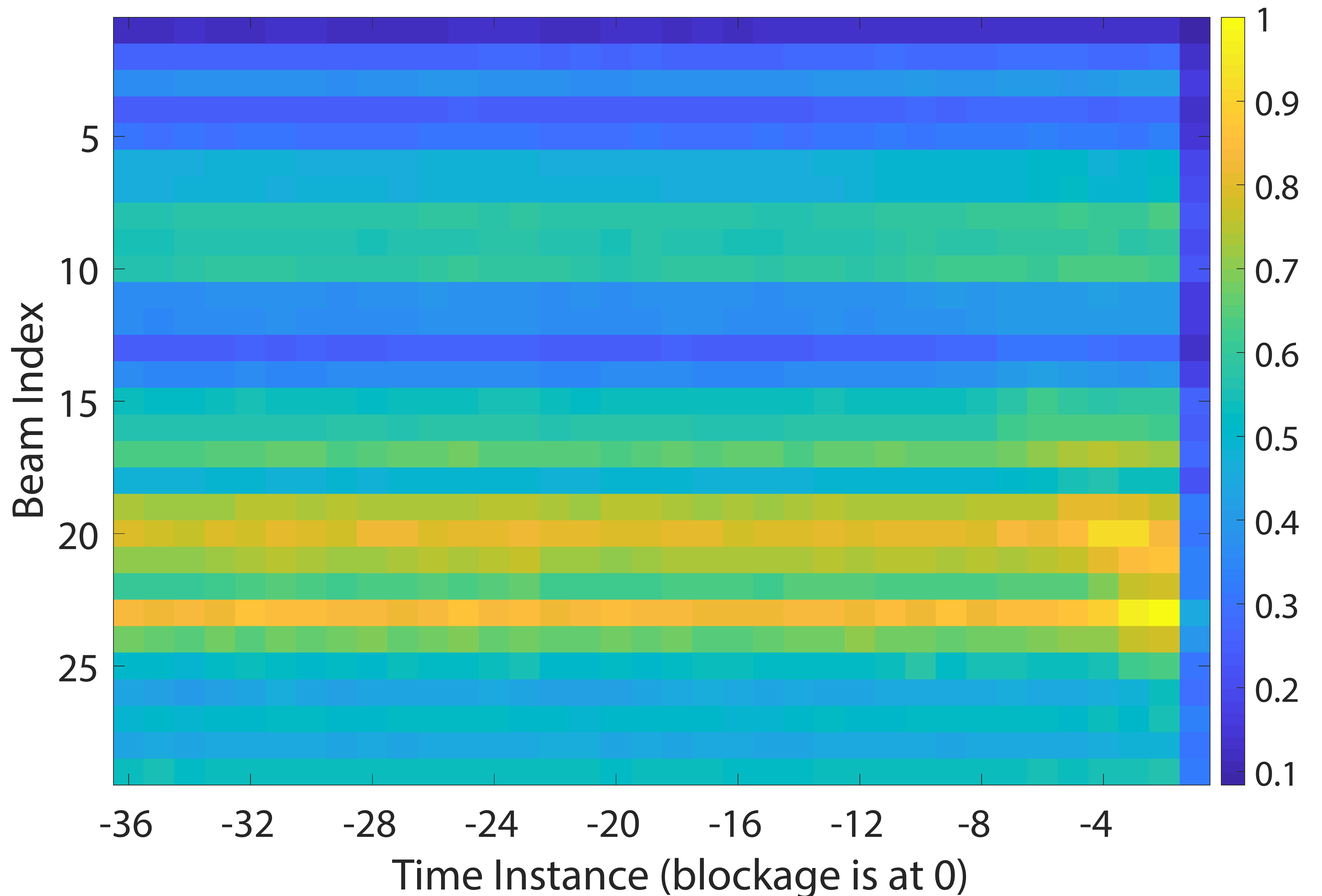}
		\caption{Mean}	
		\label{fig:out_mean_pow}
	\end{subfigure}
	\begin{subfigure}{0.5\textwidth}
		\centering
		\includegraphics[width=\linewidth]{ 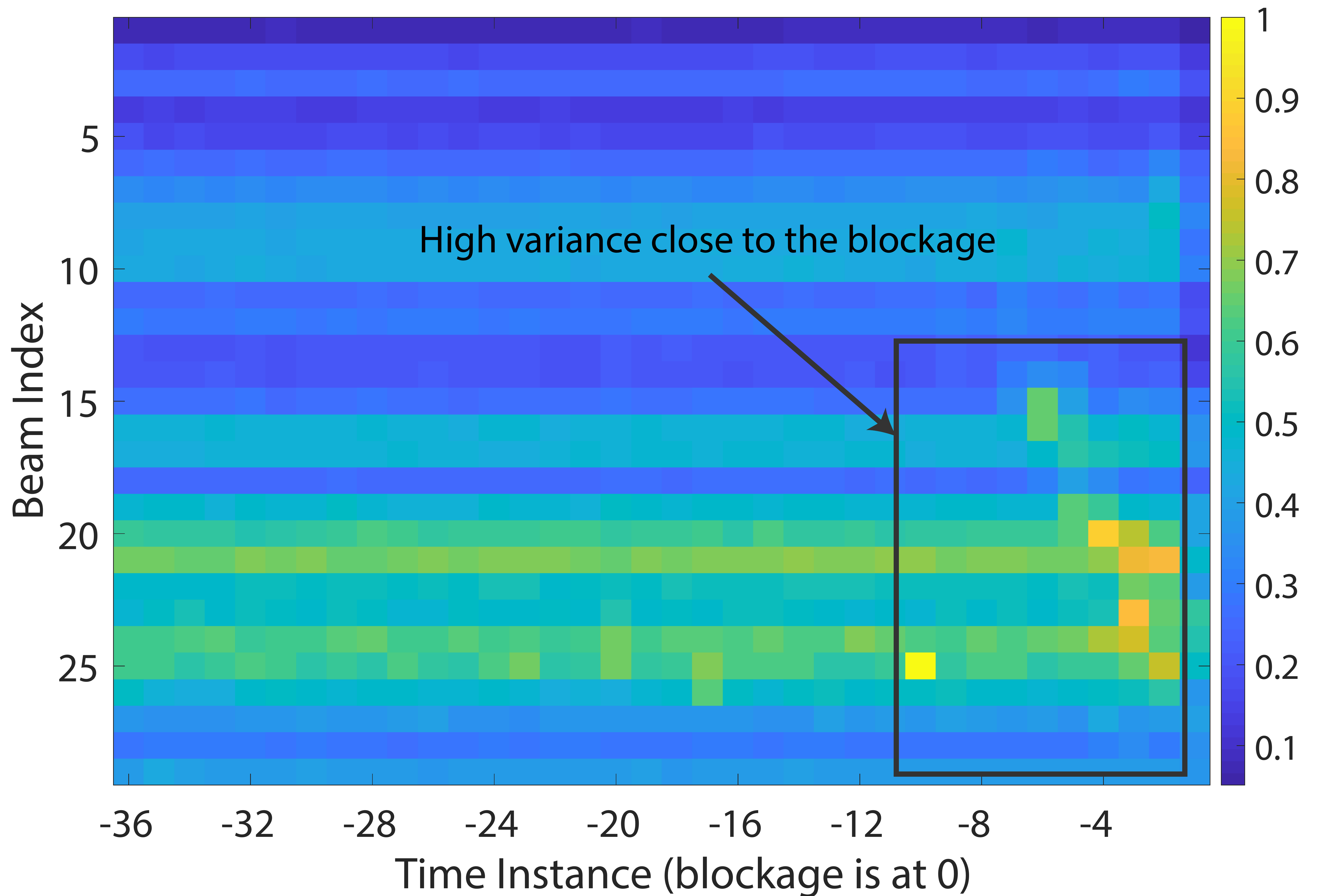}
		\caption{Standard deviation}
		\label{fig:out_std_pow}
	\end{subfigure}
	\caption{Received power in the outdoor scenario. The upper subfigure shows the mean of received power and the bottom subfigure shows the standard deviation of received power as a function of the proximity to  blockages; the black rectangular highlights the high variance for the samples  close to  blockages.}
	\label{fig:outdoor_dataset_stat}
\end{figure}

\subsection{Development Dataset for Problem 3} \label{sec:dev_p3}
Since we only have one object in indoor scenario resulting identical blockage type in every trajectory, we only consider outdoor scenario for this problem. The second processing stage starts with labeling the object classes in every image using the labels listed in \tabref{tbl:dataset_tbl}. Each object class, as the table shows, has its average blockage time interval. We utilize that average to define the blockage severity index of this scenario, see \sref{sec:prob_formul}.  We use K-means algorithm \cite{bishop2006pattern} to quantize the average blockage time interval. Further, in order to study the effect of severity level on the performance of the proposed solutions, which will be shown in \sref{sec:results}. We set the total number of the object classes $N_\mathrm{class}$ to be 3 and 4 based on the blockage time interval. For the case $N_\mathrm{class} = 3$, we have the objects bikes, scooter, skateboards and human in time interval $T_{D_1}$ (151.7ms to 237.5ms), campus carts, sedans, SUVs, vans, pickup trucks and box trucks in time interval $T_{D_2}$ (350ms to 590ms) and commercial trucks, public shuttles, campus shuttles in time interval $T_{D_3}$ (942.5ms to 1030ms). Since blockage time interval in time interval $T_{D_1}$ is small, which has little affect on our link status, we only aim to predict the severity level of the objects in $T_{D_2}$ and $T_{D_3}$. 

Next, we label the severity level of objects $b_{\text{sev},v}$ in $T_{D_2}$ as 2 and objects in $T_{D_2}$ as 3, and pair the label with its corresponding observant received-power sequence $S_\text{ob}$. Since the number of objects in $T_{D_3}$ is smaller than the number of the objects in $T_{D_2}$ (because of the uneven distribution of the type of vehicle), our dataset is imbalanced. To make the dataset balanced, we use data augmentation for the objects in $T_{D_3}$ by duplicating the pairs and adding 10dB white Gaussian noise to observation sequences. Finally, we get 1152 sequence pairs whose average blockage time interval lies in $T_{D_3}$ and 1215 sequence pairs with average blockage time interval lying in $T_{D_2}$. The dataset is represented by $\mathcal Y_{P3\_N3}^{(17)} = \{  \mathcal (S_{ob\_DA1},b_{sev})^{(u)} \}_{u=1}^{U_{5}}$, where $\mathcal S_\text{ob\_DA1}$ is the observation sequences after data argumentation, $b_{sev}$ is the label indicating the severity index of the blockages, and $U_{5} = 2367$ is the total number of samples that are input to our model of outdoor scenario for {Problem 3}.

The generation process is similar for the case $N_\mathrm{class} = 4$, we omit the objects lying in time interval $T_{D_1}$ (151.7ms to 237.5ms). Campus carts, sedans and SUVs belong to $T_{D_2}$ (350ms to 395.8ms), vans, pickup trucks and box trucks are in $T_{D_3}$ (512.5ms to 576.7ms), commercial trucks, public shuttles and campus shuttles lies in $T_{D_4}$ (591ms to 1030ms). We label the objects severity level based on Eqn.\ref{equ:p3_label} and pair the label with observation sequence $S_\text{ob}$. Then we do data argumentation for the objects in $T_{D_3}$ and $T_{D_4}$. Finally, we get 1016 sequence pairs whose average blockage time interval lies in $T_{D_2}$ and 995 sequence pairs with average blockage time interval lying in $T_{D_3}$ and 960 sequence pairs with average blockage time interval in $T_{D_4}$ after data argumentation. The dataset is represented by $\mathcal Y_{P3\_N4}^{(17)} = \{  \mathcal (S_{ob\_DA2},b_{sev})^{(u)} \}_{u=1}^{U_{6}}$, where $\mathcal S_\text{ob\_DA2}$ is the observation sequences after data argumentation, $b_{sev}$ is the label indicating the severity index of the vehicles, and $U_{6} = 2971$ is the total number of samples that are input to our model of outdoor scenario for {Problem 3}.

\begin{figure}[t]
	\centering
	\includegraphics[width=\linewidth]{ 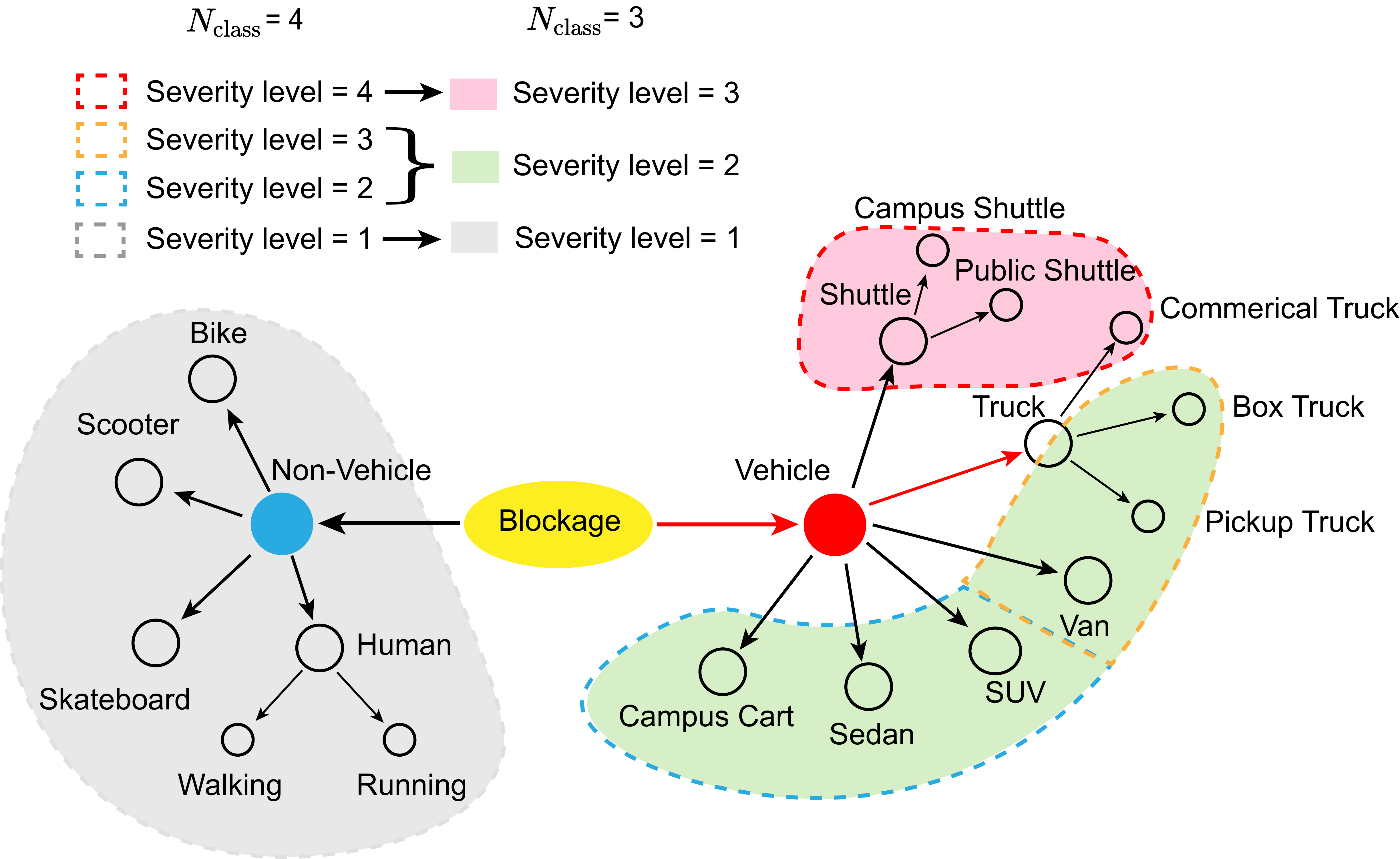}
	\caption{Object tree illustrating object classes and their relations. For $N_\mathrm{class} = 3$, the objects in each severity level are in shaded color; for $N_\mathrm{class} = 4$, the objects in each severity level are in dashed circles.}
	\label{fig:obj_tree}
\end{figure}

\subsection{Development Dataset for Problem 4} \label{sec:dev_p4}
Since the testbed in indoor scenario only have a horn antenna as the receiver, the received-power scalar is not sufficient for motion direction prediction, so we only generate development dataset for {Problem 4} for outdoor scenario. 
We label the motion direction and pair it with the observation received-power sequence, the pairs are expressed as  $(S_\text{ob},b_{dir})$. Then we duplicate the whole pairs and add 10 dB white Gaussian noise to the observation received-power sequence, the augmented pairs are denoted as $(S_\text{ob\_DA3},b_{dir})$. We get 3028 sequence pairs with directions from left to right and 2600 sequence pairs with opposite motion directions which is represented as $\mathcal Y_{P4}^{(17)} = \{  \mathcal (S_{ob\_DA3},b_{dir})^{(u)} \}_{u=1}^{U_{7}}$, where $U_{7} = 5628$ is the total number of samples that are input to our model of outdoor scenario for this problem.

\begin{figure*}[t]
	\centering
	\includegraphics[width=\linewidth]{ 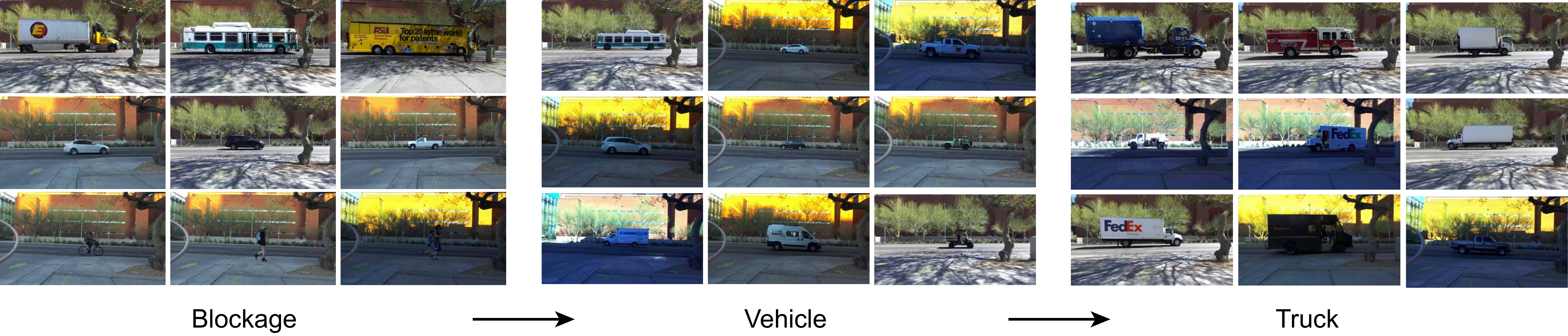}
	\caption{Dataset examples.  The image group in left panel are assorted blockages, the image group in the middle panel are vehicles, and the image group in the right panel are trucks.}
	\label{fig:dataset_ex}
\end{figure*}

\begin{table}[t]
	\centering
	\small
	\caption{Statistics of Objects in the Outdoor Scenario}
	\begin{tabular}{ccccc}
		\hline\hline
		{\textbf{Type of Objects}} & \textbf{Number} & \textbf{Avg blockage time interval}  \\ \hline
		Campus Shuttle      & 11    & 12.36 (1030ms)        \\
		Public Shuttle      & 152   & 11.43 (952.5ms)        \\
		Commercial Truck    & 29    & 11.31 (942.5ms)        \\
		Box Truck           & 36    & 7.08  (590ms)        \\
		Pickup Truck        & 112   & 6.92  (576.7ms)        \\
		Van                 & 51    & 6.15  (512.5ms)        \\
		SUV                 & 370   & 4.75  (395.8ms)        \\
		Sedan               & 631   & 4.16  (346.7ms)        \\
		Campus cart         & 15    & 4.2   (350ms)        \\
		Bike                & 79    & 2.7   (225ms)        \\
		Scooter             & 2     & 2     (166.7ms)        \\
		Skateboard          & 11    & 1.82  (151.7ms)        \\ 
		Walking Human       & 118   & 2.85  (237.5ms)        \\
		Running Human       & 14    & 2.64  (220ms)        \\ \hline\hline
	\end{tabular}
	\label{tbl:dataset_tbl}
\end{table}

\begin{figure}[t]
	\centering
	\includegraphics[width=1\linewidth]{ 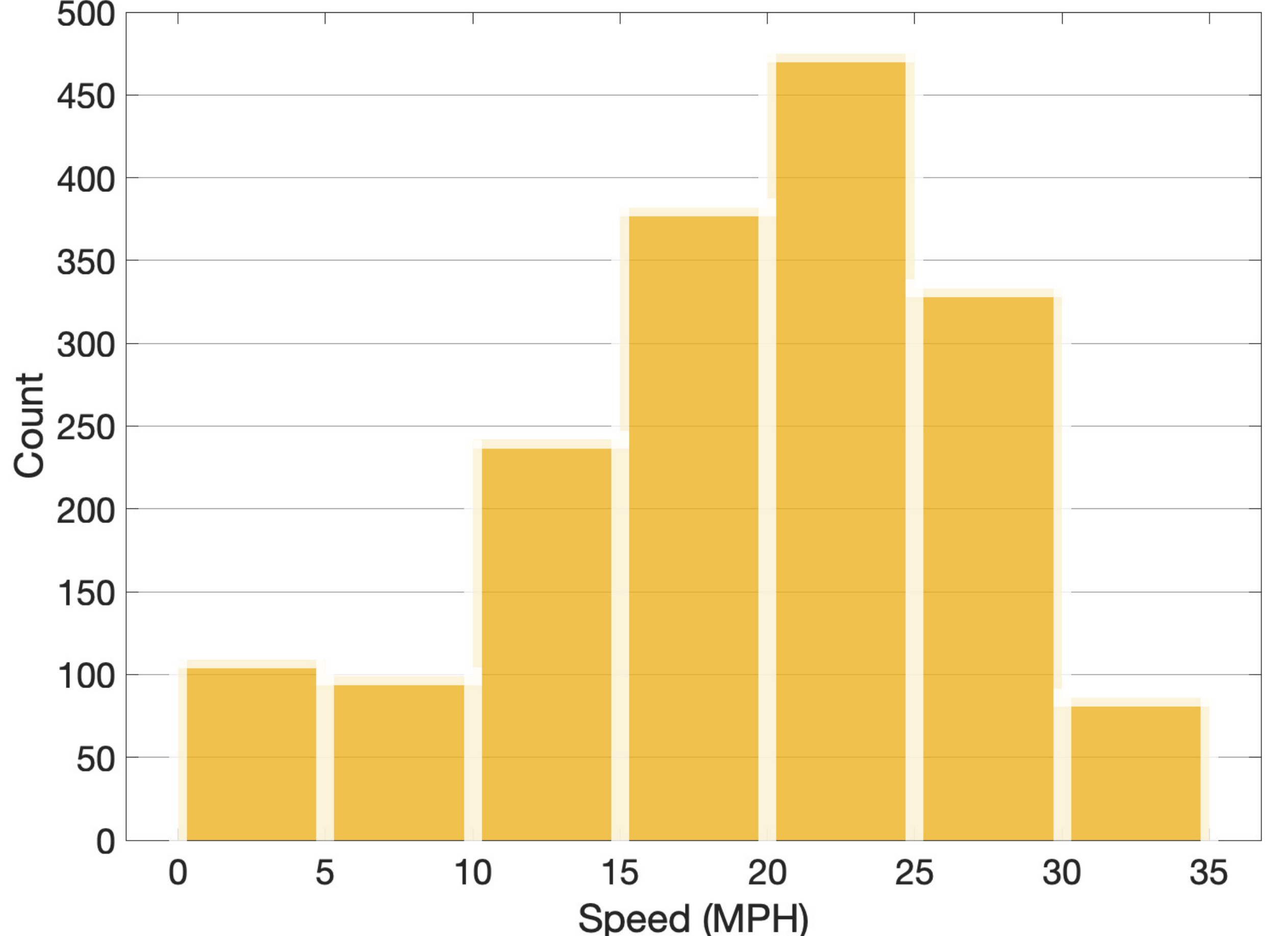}
	\caption{Histogram of speed of moving blockages, majority of blockages move at a speed at 20-25 miles per hour (MPH).}
	\label{fig:speed_hist}
\end{figure}

\section{Analysis of the Development Datasets}\label{sec:anal_dev_data}
The considered scenarios represent two different wireless environments, and, as such, their development datasets reflect various dynamics for scatterers and reflectors. Before we move on and discuss the performances of the proposed solutions for each problem, we present a simple statistical analysis for both scenarios. The goal of this analysis is twofold: (i) show some statistical evidence on the existence of the pre-blockage signature, and (ii) explore the various aspects of the data that make them suitable to address the four problems in Section \ref{sec:prob_formul}.

\subsection{Pre-Blockage Signature}
The core of our argument for proactively predicting LOS-blockages rests on the pre-blockage signature, see Section \ref{subsec:key_idea}. Therefore, before delving into the details of how well our proposed models detect that signature and predict incoming blockage, it is important to present some statistical evidence on the existence of that signature and its relation to various objects in the wireless environment. To do that, we randomly select two samples of data points from the time-sequence datasets, i.e., $\mathcal Y^{(16)}$ and $\mathcal Y^{(17)}$. Using the observed sequences $\mathcal S_{\text{ob}}$ in each sample, we compute the mean and standard deviation of the received power signal over a 16-instance window \textit{prior to} a blockage. Given the different number of antenna elements in the two scenarios, the mean and standard deviation of received power are plotted in two different figures, namely \figref{fig:indoor_dataset_stat} for indoor scenario and \figref{fig:outdoor_dataset_stat} for outdoor scenario.

Both figures verify the notion of pre-blockage signature, especially close to the blockage instance. In \figref{fig:indoor_dataset_stat}, the mean and standard deviation of the received power over the farthest few instances from the blockage, i.e., instances 1-5 on the figure, are almost constant. However, as we approach the blockage instance, instance 12 on the figure, the mean starts to fluctuate and the standard deviation of the received power widens. Both are indicators that a pattern (or a signature) is starting to form there. This pattern is different from observed sequence to another depending on the size, shape, and distance of the incoming blockage. 

The evidence on pre-blockage signature gets clearer when \figref{fig:outdoor_dataset_stat} is examined. Recall that in outdoor scenarios, phased arrays, that enable measuring directional received power, are deployed. Hence, \figref{fig:outdoor_dataset_stat} provides a visualization of the mean and standard deviation over both time (x-axis) and beam index (y-axis) in two separate subfigures. \figref{fig:out_mean_pow} shows that the means of all beams seem to stay constant over time far from the blockage instance, and they only exhibit little fluctuations close to the blockage instance  (similar to \figref{fig:indoor_dataset_stat}). Nonetheless, we see obvious difference in average received power across beams, which is valuable directional information. Such information becomes even more interesting when we turn to the standard deviation of the power in \figref{fig:out_std_pow}. It depicts more localized fluctuations (variations) around the last few instances (instance 30 to 35) and around beams 20 and 25. We argue that such pattern makes the pre-blockage signature more distinctive than that in indoor scenario; the different sizes, shapes, and speeds of moving blockages could produce different imprints across time and beam index.

\subsection{Data Diversity}\label{sec:blk_div}
A valuable and interesting feature in the datasets provided by outdoor scenarios is its diversity. Such diversity comes in various forms such as object classes, object instances, object speeds, and per-object average blockage time intervals. They collectively indicate that multiple development datasets could be generated for different blockage-related problems. \figref{fig:obj_tree} and \tabref{tbl:dataset_tbl} provide a break-down of the object classes in scenario 17-22, the number of instances, and average blockage time interval. The scenario provides a total of 14 object classes and exhibits average blockage time interval ranging from 1.82 to 12.36 time instances (151.7ms to 1030ms). In addition, those different objects appear in various shapes and move at varying speeds. This is evident in the sample photos in \figref{fig:dataset_ex} and the speed histogram in \figref{fig:speed_hist}.

The various moving object classes, different speeds, and varying average-blockage time intervals are important factors to construct development datasets for all four problems in \sref{sec:prob_formul}. Object classes and speeds are expected to have major impact on the pre-blockage signature exhibited in the dataset. Such impact makes it possible to study the relation between the signature and different blockage aspects. For instance, we could be interested in proactively identifying blockage severity or blockage direction using that signature. The object diversity in outdoor scenarios helps define different severity levels (intervals) and studying how distinctive the signature is in proactively identifying those levels. A similar thing could be done for blockage direction.

\section{Experimental Results}
\label{sec:results}
In this section, we introduce the evaluation metrics used to evaluate the results in Section \ref{subsec:eva_metric}. Section \ref{subsec:net_train} presents all the parameters we used for training our RNN and CNN. Finally, Section \ref{subsec:pref} analyzes the results.
\subsection{Evaluation Metrics}
\label{subsec:eva_metric}
Since {{Problems 1, 3, 4}} are prediction problems, we use Top-1 accuracy as our evaluation metric. It is defined as the compliment of the prediction error given in \cite{ImageNet}, and it could be expressed as:
\begin{equation}
	\text{Acc}_{\text{top-1}} = \frac{1}{U_{v1}}\sum_{u = 1}^{U_{v1}} \mathbbm{1}  (b^{(u)} = \hat {b}^{(u)}),
\end{equation}
where $\mathbbm{1}$ is the indicator function, $U_{v1}$ is total samples of the validation set in problem 1, $b^{(u)}$ and $\hat{b}^{(u)}$ are, respectively, the target and predicted link status for a future interval of $T_P$ instances, specifically, $b^{(u)}$ represents as $b_{T_p}^{(u)}$ in {Problem 1}, $b_{\text{type}}$ in {Problem 3} and $b_{\text{dir}}$ in {Problem 4}.

{{Problem 2}} is posed as a regression problem, and so we use Mean Absolute Error (MAE) and its standard deviation to evaluate the quality of our model predictions. The MAE is the mean absolute error between the ground-truth value and the predicted value. For each prediction interval $T_P$, we calculate MAE across the samples and the standard deviation of these absolute errors. 
\begin{equation}
	e^{(u)}_{T_P} = \lvert n^{\prime (u)} - \hat{n}^{\prime (u)} \rvert, \quad  \forall u \in \{1,\dots,U_{v2}\} ,    
\end{equation}
\vspace{-0.4cm}
\begin{equation}
	\bar {e}_{T_P} = \frac{1}{U_{v2}}\sum_{u = 1}^{U_{v2}} \lvert n^{\prime (u)} - \hat{n}^{\prime (u)}  \rvert,
\end{equation}

\begin{equation}
	\text{std}_{T_P} = \left({{\frac {1}{U_{v2}}}\sum _{i=1}^{U_{v2}}\left(e^{(u)}_{T_P}-{ \bar {e}_{T_P}}\right)^{2}}\right)^{\frac{1}{2}},
\end{equation}
where, $e^{(u)}_{T_P}$ is the absolute error for $u$th sample, $\bar {e}_{T_P}$ is the MAE, $ \text{std}_{T_P}$ is the standard deviation of absolute error, $U_{v2}$ is the total numbers of samples in validation set, $n^{\prime (u)}$ and $\hat{n}^{\prime (u)}$ are target and predicted time instances between current time and the time of blockage occurrence, assuming prediction interval is $T_P$.

\subsection{Network Training}
\label{subsec:net_train}
In this subsection, we present the hyper-parameters of the proposed network models for indoor and outdoor scenarios.

\textbf{Indoor Blockage Scenario}: We build the deep learning model described in Section~\ref{sec:deep_learning} using Pytorch~\cite{pytorch}. We input 10 successive time-instances of received signal power and the corresponding label for training. Our model consists of 1 RNN layer with 20 hidden states with a dropout rate = 0.2. The number of epochs is 1000. These parameters are chosen based on empirical experiments. The detailed parameters of our RNN for {{Problem 1}} and {{Problem 2}} are shown in \tabref{tbl:Scenario 15_RNN}. 

\begin{table}[tb]
	\centering
	\caption{Parameters of the RNN Model for the Indoor Scenario}
	\begin{tabular}{ccc}
		\cline{1-3}
		\hline\hline
		& \multicolumn{2}{c}{ \textbf{Value}}  \\ 
		\multirow{-2}{*}{\textbf{Name}} &  \textbf{Prob 1}    &  \textbf{Prob 2}     \\
		\hline
		Input sequence length       &  10   &  10      \\ 
		Predicted future time steps &  1-40 &  1-40    \\ 
		Hidden state of RNN         &  20   &  20      \\ 
		Output dimension            &  2    &  1       \\ 
		Number of RNN layer         &  1    &  1       \\ 
		Dropout rate                &  0.2  &  0.2     \\ 
		Epoch                       &  1000 &  1000    \\ 
		
		\hline\hline
	\end{tabular}
	\label{tbl:Scenario 15_RNN}
\end{table}

\begin{table*}[tb]
	\centering
	\caption{Parameters of the RNN Model for the Outdoor Scenario}
	\begin{tabular}{ccccc}
		\cline{1-5}
		\hline\hline
		& \multicolumn{4}{c}{ \textbf{Value}}  \\ 
		\multirow{-2}{*}{\textbf{Name}} &  \textbf{Prob 1}    &  \textbf{Prob 2} &  \textbf{Prob 3} &  \textbf{Prob 4}    \\
		\hline
		Input sequence length       &  16   &  16   &  16   &  16   \\ 
		Predicted future time steps &  1-10 &  1-10  &  1-10   &  1-10  \\ 
		Hidden state of RNN         &  20   &  20   &  10   &  10   \\ 
		Output dimension            &  2    &  1    &  2   &  2   \\ 
		Number of RNN layer         &  1    &  1   &  1   &  1    \\ 
		Dropout rate                &  0.2  &  0.2  &  0.2   &  0.2   \\ 
		Epoch                       &  2000 &  1800  &  1000   &  3000  \\ 
		\multirow{2}{*}{Total Parameters}   &  \multirow{2}{*}{2002} &  \multirow{2}{*}{1991}  &  2002 ($N_\mathrm{class} = 3$)   &  \multirow{2}{*}{2002}  \\ 
		&       &       & 2013 ($N_\mathrm{class} = 4$)      &       \\
		
		\hline\hline
	\end{tabular}
	\label{tbl:outdoor_RNN}
\end{table*}

\begin{table*}[tb]
	\centering
	\caption{Parameters of the CNN Model for the Outdoor Scenario}
	\begin{tabular}{cccccc}
		\cline{1-6}
		\hline\hline
		\multicolumn{2}{c}{\multirow{2}{*}{\textbf{Name}}} & \multicolumn{4}{c}{ \textbf{Value}}  \\ 
		\multicolumn{2}{c}{} &  \textbf{Prob 1}    &  \textbf{Prob 2} &  \textbf{Prob 3} &  \textbf{Prob 4}    \\
		\hline
		\multicolumn{2}{c}{Input sequence dimension}       &  16 $\times$ 54   &  16 $\times$ 54   &  16 $\times$ 54   &  16 $\times$ 54   \\ 
		\multicolumn{2}{c}{Predicted future time steps} &  1-10 &  1-10  &  1-10   &  1-10  \\ 
		\multirow{3}{*}{Stack 1}    & Conv 1   &  1-4-3-1  &  1-4-3-1  &  1-4-3-1  &  1-4-3-1   \\ 
		& Conv 2   &  4-4-3-1  &  4-4-3-1  &  4-4-3-1  &  4-4-3-1  \\ 
		&Max pooling 1  &  (2,3)    &  (2,3)    &  (2,3)    &  (2,3)  \\ \hline
		\multirow{3}{*}{Stack 2}    & Conv 3   &  4-8-3-1  &  4-8-3-1  &  4-8-3-1  &  4-8-3-1 \\ 
		& Conv 4   &  8-16-3-1 &  8-16-3-1 &  8-16-3-1 &  8-16-3-1  \\ 
		& Max pooling 2  &  (2,3)    &  (2,3)    &  (2,3)    &  (2,3)   \\ \hline
		\multicolumn{2}{c}{FC}                  &  (384,2)  &  (384,1)  &  (384,2)  &  (384,2)    \\ 
		\multicolumn{2}{c}{Dropout rate}        &  0.2  &  0.2  &  0.2   &  0.2   \\ 
		\multicolumn{2}{c}{Epoch}               &  2000 &  1800  &  1000   &  3000  \\ 
		\multicolumn{2}{c}{\multirow{2}{*}{Total Parameters}}    &  \multirow{2}{*}{2422} &  \multirow{2}{*}{2037}  &  2422($N_\mathrm{class} = 3$)   &  \multirow{2}{*}{2422}  \\ 
		\multicolumn{2}{c}{}    &   &    &  2807($N_\mathrm{class} = 4$)   &    \\ 
		
		\hline\hline
	\end{tabular}
	\label{tbl:outdoor_CNN}
\end{table*}
\textbf{Outdoor Blockage Scenario}: The hyper-parameters of our RNN model  for {Problems 1, 2, 3, 4} for outdoor scenario are shown in \tabref{tbl:outdoor_RNN}. Compared with the model for indoor scenario, we input 16 successive time instances of the received signal instead of 10 and we adjust the number of epochs based on the problem. The details of the parameters for the CNN model are shown in \tabref{tbl:outdoor_CNN}. For the convolution layer we show the format of parameters as input channel - output channel - kernel size - padding, and for max-pooling layer, we show the kernel size. To make the size of RNN and CNN models comparable, we use a similar number of weights for each of the networks. All parameters are based on empirical experiments.

\subsection{Performance Evaluation}
\label{subsec:pref}
In this subsection, we evaluate the performance of the proposed blockage prediction approaches for the four problems defined in \sref{sec:prob_formul}. This is done using the real-world datasets collected for indoor and outdoor scenarios, which are described in detail in Sections \ref{subsec:setup}, \ref{sec:dev_p4}, and \ref{sec:anal_dev_data}. 

\begin{figure}[t]
	\centering
	\begin{subfigure}[]{\linewidth}
		\centering
		\includegraphics[width=1\linewidth]{ 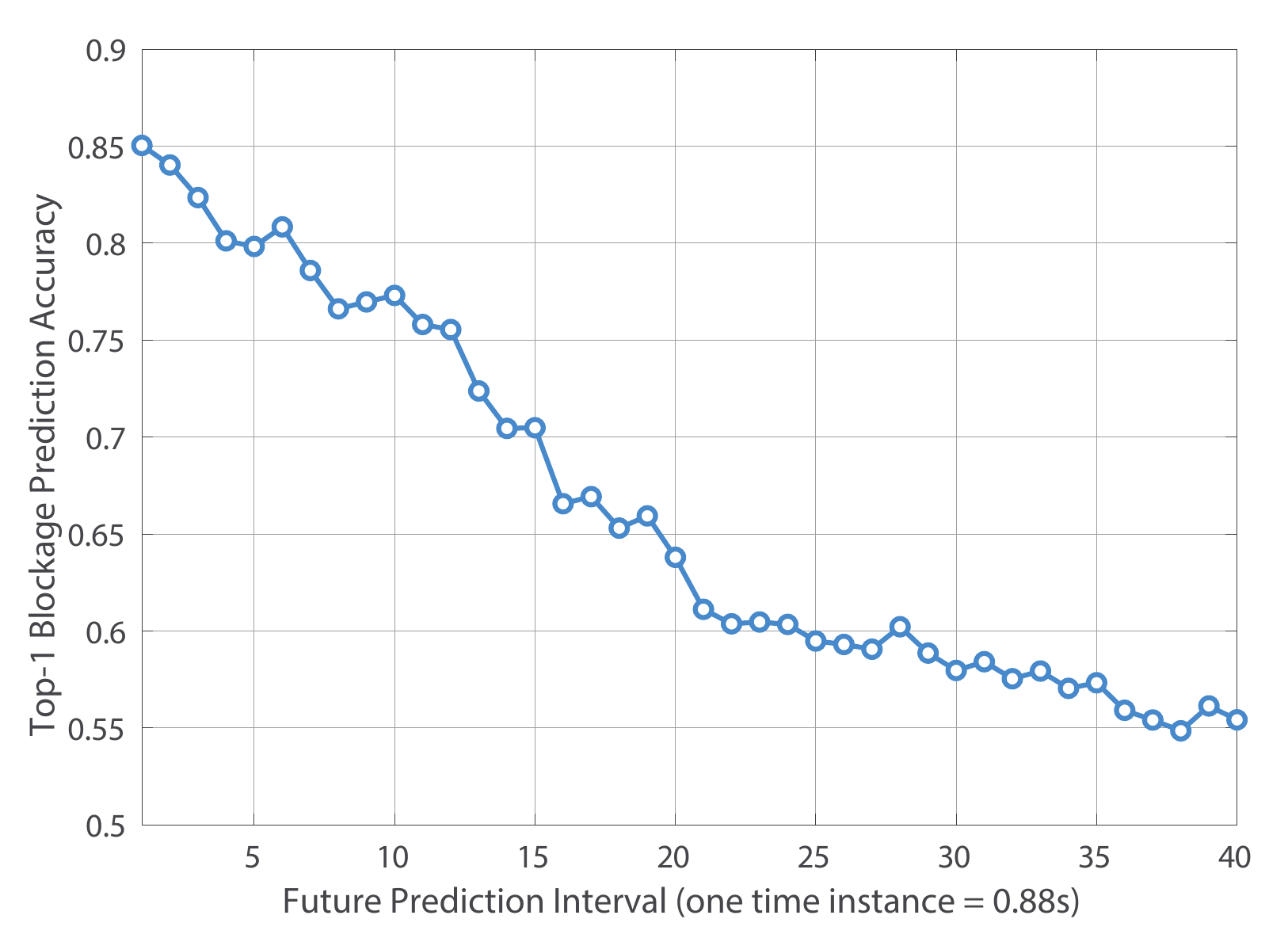}
		\caption{RNN model using the dataset indoor scenario}
		\label{fig:P1_Scenario 15}
	\end{subfigure}
	\begin{subfigure}[]{\linewidth}
		\centering
		\includegraphics[width=1\linewidth]{ 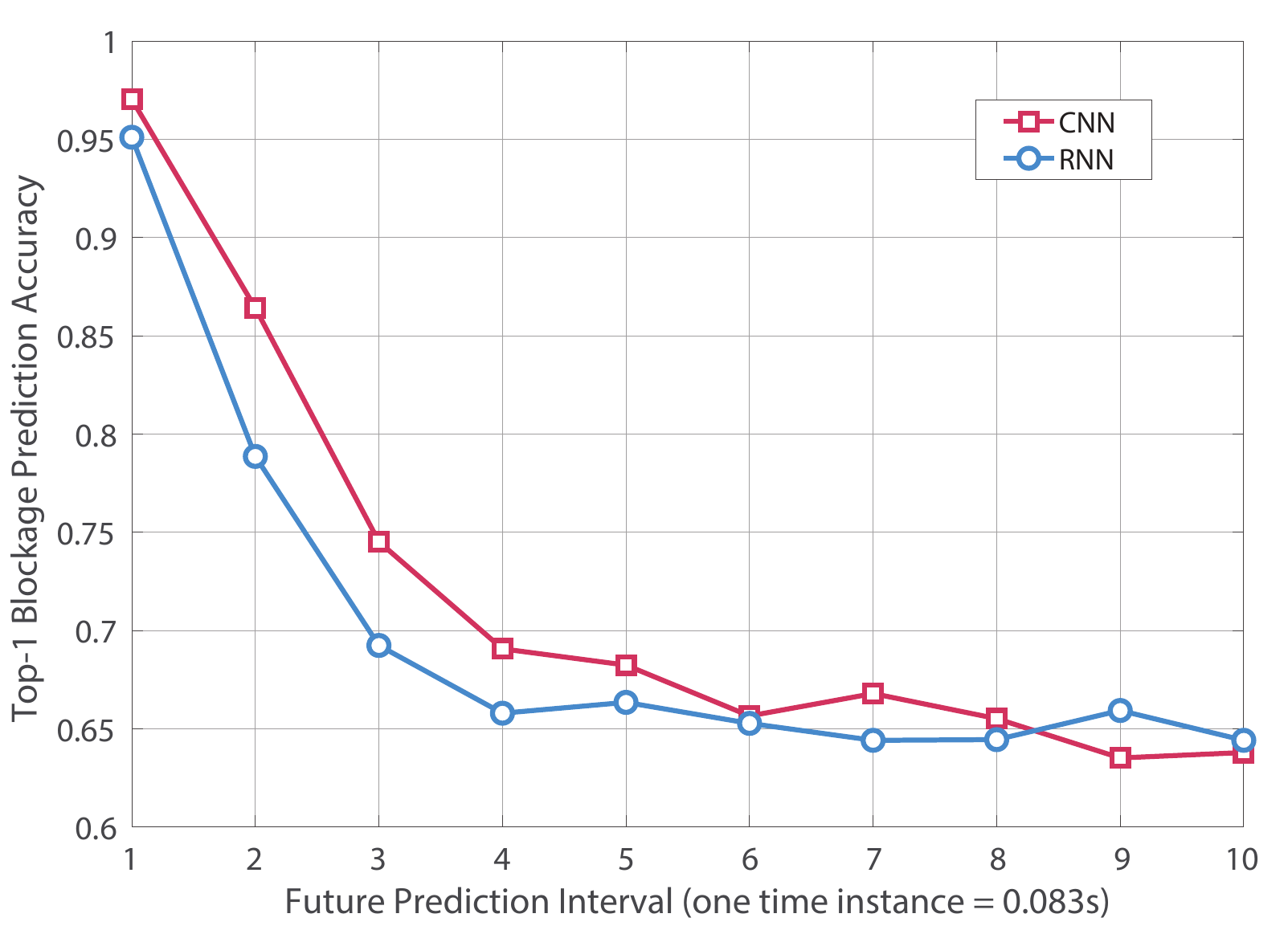}
		\caption{RNN and CNN model using dataset of outdoor scenario}
		\label{fig:P1_outdoor}
	\end{subfigure}
	\caption{Top-1 blockage prediction accuracy for different future prediction intervals for outdoor scenario. Note that 40 time instances for dataset of indoor scenario is equivalent to 3 time instances for dataset of outdoor scenario.}
\end{figure}

%\begin{figure}[t]
%	\centering
%	\includegraphics[width=1\linewidth]{ latency.pdf}
%	\caption{Average latency of the proposed proactive hand-off solutions compared to the conventional reactive approaches.}
%	\label{fig:P1_latency}
%\end{figure}

\noindent {\textbf{Problem 1. Blockage Occurrence Prediction:}} In this problem, the model predicts whether a blockage will happen or not within the next $T_p$ time instances (the future prediction interval). Next, we present the experimental results for both the indoor and outdoor scenarios.

\begin{itemize}
	\item \textbf{Indoor Blockage:} \figref{fig:P1_Scenario 15} plots the Top-1 blockage prediction accuracy as a function of the future prediction interval for the indoor scenario. For blockages occurring within 6 future time instances (around 5s in the future), our model achieves more than $80\%$ accuracy. This prediction accuracy then decreases as the future prediction interval increases, which is expected. For example,  blockages happening after 20 time instances could be predicted with around $65\%$ accuracy. The accuracy finally converges to the``random guess'' performance (i.e., approaching $50\%$) as the prediction interval exceeds 40 time instances. This is because the pre-blockage signature is less effective when the blockage is far away from the transmitter-receiver link.
	
	\item \textbf{Outdoor Blockage:} \figref{fig:P1_outdoor} plots the blockage prediction accuracy of both the developed CNN and RNN models for the outdoor scenario. As shown in this figure, our models achieve more than $95\%$ accuracy in predicting the blockages happening after 1 time instance (which corresponds to $\sim$83ms in the future). This accuracy then decreases with increasing the future interval, reaching around $65\%$ accuracy for predicting blockages happening after 10 time instances (around 0.8s).  \figref{fig:P1_outdoor} also shows that the CNN model has a slightly better performance compared to the RNN model. In general, CNN models are expected to perform well in this blockage prediction task due to their powerful capability in recognizing local features in the images, such as the pre-blockage signature in the mmWave heatmap image. 
\end{itemize}

\begin{figure}
	\centering
	\includegraphics[width=1\linewidth]{ 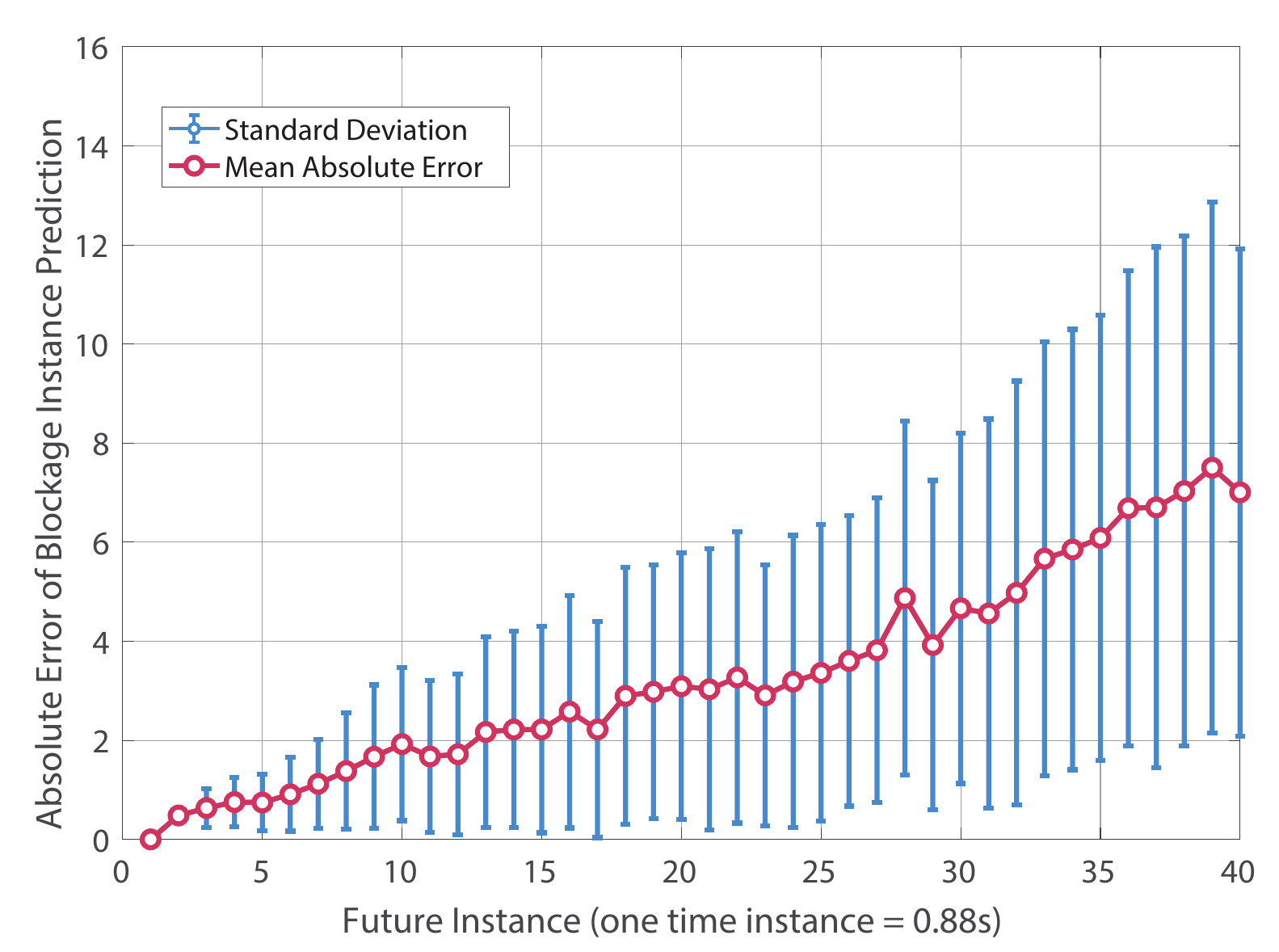}
	\caption{Mean absolute error between the target (the exact time instance where the blockage happens) and the prediction of this blockage time instance for different future prediction intervals using the indoor scenario dataset.}
	\label{fig:P2_Scenario 15}
\end{figure}

\begin{figure*}[t]
	\centering
	\begin{subfigure}[]{.479\linewidth}
		\centering
		\includegraphics[width=1\linewidth]{ 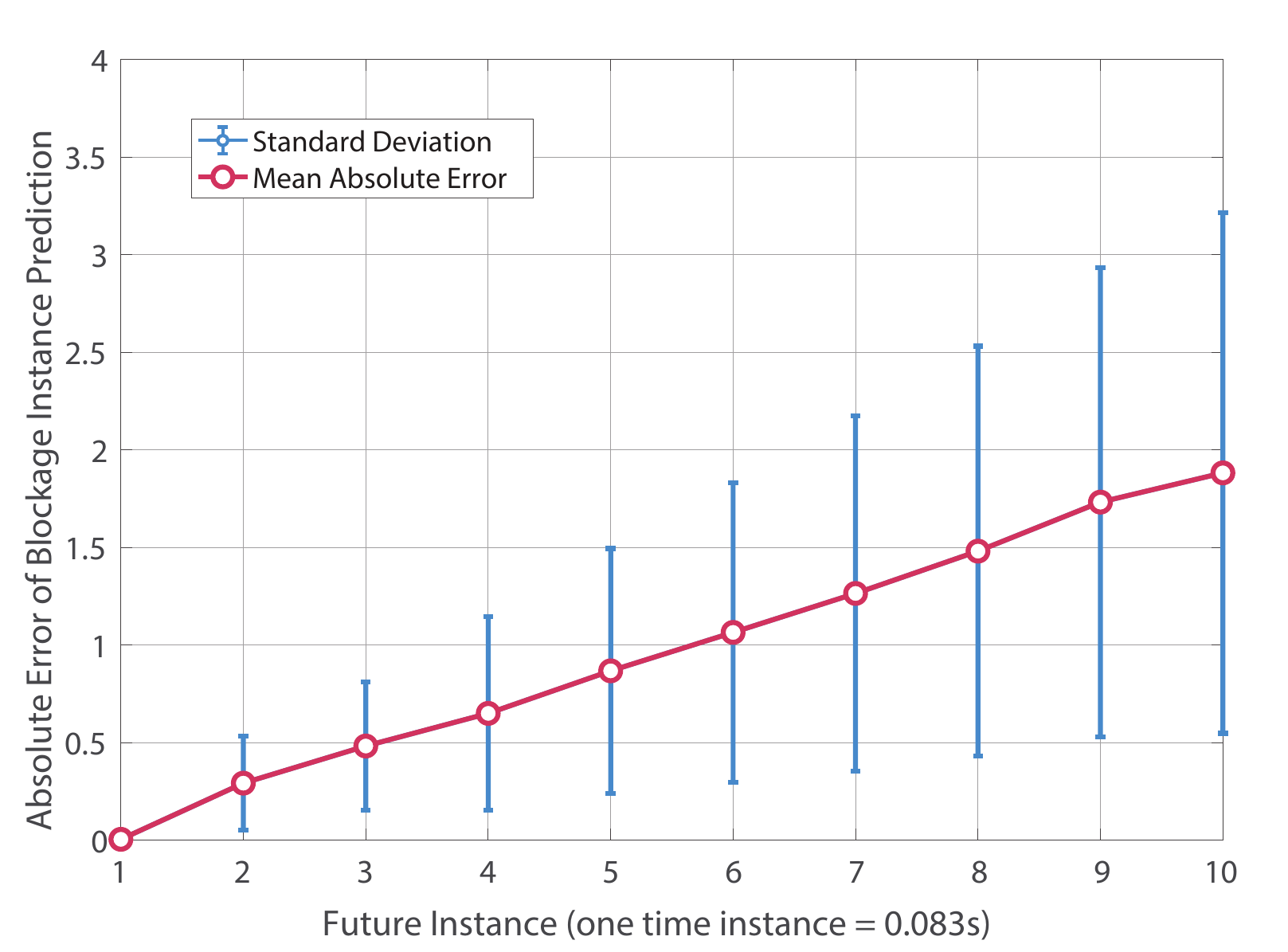}
		\caption{RNN model}
		\label{fig:P2_RNN}
	\end{subfigure}
	\begin{subfigure}[]{0.479\linewidth}
		\centering
		\includegraphics[width=1\linewidth]{ 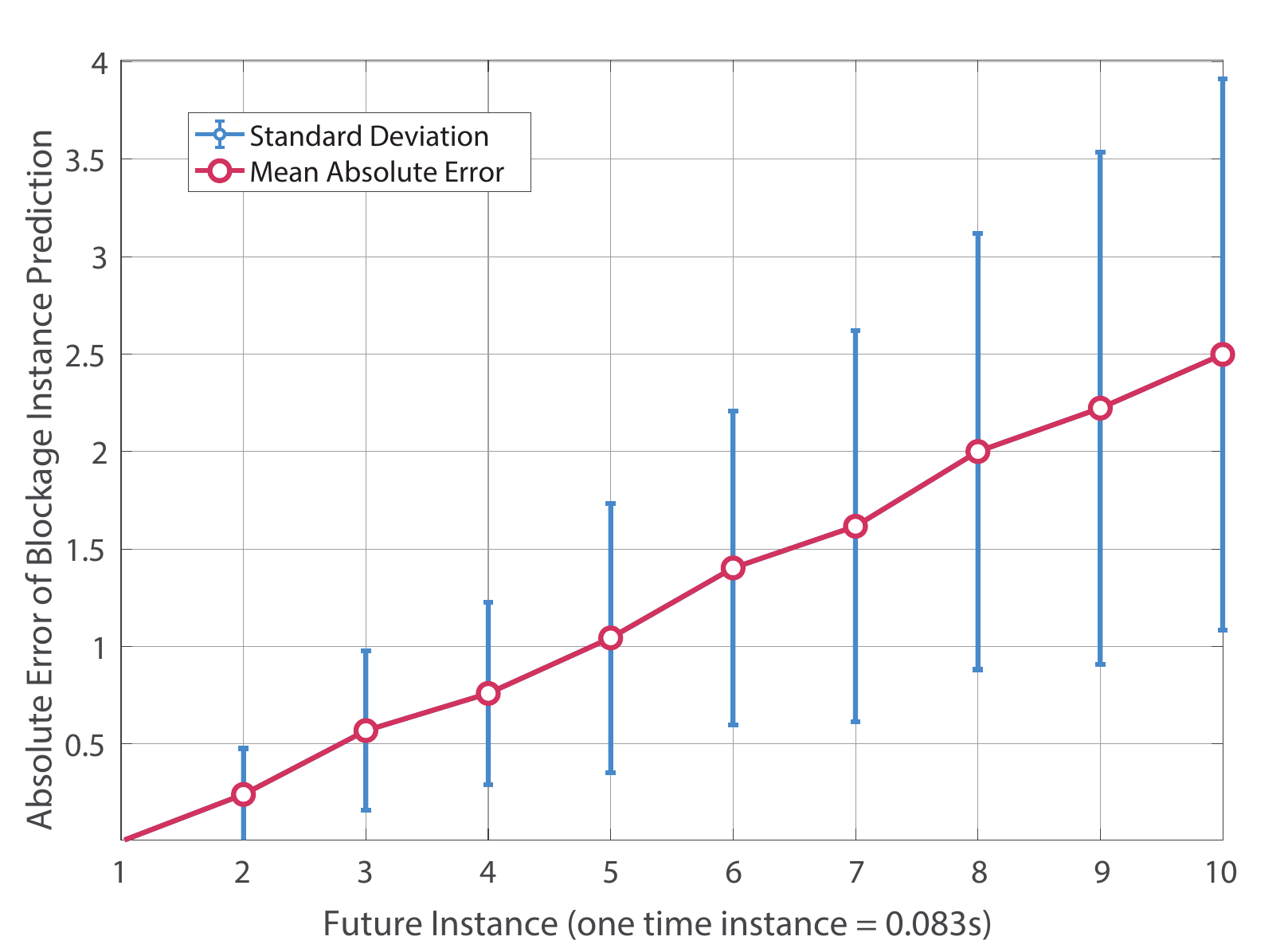}
		\caption{CNN model}
		\label{fig:P2_CNN}
	\end{subfigure}
	\caption{Mean absolute error between the target (the exact time instance where the blockage happens) and the prediction of this blockage time instance for different future prediction intervals using the outdoor scenario dataset.}
\end{figure*}

When comparing  the results of the indoor and outdoor blockage prediction scenarios, it is important to emphasize the following points: (i) The link blockage in the indoor setup is caused by a robot moving with a low speed (0.0625m/s) while the outdoor link blockages are due to high-speed moving vehicles (with speeds up to 35mph). Further, the sampling time in the indoor scenario is 0.88s compared to 0.083s in the outdoor setup. Therefore, predicitng a blockage happening after 10 samples in the indoor setup means predicting a blockage that is $8.8$ seconds in the future and $0.625$m away. For the outdoor blockage scenario, predicting a blockage happening after 10 samples means predicting a blockage $.83$ seconds before it happens while it is roughly $\sim 9$m away from the LOS link (for a 25mph vehicle). (ii) The receiver in the indoor blockage scenario uses a fixed horn antenna that is steered towards the transmitter. Hence, it only captures the receive power using one beamforming direction. For the outdoor setup, the receiver measures  the power across 64 beams in every time instance. Hence the pre-blockage wireless signature in the outdoor setup is generally richer and  is capable of addressing harder blockage prediction scenarios. This performance might be further improved by optimizing the design of the beamforming vectors/codebooks \cite{Taha2021, Zhang2021}.

\noindent {\textbf{Problem 2. Blockage Instance Prediction:}} In this problem, the model attempts to predict the exact future instance where the blockage will happen. Next, we present the experimental results for both the indoor and outdoor blockage scenarios. 

\begin{itemize}
	\item \textbf{Indoor Blockage:} To evaluate the performance of our model in predicting the future blockage instance, we plot the mean absolute error between predicted blockage instance and the ground-truth instance in  \figref{fig:P2_Scenario 15}. We also show the standard deviation as error bars on the same figure. First, this figure shows that our model can successfully predict the blockage instances with an average error of 2 samples for the blockages happening within 15 future time samples/instances.  This prediction also happens with relatively low volatility ($\pm1.5$). As the blockage happens further in the future, the blockage instance prediction error increases. Nevertheless, with blockages  40 time instance ahead in the future, our approach can still predict the exact blockage occurrence time with less than 8 instances in mean absolute error.
	
	\item \textbf{Outdoor Blockage:} The performance of our RNN and CNN models in predicting the future blockage instance  are shown in Figures  \ref{fig:P2_RNN}  and \ref{fig:P2_CNN}  for the outdoor setup. For the RNN model, it achieves a low mean absolute error under 1.9 time instances and volatility of $\pm1.2$ when the blockages are within 10 future time instances. The performance is slightly worse using the CNN model where the mean absolute error is under 2.5 with volatility $\pm1.5$ for the same scenario. 
\end{itemize}

\begin{figure}[t]
	\centering
	\begin{subfigure}[]{\linewidth}
		\centering
		\includegraphics[width=1\linewidth]{ 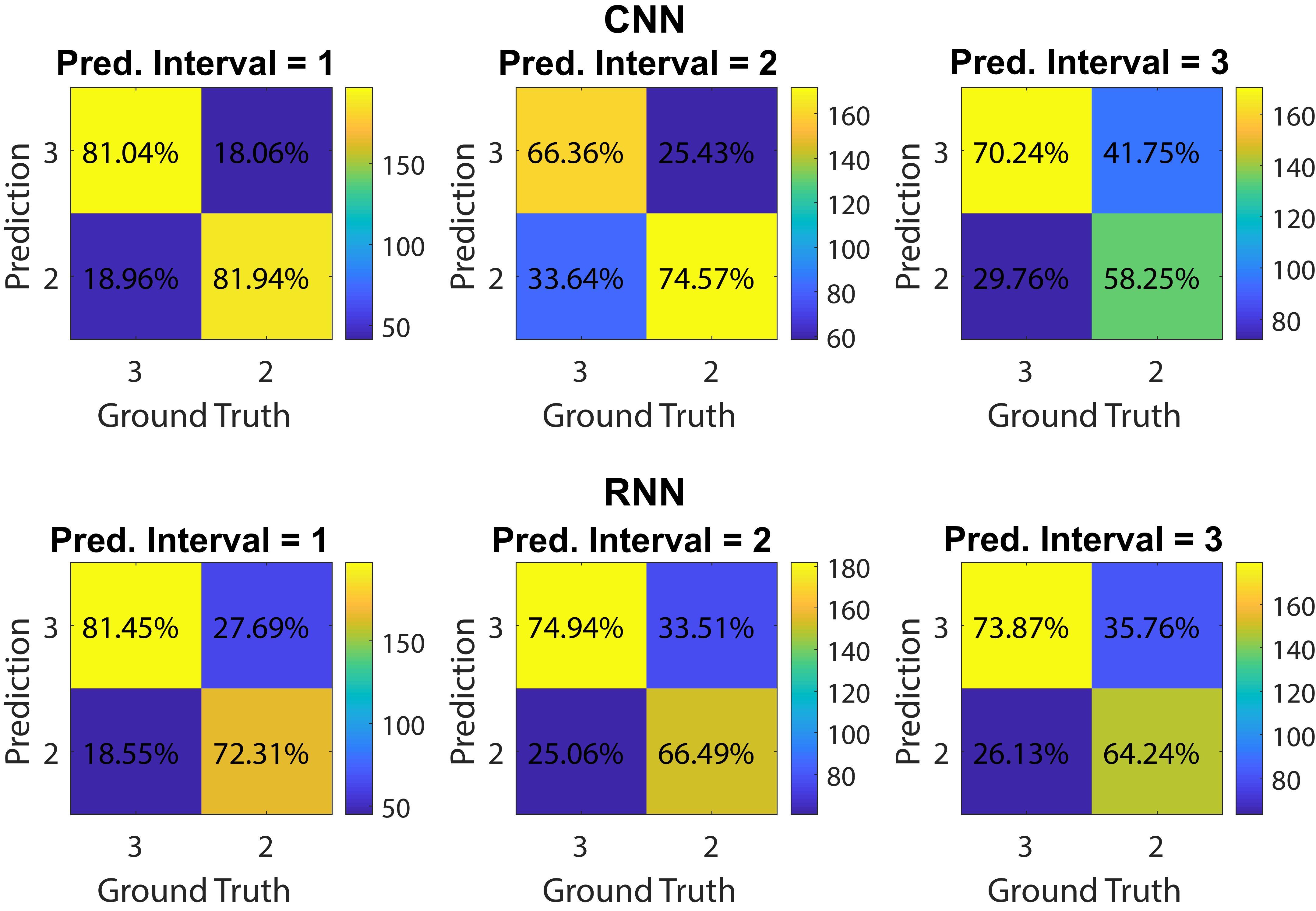}
		\caption{Blockage prediction with 2 severity levels}
		\label{fig:P3_2cls}
	\end{subfigure}
	\begin{subfigure}[]{\linewidth}
		\centering
		\includegraphics[width=1\linewidth]{ 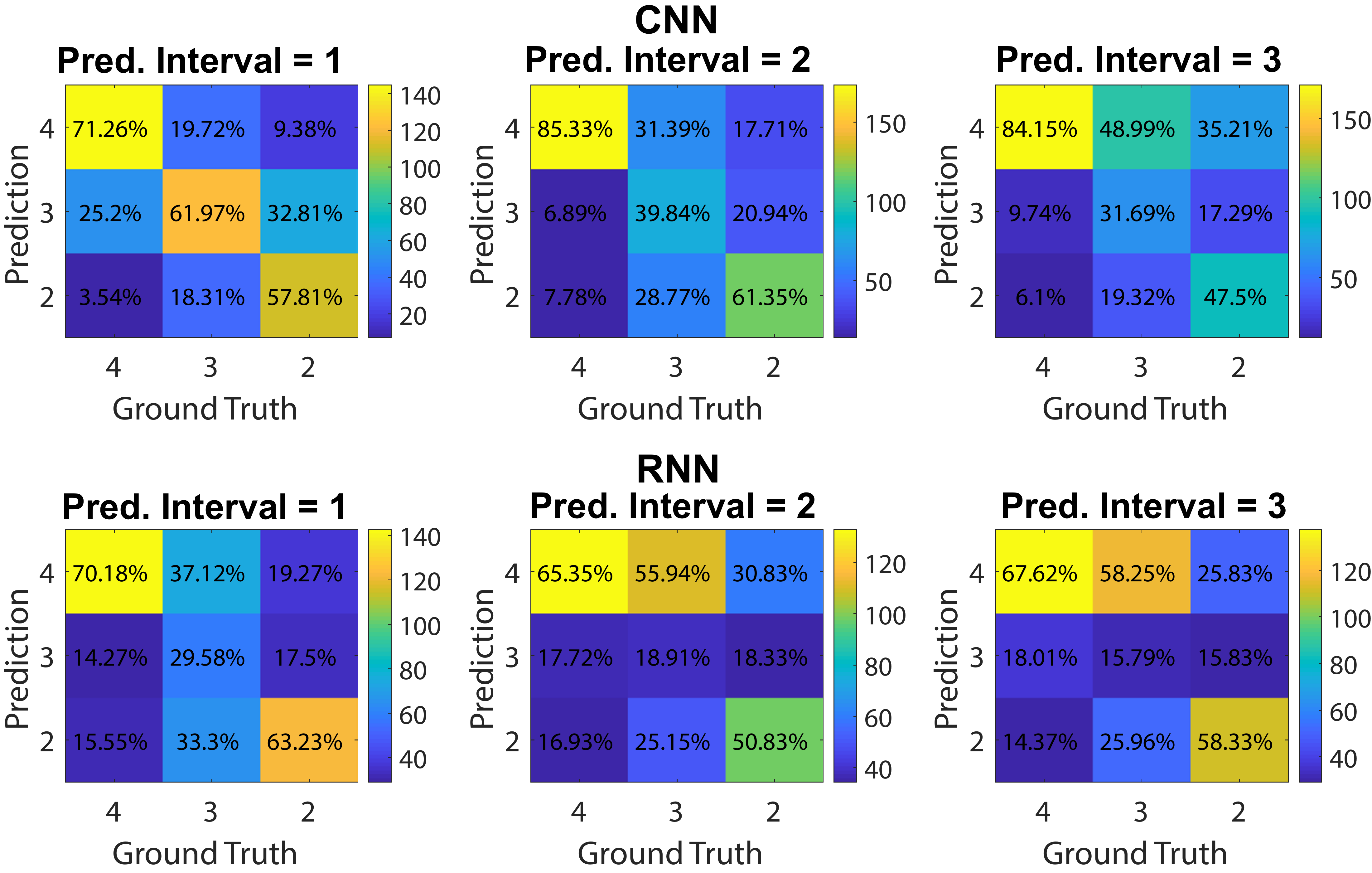}
		\caption{Blockage prediction with 3 severity levels}
		\label{fig:P3_3type}
	\end{subfigure}
	\caption{This figure shows confusion matrices for the severity level prediction with the RNN and CNN models using the outdoor scenario dataset in first 3 prediction time instances. Subfigure (a) illustrates the case where we consider the objects in the classes with severity levels 2 and 3 and Subfigure (b) illustrates the case when considering the objects in the classes with severity levels 2, 3 and 4.}
\end{figure}

Overall, these results highlight the feasibility of using the proposed approach to not just predict whether a blockage will happen in the near future or not, but to  also predict when this blockage will happen with a relatively good accuracy.

\begin{figure}[t]
	\centering
	\begin{subfigure}[]{\linewidth}
		\centering
		\includegraphics[width=1\linewidth]{ 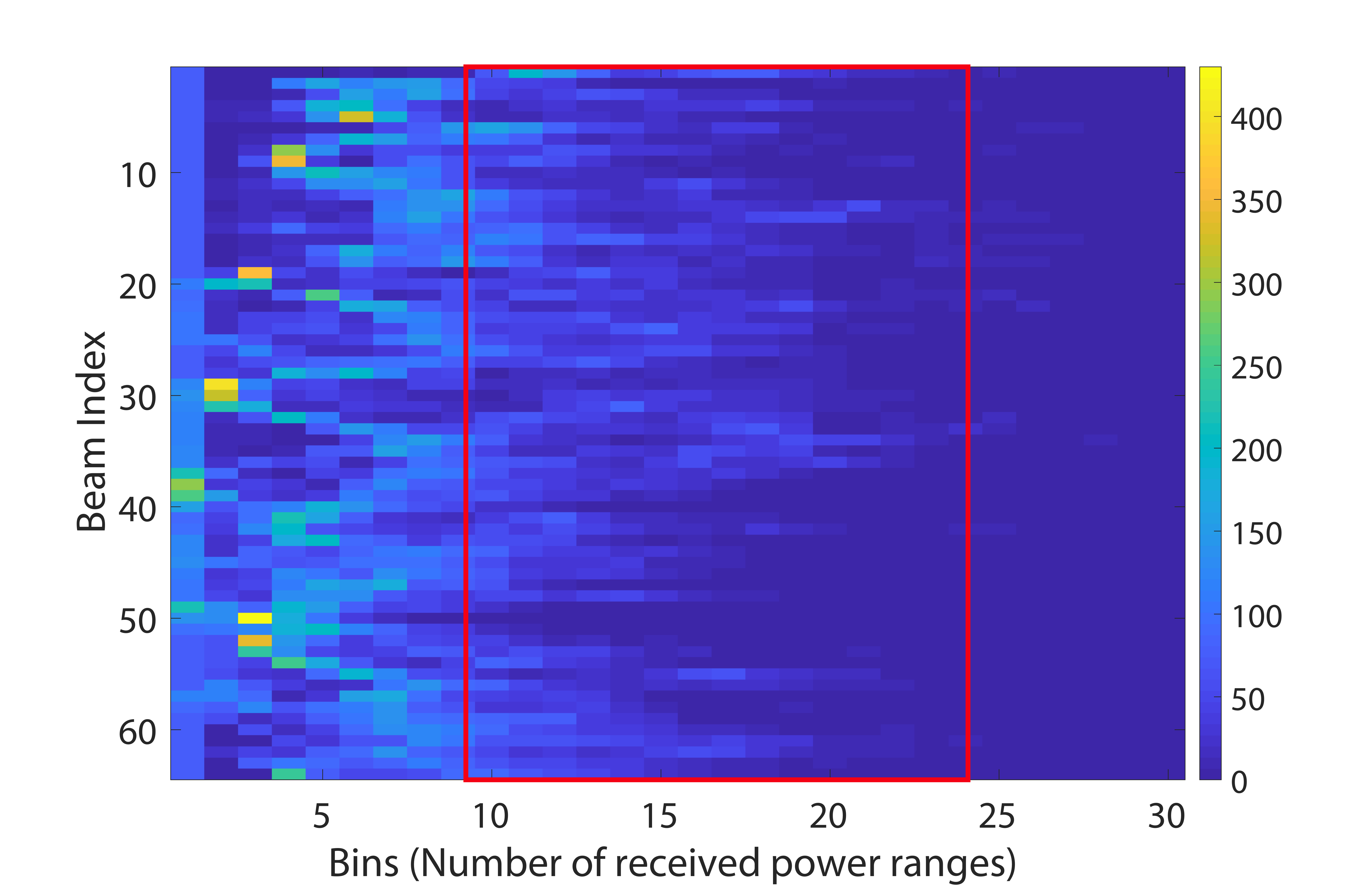}
		\caption{Histogram of the observation power  for objects with severity level 4}
		\label{fig:big_hist}
	\end{subfigure}
	\begin{subfigure}[]{\linewidth}
		\centering
		\includegraphics[width=1\linewidth]{ 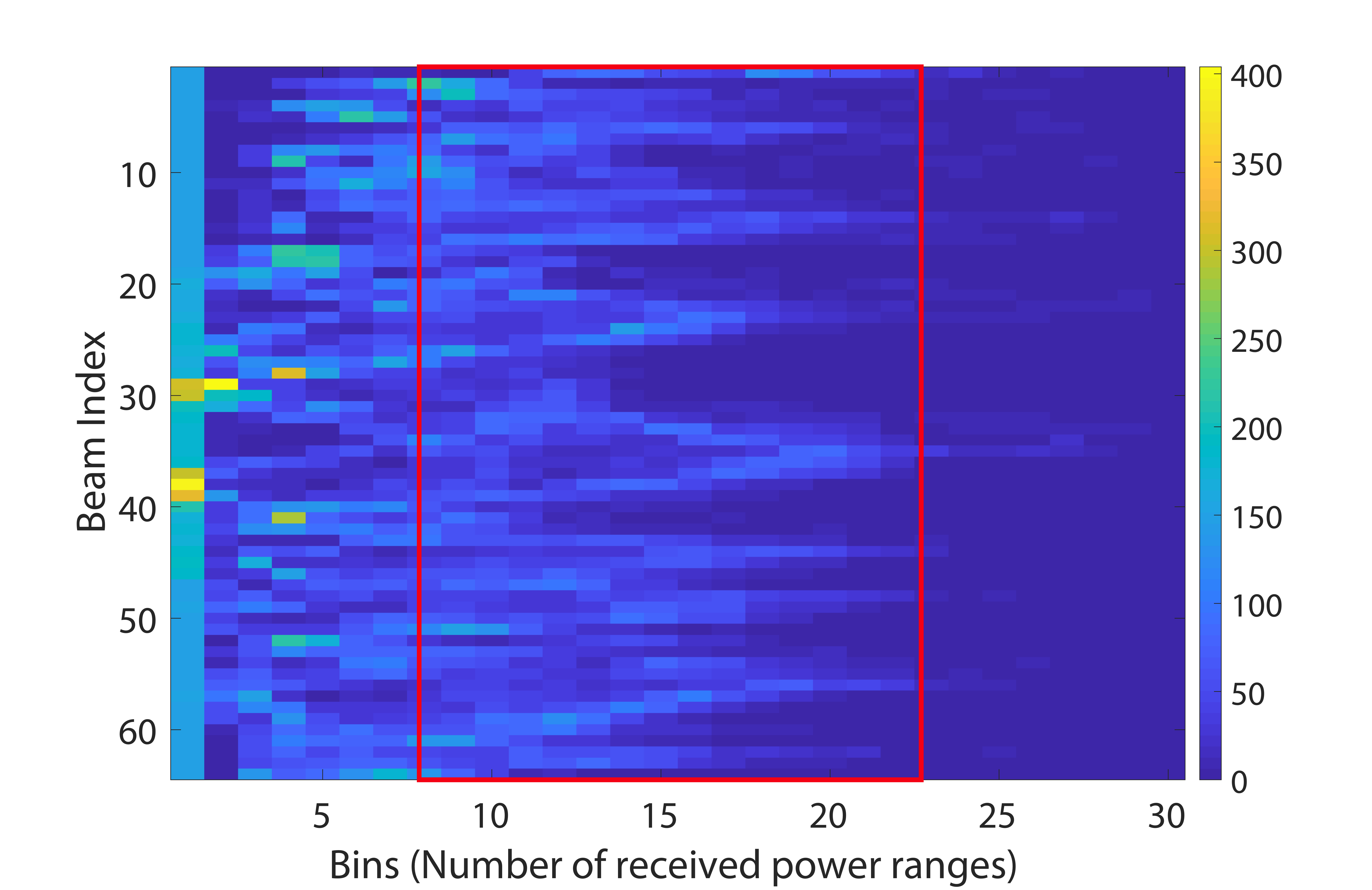}
		\caption{Histogram of the observation power  for objects with severity level 3}
		\label{fig:medium_hist}
	\end{subfigure}
	\begin{subfigure}[]{\linewidth}
		\centering
		\includegraphics[width=1\linewidth]{ 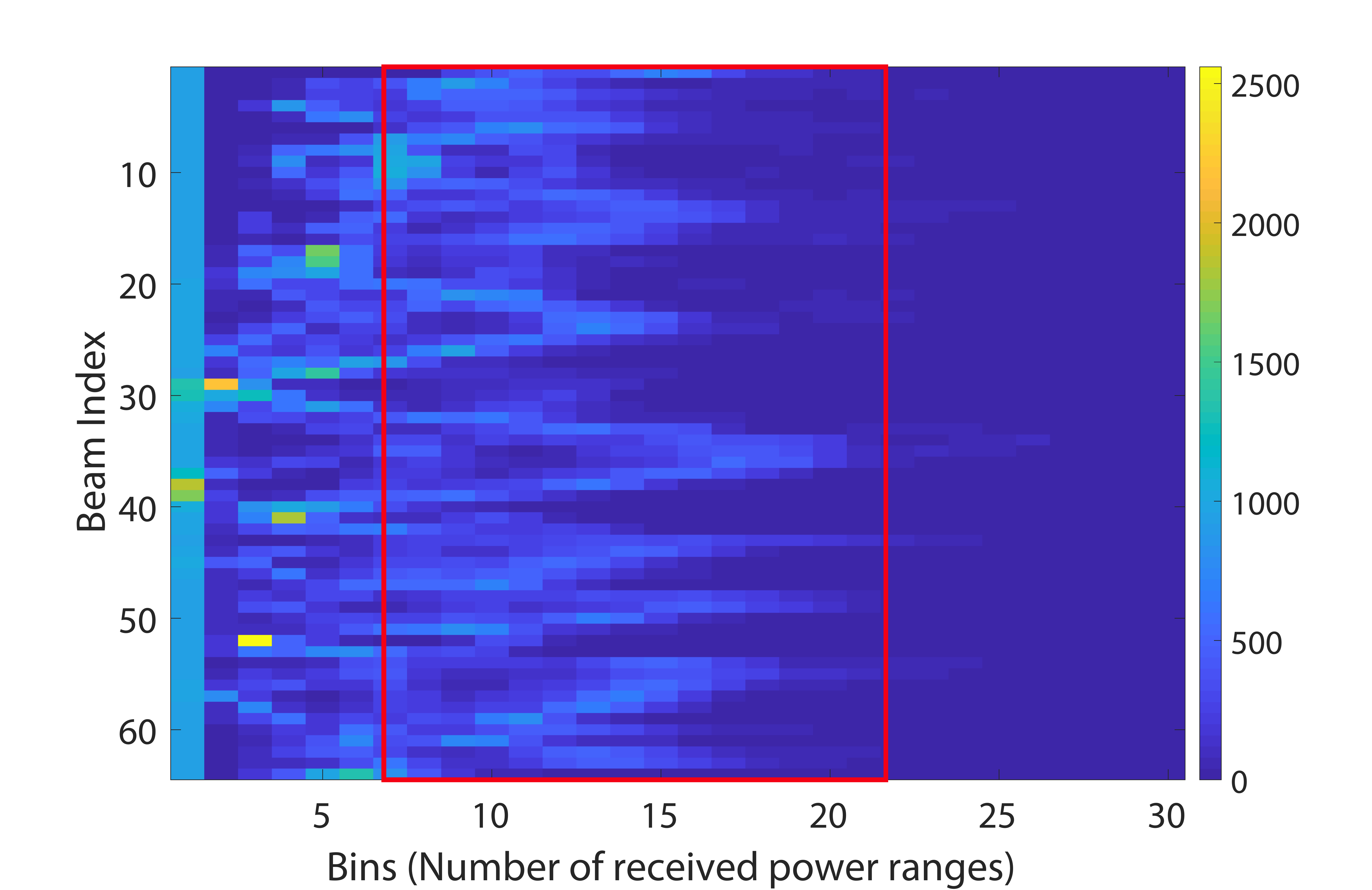}
		\caption{Histogram of the observation power  for objects with severity level 2}
		\label{fig:small_hist}
	\end{subfigure}
	
	\caption{Histogram of the received power sequences before the objects in severity level 2, 3, 4 block the LOS link. The difference of the three histograms patterns are shown within the red boxes. This figure shows the patterns of objects at severity level 3 and severity level 2 are similar. The pattern of objects at severity level 4 is quite different from the other two levels.}
	\label{fig:hist}
\end{figure}
% The bins are usually specified as consecutive, non-overlapping intervals of a variable.

\textbf{\textbf{Problem 3. Blockage Severity Prediction:}} The performance of  predicting the blockage severity is shown by the confusion matrix in \figref{fig:P3_2cls}. The vertical label is the prediction class and the horizontal label is the ground truth class.  We present results for future prediction interval of $t = $1, 2 and 3 time instances for the case $N_\mathrm{class} = 3$. The confusion matrices in the upper row are generated using the CNN model and the ones in the bottom row are generated using the RNN model. As the prediction time interval increases, the prediction accuracy decreases. However the accuracy is still above $90\%$ even when prediction time interval $t$ = 5 using the CNN model. For the RNN model, the accuracy is $91\%$, but it degrades to $86\%$ as the prediction time interval increases to 3. Both CNN and RNN models have the ability to distinguish between the objects in $T_{D2}$ and $T_{D3}$ with high accuracy. Since CNN detects the local features which is the main difference between the objects in  $T_{D2}$ and $T_{D3}$, it is reasonable that CNN performs better than RNN for solving this problem.

\begin{figure}[t]
	\centering
	\includegraphics[width=1\linewidth]{ 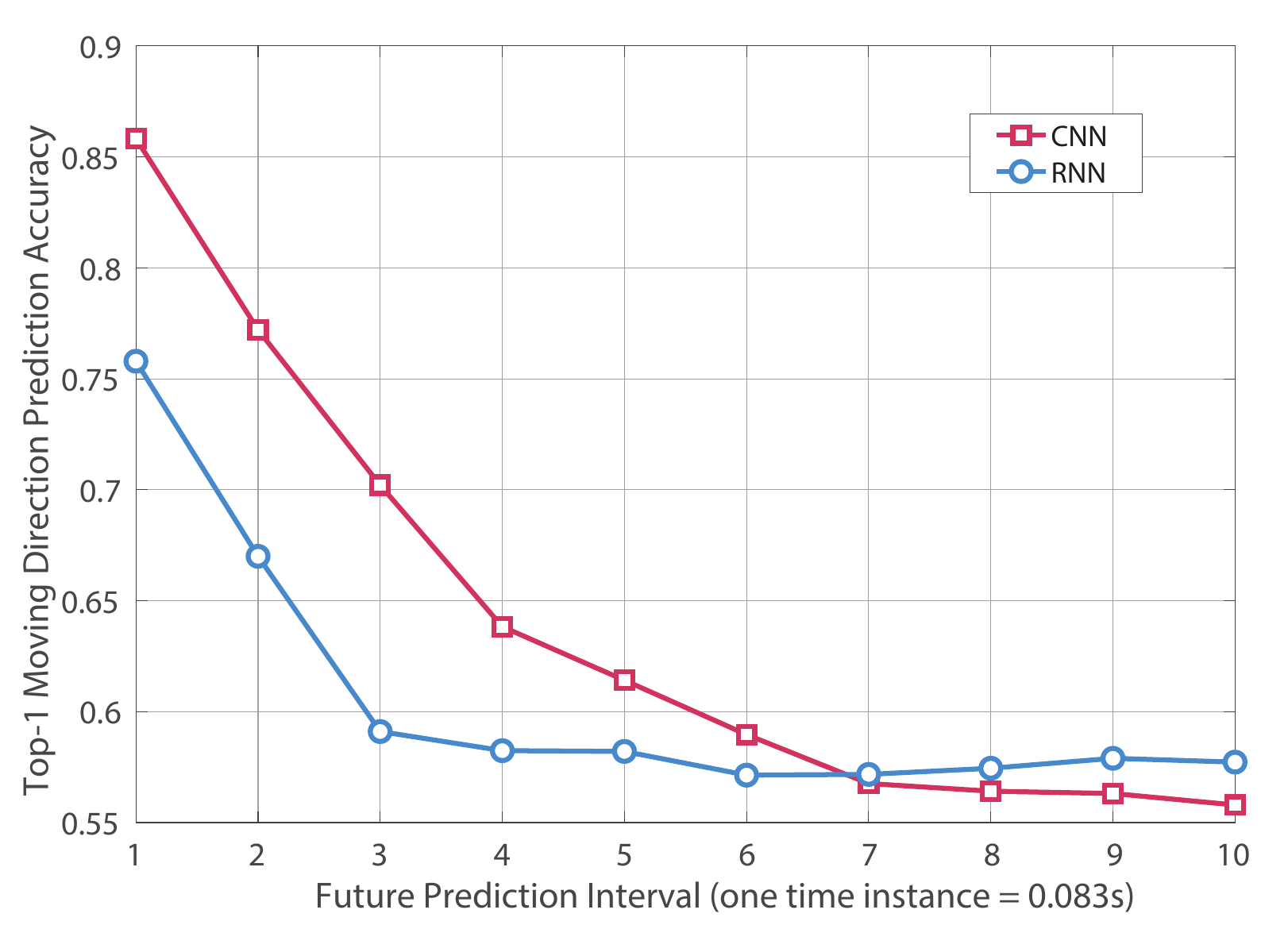}
	\caption{This figure plots the top-1 blockage direction prediction accuracy for different future prediction intervals.}
	\label{fig:P4}
\end{figure}

To better understand the blockage prediction problem, we take a step further. Since both RNN and CNN address the case $N_\mathrm{class} = 3$ with high accuracy, we extend it to the case that $N_\mathrm{class} = 4$  (presented in \sref{sec:dev_p3} and in \figref{fig:obj_tree}). The result of the case $N_\mathrm{class} = 4$ prediction problem is shown in \figref{fig:P3_3type}. Both RNN and CNN models can distinguish the objects in $T_{D4}$ with high accuracy of $89\%$. However, when it comes to  $T_{D2}$ and $T_{D3}$, the model cannot classify them very well. Since the shape of the front face of objects in $T_{D4}$ is quite different from objects in $T_{D2}$ and $T_{D3}$, the corresponding pre-blockage signature patterns of $T_{D4}$ vs $T_{D2}$ and $T_{D3}$ are different. However, the difference in the front face of objects in $T_{D2}$ and $T_{D3}$ is not significant, which results in similar patterns in the Pre-blockage signature. 

To further understand this, histograms of the observation power matrix of each class are shown in \figref{fig:hist}. Each received power bin is specified as consecutive, non-overlapping, equal intervals of received power, which is calculated by $\delta = (P_{max}-P_{min})/N$, $P_{max}$ and $P_{min}$ are the maximum and minimum value of the observation power matrix, $N$ is the total number of bins. Using $n$ as the index of the bins (30 in our case), the bound of $n$ bin is expressed as $[(n-1)*\delta, n*\delta]$ By comparing the three histograms, we see that the histogram of objects in $T_{D4}$ is quite different from objects in $T_{D2}$ and $T_{D3}$. However, the histograms of objects in $T_{D2}$ and $T_{D3}$ are similar. By analyzing both physical shapes and histograms of the three classes, we conclude that our model can distinguish the objects in $T_{D4}$ from the other two severity levels successfully, but it still needs additional effort to classify objects in $T_{D2}$ and $T_{D3}$.

\textbf{\textbf{Problem 4. Blockage Direction Prediction:}} The results for blockage moving direction prediction are shown in \figref{fig:P4}. We plot Top-1 moving blockage direction prediction accuracy as a function of the future prediction interval. The performance of the CNN model can achieve $85\%$ accuracy for the first time instance; however the accuracy degrades as the prediction interval increases from 1 to 4. It continues degrading with a smoother slope and flattens out at $56\%$ after 7 time instances. The performance of the RNN model drops during the first 3 time instances from $75\%$ to $58\%$, then it flattens out and stabilizes at $57\%$. CNN model performs better than RNN model in first 7 time instances and after that the performance is similar. Recall that for larger prediction intervals, there is less pre-blockage signature in the training dataset. This could explain the degradation in the first 3 time instances for the RNN model and the first 4 time instances for the CNN model. From 3 to 7 time instances, RNN can barely predict the moving direction of the objects, however, CNN can capture some pre-blockage signatures even if they are not obvious. As the future prediction interval continues increasing, the performance gap between CNN and RNN gradually narrows, because the pre-blockage signature becomes less effective.

\section{Conclusion}
\label{sec:con}
In this paper, we explored the potential of utilizing  mmWave received power data to proactively predict dynamic blockages in mmWave systems, thereby allowing the network to proactively manage hand-off/beam-switching decisions. We formulated the wireless signature blockage prediction problem and developed an efficient machine learning model for this task based on both CNN and RNN architectures. To validate the feasibility of the proposed approach, we constructed a large-scale real-world mmWave dataset and generated the development dataset for our neural network models using the collected data. Evaluation on this real-world dataset yields the following main takeaways: 
\begin{itemize}
	\item The mmWave pre-blockage signature approach achieves high accuracy ($85\%$ for indoor scenario and $97\%$ for outdoor scenario) for predicting future moving blockages that are within a short window, using our proposed machine learning models.
	\item Our approaches are able to predict incoming blockage above 75\% accuracy for outdoor scenario 3 time instance (249ms) ahead and for indoor scenario first 12 time instance (10.56s) ahead. 
	\item Our models are able to classify the shape of the future blockages with more than $90\%$ accuracy and classify the direction of the moving blockages with more than $85\%$ accuracy. 
\end{itemize}
These results highlight a promising solution for overcoming the blockage challenges and enhancing the reliability,  latency, and resource utilization efficiency of future mmWave/THz networks. In the future work, it is interesting to develop proactive blockage prediction approaches with the fusion of multiple data modalities (such as sub-6GHz channels \cite{Alrabeiah2020} and visual/radar data \cite{charan2021visionposition,Alrabeiah_URLL}). It is also interesting to extend the proposed solutions to scenarios with distributed sensing at both the infrastructure and user terminals.

%------------------------------------------------------------------------------------------------------------
%\linespread{1.3}
\balance

\end{document}

%% file: input.tex
\usepackage{amsfonts}
\usepackage{times}
\usepackage{graphicx}
\usepackage{latexsym}
\usepackage{dsfont}
\usepackage{amssymb}
\usepackage{amsmath}
\usepackage{cite}
\usepackage{verbatim}

\newcommand{\figref}[1]{{Fig.}~\ref{#1}}
\newcommand{\tabref}[1]{{Table}~\ref{#1}}

\def\bb0{{\mathbb{0}}}

\def\ba{{\mathbf{a}}}
\def\bb{{\mathbf{b}}}

\def\bff{{\mathbf{f}}}

\def\bh{{\mathbf{h}}}

\def\bm{{\mathbf{m}}}

\def\b0{{\mathbf{0}}}

\def\bA{{\mathbf{A}}}
\def\bB{{\mathbf{B}}}

\def\bI{{\mathbf{I}}}

\def\bR{{\mathbf{R}}}

\def\bbE{{\mathbb{E}}}

\def\cA{\mathcal{A}}

\def\cN{\mathcal{N}}

\def\sf0{{\mathsf{0}}}

%% file: arXiv.bbl
\begin{thebibliography}{10}
	\providecommand{\url}[1]{#1}
	\csname url@samestyle\endcsname
	\providecommand{\newblock}{\relax}
	\providecommand{\bibinfo}[2]{#2}
	\providecommand{\BIBentrySTDinterwordspacing}{\spaceskip=0pt\relax}
	\providecommand{\BIBentryALTinterwordstretchfactor}{4}
	\providecommand{\BIBentryALTinterwordspacing}{\spaceskip=\fontdimen2\font plus
		\BIBentryALTinterwordstretchfactor\fontdimen3\font minus
		\fontdimen4\font\relax}
	\providecommand{\BIBforeignlanguage}[2]{{%
			\expandafter\ifx\csname l@#1\endcsname\relax
			\typeout{** WARNING: IEEEtran.bst: No hyphenation pattern has been}%
			\typeout{** loaded for the language `#1'. Using the pattern for}%
			\typeout{** the default language instead.}%
			\else
			\language=\csname l@#1\endcsname
			\fi
			#2}}
	\providecommand{\BIBdecl}{\relax}
	\BIBdecl
	
	\bibitem{Wu2021}
	S.~Wu, M.~Alrabeiah, A.~Hredzak, C.~Chakrabarti, and A.~Alkhateeb, ``Deep
	learning for moving blockage prediction using real millimeter wave
	measurements,'' \emph{arXiv preprint arXiv:2101.06886}, 2021.
	
	\bibitem{Rappaport2019}
	T.~S. Rappaport, Y.~Xing, O.~Kanhere, S.~Ju, A.~Madanayake, S.~Mandal,
	A.~Alkhateeb, and G.~C. Trichopoulos, ``Wireless communications and
	applications above 100 {GHz}: Opportunities and challenges for {6G} and
	beyond,'' \emph{IEEE Access}, vol.~7, pp. 78\,729--78\,757, 2019.
	
	\bibitem{Rappaport2013a}
	T.~Rappaport, S.~Sun, R.~Mayzus, H.~Zhao, Y.~Azar, K.~Wang, G.~Wong, J.~Schulz,
	M.~Samimi, and F.~Gutierrez, ``Millimeter wave mobile communications for {5G}
	cellular: It will work!'' \emph{IEEE Access}, vol.~1, pp. 335--349, May 2013.
	
	\bibitem{Alkhateeb2018}
	A.~{Alkhateeb}, I.~{Beltagy}, and S.~{Alex}, ``Machine learning for reliable
	mmwave systems: Blockage prediction and proactive handoff,'' in \emph{in
		{Proc.} of IEEE Global Conference on Signal and Information Processing
		(GlobalSIP)}, Nov 2018, pp. 1055--1059.
	
	\bibitem{Alrabeiah2020a}
	M.~{Alrabeiah}, A.~{Hredzak}, and A.~{Alkhateeb}, ``Millimeter wave base
	stations with cameras: Vision-aided beam and blockage prediction,'' in
	\emph{2020 IEEE 91st Vehicular Technology Conference (VTC2020-Spring)}, 2020,
	pp. 1--5.
	
	\bibitem{giordani2016multi}
	M.~Giordani, M.~Mezzavilla, S.~Rangan, and M.~Zorzi, ``Multi-connectivity in
	{5G} mmwave cellular networks,'' in \emph{2016 Mediterranean Ad Hoc
		Networking Workshop (Med-Hoc-Net)}.\hskip 1em plus 0.5em minus 0.4em\relax
	IEEE, 2016, pp. 1--7.
	
	\bibitem{polese2017improved}
	M.~Polese, M.~Giordani, M.~Mezzavilla, S.~Rangan, and M.~Zorzi, ``Improved
	handover through dual connectivity in {5G} mmwave mobile networks,''
	\emph{IEEE Journal on Selected Areas in Communications}, vol.~35, no.~9, pp.
	2069--2084, 2017.
	
	\bibitem{petrov2017dynamic}
	V.~Petrov, D.~Solomitckii, A.~Samuylov, M.~A. Lema, M.~Gapeyenko,
	D.~Moltchanov, S.~Andreev, V.~Naumov, K.~Samouylov, M.~Dohler \emph{et~al.},
	``Dynamic multi-connectivity performance in ultra-dense urban mmwave
	deployments,'' \emph{IEEE Journal on Selected Areas in Communications},
	vol.~35, no.~9, pp. 2038--2055, 2017.
	
	\bibitem{aziz2016architecture}
	D.~Aziz, J.~Gebert, A.~Ambrosy, H.~Bakker, and H.~Halbauer, ``Architecture
	approaches for 5g millimetre wave access assisted by 5g low-band using
	multi-connectivity,'' in \emph{2016 IEEE Globecom Workshops (GC
		Wkshps)}.\hskip 1em plus 0.5em minus 0.4em\relax IEEE, 2016, pp. 1--6.
	
	\bibitem{Mismar2021}
	F.~B. Mismar, A.~Alammouri, A.~Alkhateeb, J.~G. Andrews, and B.~L. Evans,
	``Deep learning predictive band switching in wireless networks,'' \emph{IEEE
		Transactions on Wireless Communications}, vol.~20, no.~1, pp. 96--109, 2021.
	
	\bibitem{alrabeiah2020deep}
	M.~Alrabeiah and A.~Alkhateeb, ``Deep learning for {mmWave} beam and blockage
	prediction using {sub-6 GHz} channels,'' \emph{IEEE Transactions on
		Communications}, vol.~68, no.~9, pp. 5504--5518, 2020.
	
	\bibitem{Ali2020}
	Z.~{Ali}, A.~{Duel-Hallen}, and H.~{Hallen}, ``Early warning of mmwave signal
	blockage and aoa transition using {sub-6 GHz} observations,'' \emph{IEEE
		Communications Letters}, vol.~24, no.~1, pp. 207--211, 2020.
	
	\bibitem{Alrabeiah2019a}
	M.~Alrabeiah and A.~Alkhateeb, ``Deep learning for {TDD} and {FDD} massive
	mimo: Mapping channels in space and frequency,'' in \emph{in Proc. of
		Asilomar Conference on Signals, Systems, and Computers}, 2019, pp.
	1465--1470.
	
	\bibitem{charan2020}
	G.~Charan, M.~Alrabeiah, and A.~Alkhateeb, ``Vision-aided {6G} wireless
	communications: Blockage prediction and proactive handoff,'' \emph{IEEE
		Transactions on Vehicular Technology}, vol.~70, no.~10, pp. 10\,193--10\,208,
	2021.
	
	\bibitem{choi2017deep}
	J.-S. Choi, W.-H. Lee, J.-H. Lee, J.-H. Lee, and S.-C. Kim, ``Deep learning
	based nlos identification with commodity wlan devices,'' \emph{IEEE
		Transactions on Vehicular Technology}, vol.~67, no.~4, pp. 3295--3303, 2017.
	
	\bibitem{huang2020machine}
	C.~Huang, A.~F. Molisch, R.~He, R.~Wang, P.~Tang, B.~Ai, and Z.~Zhong,
	``Machine learning-enabled los/nlos identification for mimo system in dynamic
	environment,'' \emph{IEEE Transactions on Wireless Communications}, 2020.
	
	\bibitem{ImageNet}
	O.~Russakovsky, J.~Deng, H.~Su, J.~Krause, S.~Satheesh, S.~Ma, Z.~Huang,
	A.~Karpathy, A.~Khosla, M.~Bernstein \emph{et~al.}, ``Imagenet large scale
	visual recognition challenge,'' \emph{International journal of computer
		vision}, vol. 115, no.~3, pp. 211--252, 2015.
	
	\bibitem{goldsmith2005}
	A.~Goldsmith, \emph{Wireless communications}.\hskip 1em plus 0.5em minus
	0.4em\relax Cambridge university press, 2005.
	
	\bibitem{resnet}
	K.~He, X.~Zhang, S.~Ren, and J.~Sun, ``Deep residual learning for image
	recognition,'' in \emph{Proceedings of the IEEE conference on computer vision
		and pattern recognition}, 2016, pp. 770--778.
	
	\bibitem{Liu2019}
	Z.~{Liu}, B.~{Xiao}, M.~{Alrabeiah}, K.~{Wang}, and J.~{Chen}, ``Single image
	dehazing with a generic model-agnostic convolutional neural network,''
	\emph{IEEE Signal Processing Letters}, vol.~26, no.~6, pp. 833--837, 2019.
	
	\bibitem{Yolo}
	J.~Redmon, S.~Divvala, R.~Girshick, and A.~Farhadi, ``You only look once:
	Unified, real-time object detection,'' in \emph{Proceedings of the IEEE
		Conference on Computer Vision and Pattern Recognition (CVPR)}, June 2016.
	
	\bibitem{SpeechRecog}
	W.~{Chan}, N.~{Jaitly}, Q.~{Le}, and O.~{Vinyals}, ``Listen, attend and spell:
	A neural network for large vocabulary conversational speech recognition,'' in
	\emph{2016 IEEE International Conference on Acoustics, Speech and Signal
		Processing (ICASSP)}, 2016, pp. 4960--4964.
	
	\bibitem{SpeechRecog2}
	A.~Graves and N.~Jaitly, ``Towards end-to-end speech recognition with recurrent
	neural networks,'' in \emph{International conference on machine learning},
	2014, pp. 1764--1772.
	
	\bibitem{DLBook}
	\BIBentryALTinterwordspacing
	I.~Goodfellow, Y.~Bengio, and A.~Courville, ``Deep learning,'' 2016, book in
	preparation for MIT Press. [Online]. Available:
	\url{http://www.deeplearningbook.org}
	\BIBentrySTDinterwordspacing
	
	\bibitem{ma2016}
	X.~Ma and E.~Hovy, ``End-to-end sequence labeling via bi-directional
	lstm-cnns-crf,'' \emph{arXiv preprint arXiv:1603.01354}, 2016.
	
	\bibitem{2014gru}
	K.~Cho, B.~Van~Merri{\"e}nboer, D.~Bahdanau, and Y.~Bengio, ``On the properties
	of neural machine translation: Encoder-decoder approaches,'' \emph{arXiv
		preprint arXiv:1409.1259}, 2014.
	
	\bibitem{simonyan2014CNN}
	K.~Simonyan and A.~Zisserman, ``Very deep convolutional networks for
	large-scale image recognition,'' \emph{arXiv preprint arXiv:1409.1556}, 2014.
	
	\bibitem{EffBackProp}
	Y.~A. LeCun, L.~Bottou, G.~B. Orr, and K.-R. M{\"u}ller, ``Efficient
	backprop,'' in \emph{Neural networks: Tricks of the trade}.\hskip 1em plus
	0.5em minus 0.4em\relax Springer, 2012, pp. 9--48.
	
	\bibitem{DeepSense}
	\BIBentryALTinterwordspacing
	A.~Alkhateeb, G.~Charan, M.~Alrabeiah, T.~Osman, A.~Hredzak, N.~Srinivas, and
	M.~Seth, ``{DeepSense 6G}: Real-world multi-modal sensing and {CSI} dataset
	for {6G} deep learning research,'' 2021. [Online]. Available:
	\url{https://www.DeepSense6G.net}
	\BIBentrySTDinterwordspacing
	
	\bibitem{bishop2006pattern}
	C.~M. Bishop, ``Pattern recognition,'' \emph{Machine learning}, vol. 128,
	no.~9, 2006.
	
	\bibitem{pytorch}
	A.~Paszke \emph{et~al.}, ``Pytorch: An imperative style, high-performance deep
	learning library,'' in \emph{Advances in Neural Information Processing
		Systems 32}.\hskip 1em plus 0.5em minus 0.4em\relax Curran Associates, Inc.,
	2019, pp. 8024--8035.
	
	\bibitem{Taha2021}
	A.~Taha, Q.~Qu, S.~Alex, P.~Wang, W.~L. Abbott, and A.~Alkhateeb, ``Millimeter
	wave {MIMO}-based depth maps for wireless virtual and augmented reality,''
	\emph{IEEE Access}, vol.~9, pp. 48\,341--48\,363, 2021.
	
	\bibitem{Zhang2021}
	Y.~Zhang, M.~Alrabeiah, and A.~Alkhateeb, ``Reinforcement learning of beam
	codebooks in millimeter wave and terahertz {MIMO} systems,'' \emph{IEEE
		Transactions on Communications}, pp. 1--1, 2021.
	
	\bibitem{Alrabeiah2020}
	M.~{Alrabeiah} and A.~{Alkhateeb}, ``Deep learning for mmwave beam and blockage
	prediction using {Sub-6GHz} channels,'' \emph{IEEE Transactions on
		Communications}, pp. 1--1, 2020.
	
	\bibitem{charan2021visionposition}
	G.~Charan, T.~Osman, A.~Hredzak, N.~Thawdar, and A.~Alkhateeb,
	``Vision-position multi-modal beam prediction using real millimeter wave
	datasets,'' \emph{arXiv preprint arXiv:2111.07574}, 2021.
	
	\bibitem{Alrabeiah_URLL}
	M.~Alrabeiah, U.~Demirhan, A.~Hredzak, and A.~Alkhateeb, ``Vision aided {URLL}
	communications: Proactive service identification and coexistence,'' in
	\emph{2020 54th Asilomar Conference on Signals, Systems, and Computers},
	2020, pp. 174--178.
	
\end{thebibliography}
